\documentclass[final,3p,fleqn,11pt]{elsarticle}

\usepackage{amssymb}
\usepackage{amsmath}
\allowdisplaybreaks
\usepackage{multicol}
\usepackage{pifont}
\usepackage[dvipsnames]{xcolor}
\usepackage{colortbl}
\usepackage{textcomp}

\journal{Computer Methods in Applied Mechanics and Engineering}

\usepackage{amsmath}
\usepackage{setspace}
\usepackage{amsfonts}
\usepackage{amsthm}
\usepackage{mathtools}
\usepackage{subfig}
\usepackage{lineno}
\usepackage{algorithm}
\usepackage{algpseudocode}
\usepackage{hyperref}
\usepackage{float}
\usepackage{array,multirow}
\usepackage{color}
\usepackage{tikz}
\usepackage{graphicx}
\usepackage[flushleft]{threeparttable}
\usepackage{blindtext}
\usepackage[colorinlistoftodos,prependcaption,textsize=tiny]{todonotes}
\usepackage{regexpatch}
\usepackage[export]{adjustbox}
\usepackage{xcolor}

\definecolor{lightgray}{gray}{0.80}

\usetikzlibrary{decorations.pathreplacing}
\usepackage[listings,theorems,skins,breakable]{tcolorbox}
\tcbset{skin=enhanced}
\newtcolorbox{lbracebox}[1][Word]{%
   frame hidden,enlarge left by=2cm,width=\linewidth-2cm,%
  overlay unbroken = {\draw [decorate,decoration={brace,amplitude=10pt},]%
                     (frame.south west)-- (frame.north west)
                    node [black,midway,left,xshift=-.6cm] {#1};},%
}

\usepackage{scalerel,amssymb}

\usepackage{svg}
\graphicspath{{./sections/figures/svg/}}

\usepackage{tabularx}
\newcommand{\Bezier}{B\'ezier~}

\definecolor{Pred}{RGB}{255,105,97}
\definecolor{blue1}{RGB}{65,105,255}
\definecolor{red1}{RGB}{217,83,25}
\definecolor{green1}{RGB}{119,172,48}
\definecolor{purple1}{RGB}{176,156,217}
\definecolor{orange1}{RGB}{237,177,32}

\usepackage{color, soul}
\definecolor{blue2}{RGB}{125, 249, 255}
\definecolor{green2}{RGB}{90,250,90}

\DeclareRobustCommand{\reviewerI}[1]{{\sethlcolor{pink}\hl{#1}}}

\DeclareRobustCommand{\reviewerII}[1]{{\sethlcolor{yellow}\hl{#1}}}

\DeclareRobustCommand{\reviewerIII}[1]{{\sethlcolor{green2}\hl{#1}}}

\DeclareRobustCommand{\reviewerIV}[1]{{\sethlcolor{blue2}\hl{#1}}} 

\soulregister\reviewerI{1}
\soulregister\reviewerII{1}
\soulregister\reviewerIII{1}
\soulregister\reviewerIV{1}

\soulregister\ref7
\soulregister\pageref7
\soulregister\eqref7
\soulregister\cite7

\makeatletter
\xpatchcmd{\@todo}{\setkeys{todonotes}{#1}}{\setkeys{todonotes}{inline,#1}}{}{}
\makeatother

\theoremstyle{plain}
\newtheorem{theorem}{Theorem}[section]

\newtheorem{remark}[theorem]{Remark}

\theoremstyle{definition}

\newcommand{\mat}[1]{\mathbf{#1}} 
\newcommand{\mmat}[1]{\mathsf{#1}}

\newcommand{\eigenvec}{U}    

\modulolinenumbers[5]

\begin{document}

\begin{frontmatter}

\title{Leveraging spectral analysis to elucidate membrane locking and unlocking in isogeometric finite element formulations of the curved Euler-Bernoulli beam} 
\corref{cor1}
\author[ibnm]{Thi-Hoa Nguyen}
\ead{hoa.nguyen@ibnm.uni-hannover.de}

\author[ibnm]{Ren\'e R. Hiemstra}
\ead{rene.hiemstra@ibnm.uni-hannover.de}

\author[ibnm]{Dominik Schillinger}
\ead{schillinger@ibnm.uni-hannover.de}

\cortext[cor1]{Corresponding author}
\address[ibnm]{Institute of Mechanics and Computational Mechanics, Leibniz University Hannover, Germany}

\begin{abstract}
In this paper, we take a fresh look at using spectral analysis for assessing locking phenomena in finite element formulations. We propose to ``measure'' locking by comparing the difference between eigenvalue and mode error curves computed on coarse meshes with ``asymptotic'' error curves computed on ``overkill'' meshes, both plotted with respect to the normalized mode number.
To demonstrate the intimate relation between membrane locking and spectral accuracy, we focus on the example of a circular ring discretized with isogeometric curved Euler-Bernoulli beam elements. We show that the transverse-displacement-dominating modes are locking-prone, while the circumferential-displacement-dominating modes are naturally locking-free. We use eigenvalue and mode errors to assess five isogeometric finite element formulations in terms of their locking-related efficiency: the displacement-based formulation with full and reduced integration and three locking-free formulations based on the B-bar, discrete strain gap and Hellinger-Reissner methods. Our study shows that spectral analysis uncovers locking-related effects across the spectrum of eigenvalues and eigenmodes, rigorously characterizing membrane locking in the displacement-based formulation and unlocking in the locking-free formulations. 
\end{abstract}

\begin{highlights}
\item We propose to “measure” locking by comparing eigenvalue and mode error curves on coarse meshes with “asymptotic” error curves on “overkill” meshes.
\item We illustrate this concept for a circular ring discretized with isogeometric curved Euler-Bernoulli beam elements.
\item We assess the displacement-based formulation with full and reduced integration, and the B-bar, discrete strain gap and Hellinger-Reissner methods.
\item Our study shows that spectral analysis can rigorously characterize membrane locking and unlocking for Euler-Bernoulli beam formulations. 
\end{highlights}

\begin{keyword}
Spectral analysis \sep Membrane locking \sep Euler-Bernoulli curved beam elements \sep Closed circular ring \sep Isogeometric analysis 
\end{keyword}

\end{frontmatter}


\linenumbers


\newdefinition{rmk}{Remark}

\section{Introduction}\label{sec:introduction}

\begin{figure}[!htb]
	\centering
	\begin{minipage}[b]{0.49\textwidth}
	\centering
	\includegraphics[width=1.0\textwidth]{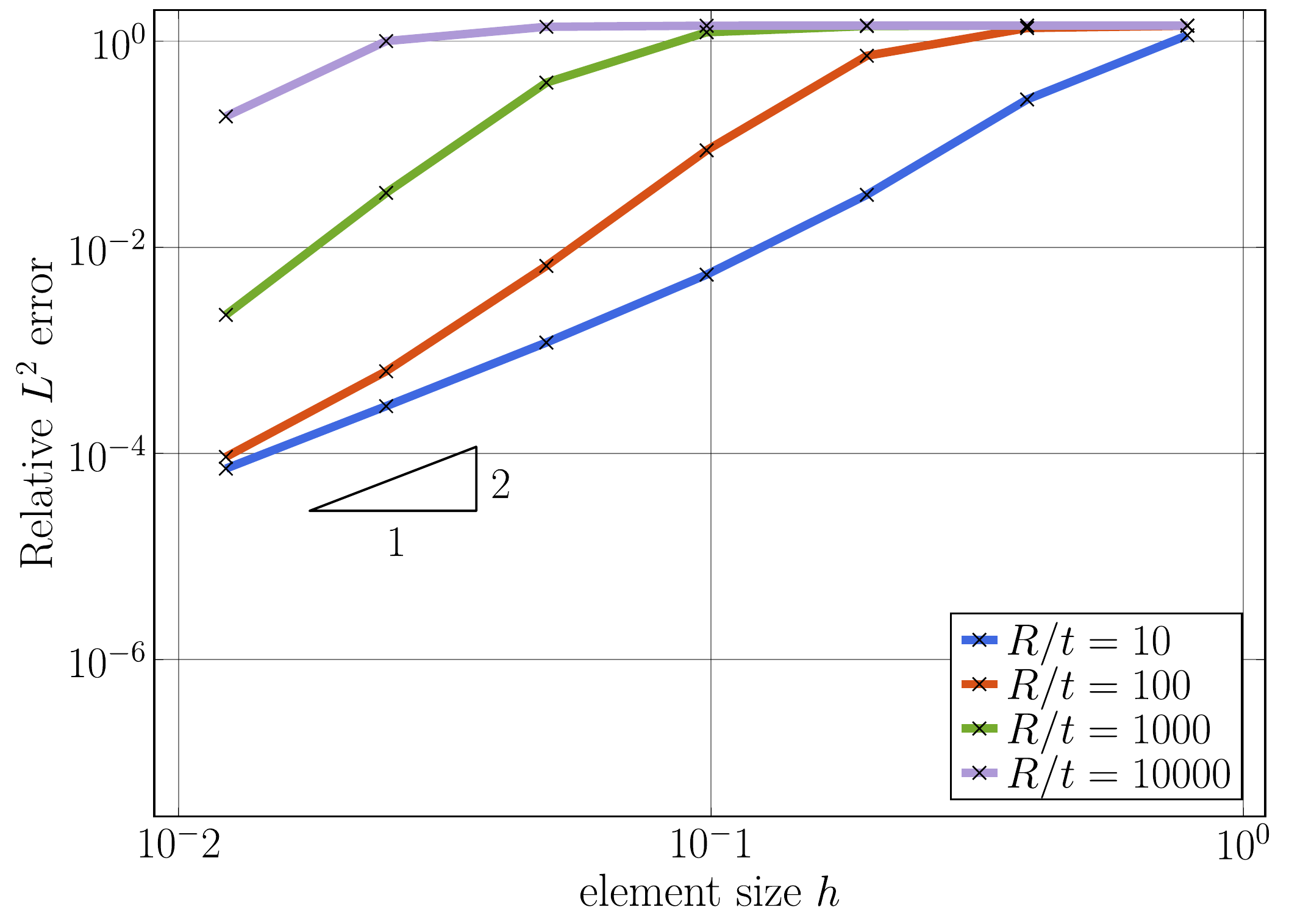}
	\footnotesize{(a) Increasing slenderness $R/t$, $p = 2$}
	\end{minipage}
	\begin{minipage}[b]{0.49\textwidth}
	\centering
	\includegraphics[width=1.0\textwidth]{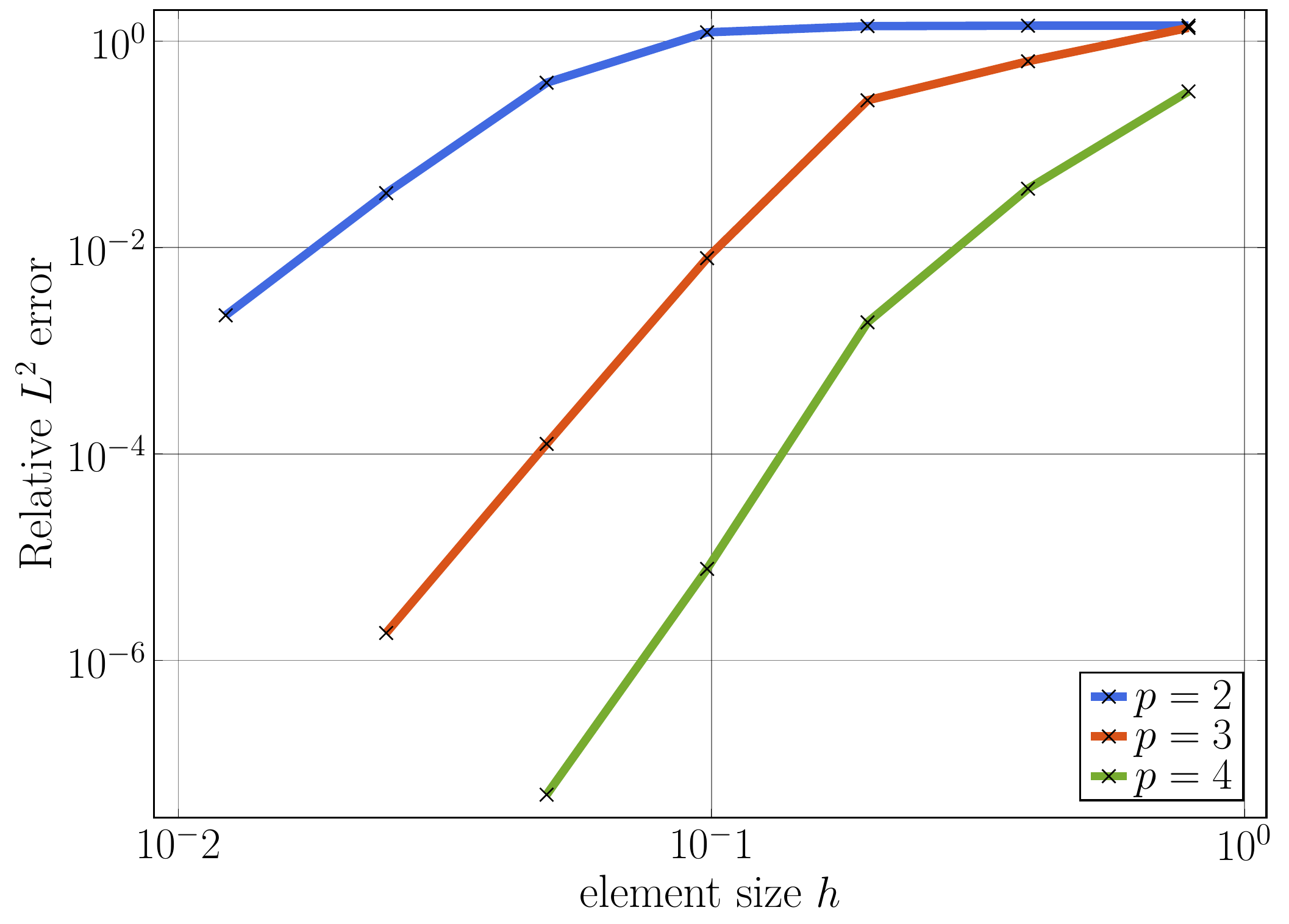}
	\footnotesize{(b) Increasing $p$, $R/t = 1\cdot10^3$}
	\end{minipage}
	\caption{Convergence of the relative $L^2$-norm errors of the displacements field of a curved Euler-Bernoulli beam (quarter circle cantilever, unit shear force at the free end, radius $R$, thickness $t$), discretized with B-splines of degree $p$ and uniform $h$-refinement.}
	\label{fig:cantilever_convergence}
\end{figure}

In finite element discretizations of curved beam and shell models, membrane locking refers to the phenomenon of artificial bending stiffness due to the coupling of the bending response and membrane response caused by the local curvature \cite{Stolarski1982,bischoff2018models}. 
Membrane locking negatively affects accuracy and convergence, illustrated in Fig.~\ref{fig:cantilever_convergence} for an isogeometric finite element discretization of a curved Euler-Bernoulli cantilever \cite{Cazzani2016}. We observe that for practically relevant coarse meshes, the accuracy of the displacement solution measured via the relative error in the $L^2$ norm does not improve when the mesh is refined. The size of the resulting plateau in the convergence curve indicates the severity of membrane locking.  
We can also see that locking becomes more severe with increasing slenderness of the beam, and seems to reduce with increasing polynomial degree of the basis functions. We note that purely displacement-based finite element formulations of the Euler-Bernoulli beam model require basis functions in the Sobolev space $H^3$ to achieve optimal convergence rates $\mathcal{O}(p+1)$ in the $L^2$ norm \cite{engel2002continuous,Tagliabue2014}. In Fig.~\ref{fig:cantilever_convergence}a, we therefore observe convergence order two for quadratic B-splines that are only in $H^2$.

\begin{table}[ht]
\begin{tabularx}{\linewidth}{| >{\hsize=.14\hsize}X | >{\hsize=.45\hsize}X | >{\hsize=.25\hsize}X | >{\hsize=.2\hsize}X | }
    \hline
    \multicolumn{2}{|c|} { \textbf{Approach, concept} }  &  \multicolumn{2}{ c |}{\textbf{Application in}} \\\cline{3-4}
    \multicolumn{2}{|c|}{derivation via principle of virtual work ($\star$)} & nodal finite& isogeometric \\
    \multicolumn{2}{|c|}{ or as a mixed method ($\diamond$)} & elements & analysis \\
    \hline \hline
    \multicolumn{2}{|c|}{Higher-order basis ($\star$)} & \cite{Ashwell1976, Rank1998} & \cite{Elguedj2007, Benson2010} \\
    \hline
    \multicolumn{2}{|c|}{Field consistent approach ($\star$)} & \cite{Babu1986, Prathap1993, Isha2012} & \cite{KanokNukulchai2001, Bouclier2012, BeiraodaVeiga2012} \\
    \hline
    \multirow{2}{\hsize}{Reduced integration} & Selective reduced integration ($\star$) & \cite{Zienkiewicz1971, Malkus1978, Hughes1977, Noor1981, Schwarze2009} & \cite{Elguedj2007, Bouclier2012, Adam2015, Leonetti2018,Zou_quadrature2021} \\\cline{2-4} 
     & Hourglass mode control ($\star$) & \cite{Flanagan1981, Belytschko1984, Reese2007, Li2015} & \cite{Bouclier2012, Adam2015b} \\
    \hline 
    \multirow{2}{\hsize}{Strain modification} & B-bar method ($\star$)  & \cite{Hughes1977, Hughes1980, Recio2006} & \cite{Elguedj2007, Bouclier2012, Antolin2019, Bouclier2013} \\\cline{2-4}
     & Assumed natural strain (ANS) ($\star$) & \cite{Simo1986, Bathe1985, Bucalem1993} &  \cite{Antolin2019, Caseiro2014, Caseiro2015,Greco_ANS_2018} \\\cline{2-4}
     & Enhanced assumed strain (EAS) ($\diamond$) & \cite{Andelfinger1993, Bischoff1999, Alves2005, Cesar2002} & \cite{Cardoso2012} \\\cline{2-4} 
     & Discrete shear/strain gap (DSG) ($\star$) & \cite{Bletzinger2000, Bischoff2001, Koschnick2005} & \cite{Bouclier2012, Echter2010, Echter2013} \\
    \hline
    \multirow{2}{\hsize}{Variational principles} & Hu-Washizu ($\diamond$) & \cite{Malkus1978, Noor1981, Kim1998, Wagner_mixed_2008} & \cite{Echter2013, Taylor2011}  \\\cline{2-4}
     & Hellinger-Reissner ($\diamond$) & \cite{Bischoff1999, Stolarski1986, Lee2012} & \cite{BeiraodaVeiga2012, Echter2013, Taylor2011,Zou_locking_2020} \\\cline{2-4}
    \hline 
\end{tabularx}
\caption{Overview of locking-preventing finite element technology, developed in the context of standard finite element and isogeometric analysis, and associated literature (without claim to completeness).}
    \label{tab:locking_methods}
\end{table}

Locking-free finite element discretizations do not show any pre-asymptotic plateau, but converge right away with the optimal rate on coarse meshes. For completeness, we note that membrane locking is only one of several sources of locking, the most well-known being transverse shear locking in beam, plate and shell elements \cite{bischoff2018models} and volumetric locking due to incompressibility in solid elements \cite{Hughes2000}.
The development of locking-preventing discretization technology has a history of more than $40$ years, first within classical finite elements and then in isogeometric analysis. Without any claim to completeness, Table~\ref{tab:locking_methods} summarizes the most important locking-preventing formulations that can be applied against membrane locking, using the standard classifications into higher-order methods, the field-consistent approach, reduced integration, strain modification, and variational principles. 
For details on the underlying ideas and derivations, we refer the interested reader to the pertinent literature, also given in Table~\ref{tab:locking_methods} for nodal finite elements and isogeometric analysis. 

\begin{figure}[h!]
	\centering
	\begin{minipage}[b]{0.49\textwidth}
	\centering
	\includegraphics[width=1.0\textwidth]{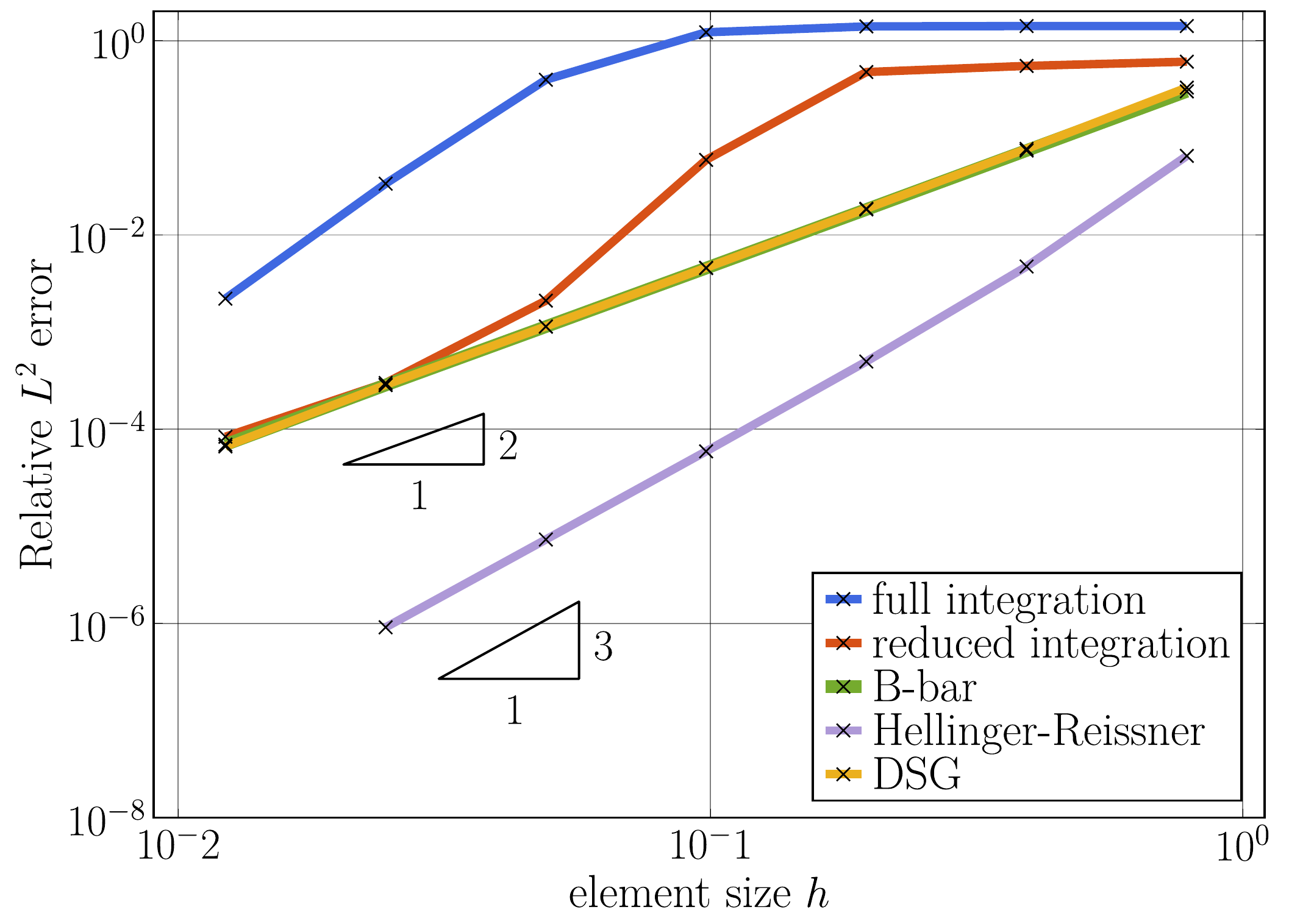}
	\footnotesize{(a) Quadratic B-splines ($p = 2$)}
	\end{minipage}
	\begin{minipage}[b]{0.49\textwidth}
	\centering
	\includegraphics[width=1.0\textwidth]{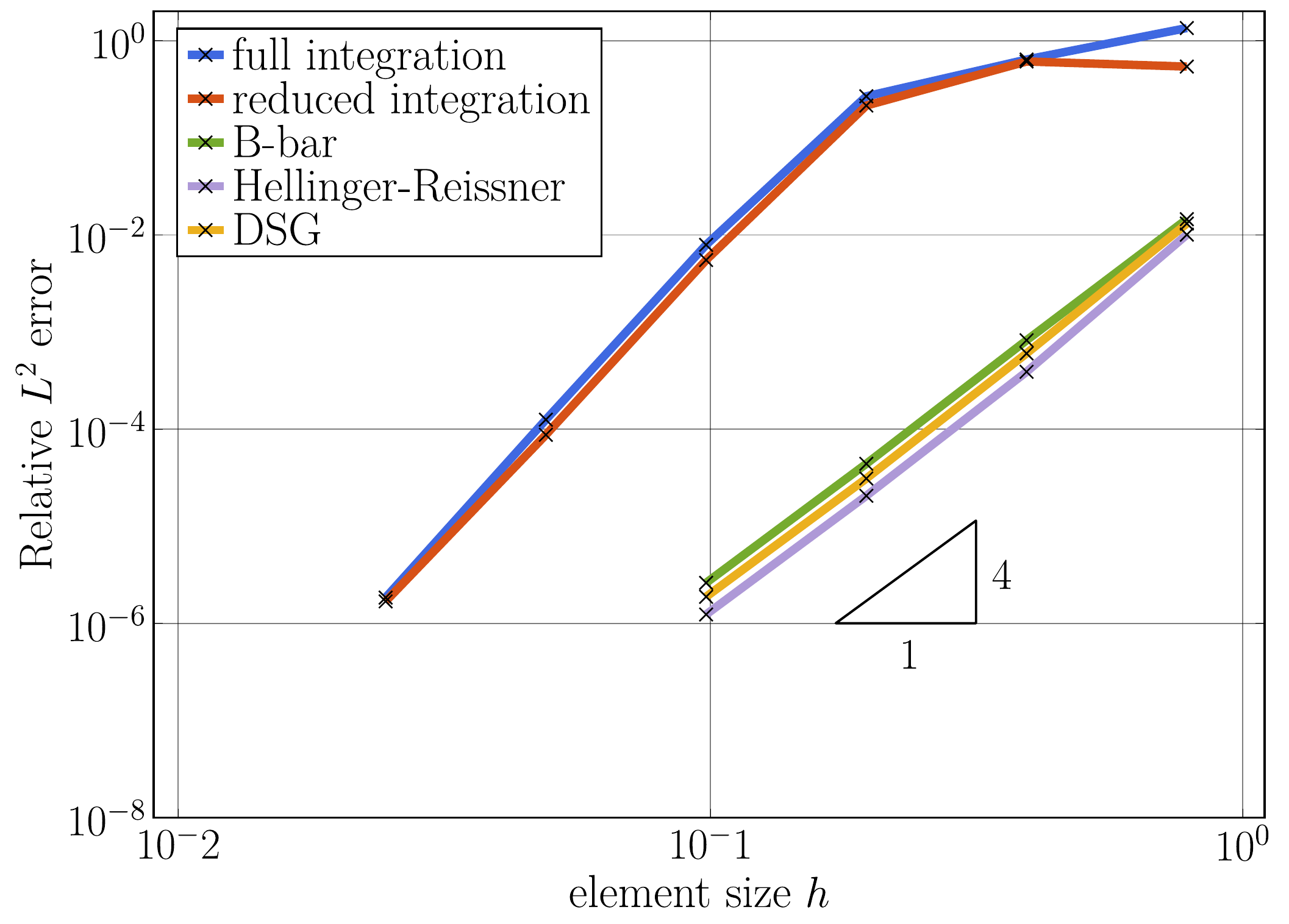}
	\footnotesize{(b) Cubic B-splines ($p = 3$)}
	\end{minipage}
	\begin{minipage}[b]{0.49\textwidth}
	\centering
	\vspace{0.26cm}
	\includegraphics[width=1.0\textwidth]{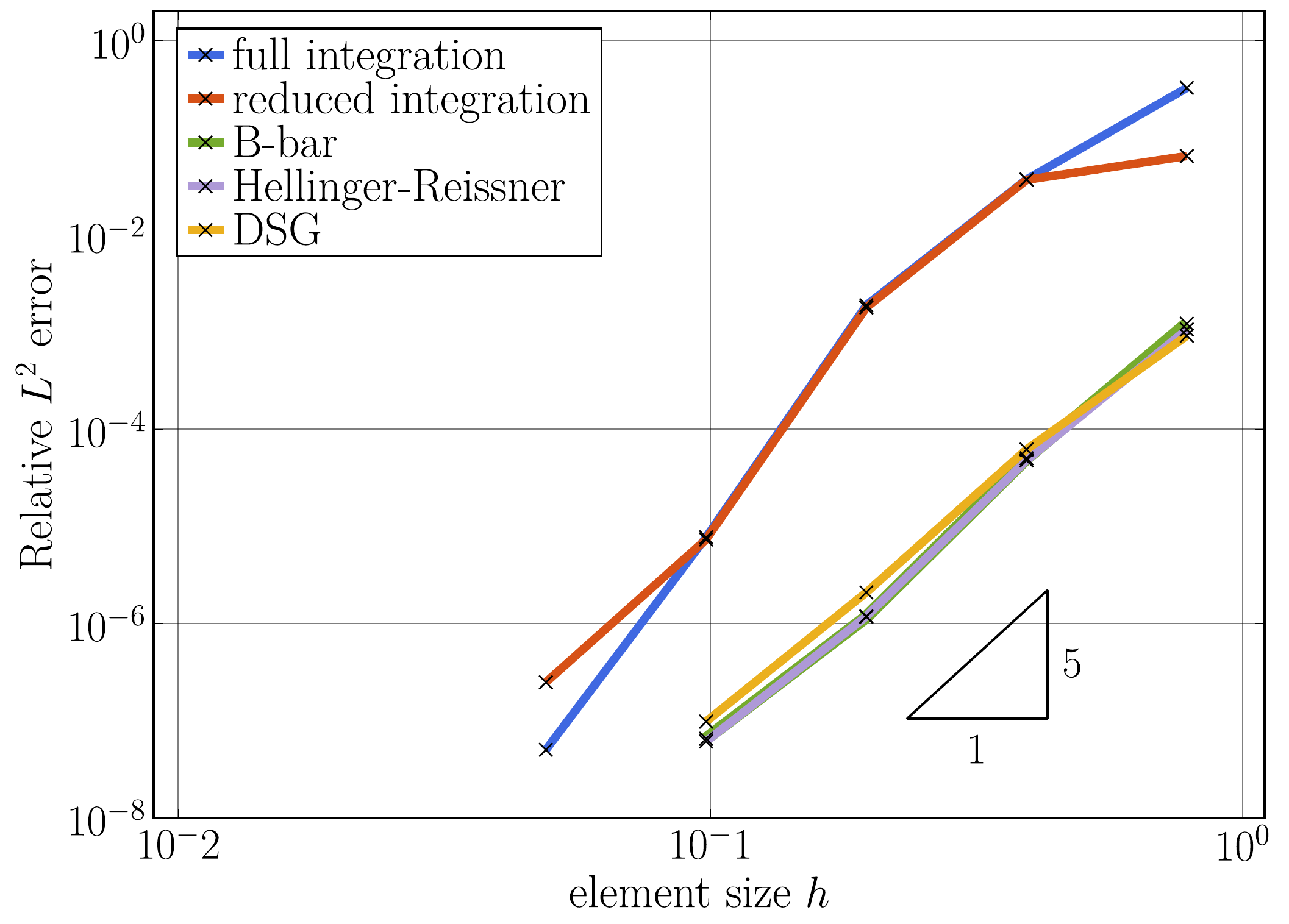}
	\footnotesize{(c) Quartic B-splines ($p = 4$)}
	\end{minipage}
	\begin{minipage}[b]{0.49\textwidth}
	\centering
	\vspace{0.26cm}
	\includegraphics[width=1.0\textwidth]{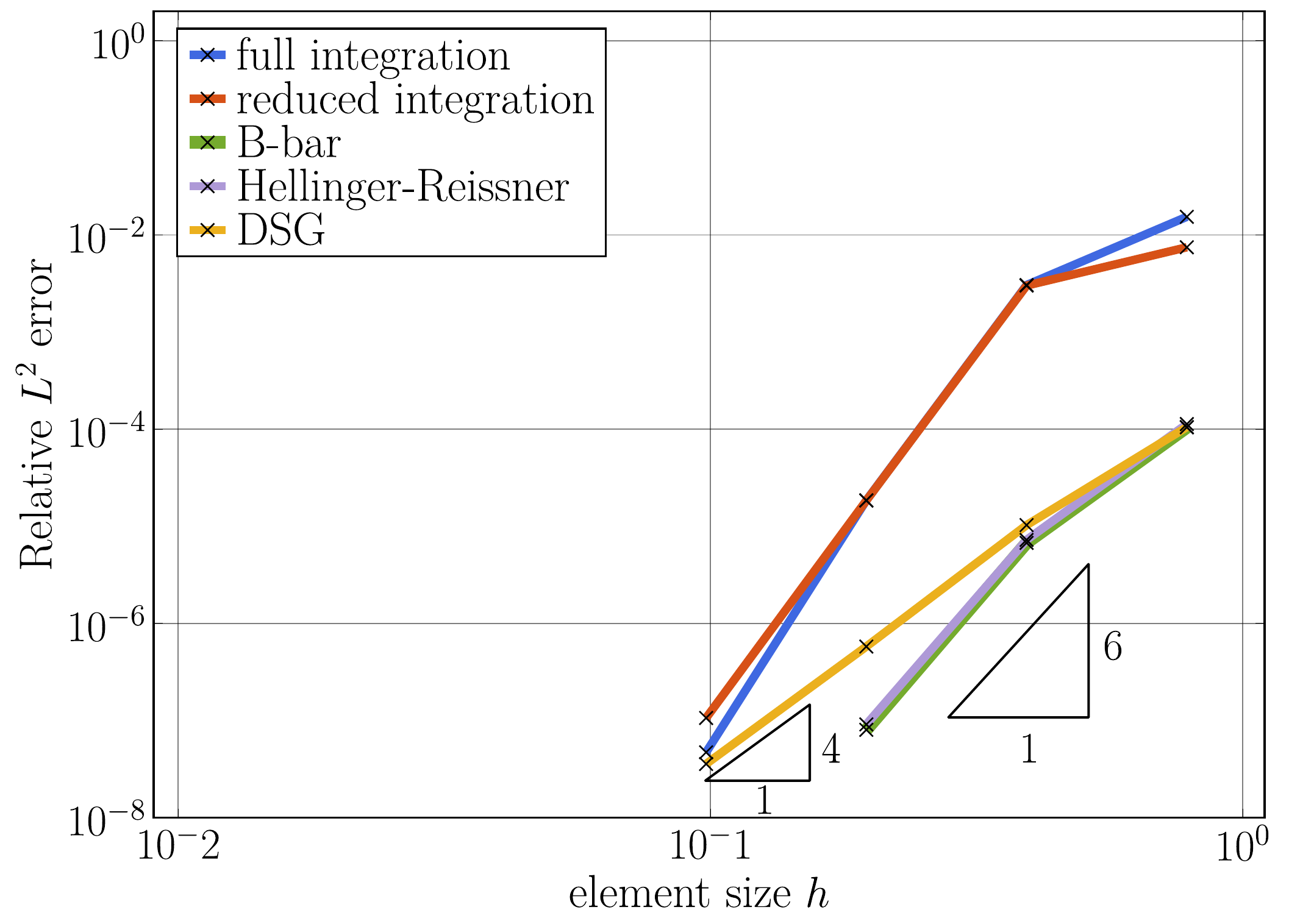}
	\footnotesize{(d) Quintic B-splines ($p = 5$)}
	\end{minipage}
	\caption{Relative $L^2$-norm errors of the displacements field of a curved Euler-Bernoulli beam (quarter circle cantilever, unit shear force at the free end, radius $R$, thickness $t$, slenderness $R/t = 1 \cdot 10^3$), obtained with different locking-free formulations via uniform $h$-refinement of quadratic, cubic, quartic and quintic B-spline basis functions.}
\label{fig:cantilever_convergence_methods}
\end{figure}

To illustrate the effect of locking-free formulations, we compare the results obtained with the standard finite element formulation, using full and reduced integration,  
to the results obtained with three representative variants,  
namely B-bar strain projection, the discrete strain gap (DSG) method and a Hellinger-Reissner approach. For the curved cantilever problem described above, Figure~\ref{fig:cantilever_convergence_methods} plots the resulting convergence curves in terms of the $L^2$-norm errors of the displacements for quadratic, cubic, quartic and quintic B-spline basis functions. 
We observe that all three locking-free formulations mitigate the effect of membrane locking with respect to the standard finite element formulation that is affected significantly by membrane locking. For quadratic basis functions on finer meshes, reduced integration mitigates the effect of membrane locking, but does not have any effect for $p\ge3$. We also see that the DSG method does not consistently perform well for all polynomial degrees. 
This example illustrates that convergence studies of simple benchmark problems do not constitute a satisfactory way of assessing discretization methods in terms of their locking-related robustness and accuracy. Given the multitude of formulations addressing membrane locking, the question arises how to best compare and assess their accuracy and effectivity.  

The analysis of the discrete spectrum of eigenvalues and eigenmodes constitutes an alternative way of assessing the accuracy of a discretization scheme. Eigenvalues and eigenmodes are computed from a discrete generalized eigenvalue problem that for models in structural mechanics corresponds to the discretized variational formulation of the associated free vibration problem without damping. Spectral accuracy directly relates to the accuracy of a discretized boundary value problem, as the solution of the latter can be represented in terms of eigenvalues and eigenmodes. 
For instance, spectral analysis has been recently used to explain the superior per-degree-of-freedom accuracy and robustness that is achieved by isogeometric analysis with smooth spline functions \cite{Hughes2014}.

In this paper, we take first steps towards establishing spectral analysis as a tool for identifying locking behavior and assessing the effectiveness of locking-free formulations. Our approach is global, that is, we look at the complete spectrum and modal behavior, and it thus goes far beyond existing work \cite{Zou_quadrature2021,Lee_MITC3_2015}. On the one hand, the solution to the discrete eigenvalue problem depends on the terms that appear in the variational formulation, their evaluation, and the solution space spanned by the basis functions of the finite element discretization. Therefore, all aspects of the various locking-preventing formulations given in Table~\ref{tab:locking_methods} are reflected in the discrete spectrum. On the other hand, spectral analysis provides access to information that cannot be obtained via standard convergence measures based on error norms.
Moreover, we propose a practical way of ``measuring'' locking (or unlocking) in the spectrum by comparing normalized spectra computed on coarse and ``overkill'' meshes. We define a method as locking-free if the normalized spectra are converged on coarse meshes, that is, the spectra obtained from coarse and ``overkill'' meshes do not differ. Accordingly, we define a method as locking if the normalized spectra are not converged on coarse meshes, that is, the spectra obtained from coarse and ``overkill'' meshes differ significantly.
We illustrate the validity and significance of spectral analysis in this context via the example of a circular ring discretized with isogeometric curved Euler-Bernoulli beam elements susceptible to membrane locking. 
We then compare and assess the effectivity of the three representative locking-free formulations selected above 
in terms of their impact on the accuracy of the eigenvalues and eigenmodes. 

The structure of the paper is as follows: in Section~\ref{sec:theory_background}, we briefly review the generalized eigenvalue problem and associated error measures in spectral analysis. In Section~\ref{sec:formulations1}, we introduce the circular ring problem and its discretization via isogeometric Euler-Bernoulli beam elements. 
In Section~\ref{sec:formulations2}, we review the three representative locking-free formulations. 
In Section~\ref{sec:spectra_beam}, we present detailed spectral analysis carried out for isogeometric discretizations of the thin circular ring and various locking and locking-free formulations. In addition, we provide an in-depth discussion of the validity and strengths of spectral analysis to interpret membrane locking in this context.  
In Section \ref{sec:conclusion}, we summarize our results and conclusions.

\section{Generalized eigenvalue problem and error measures in spectral analysis}\label{sec:theory_background}

\subsection{Generalized eigenvalue problem}

We recall the generalized eigenvalue problem that governs free vibrations of an undamped linear structural system in the continuous setting. Each continuous eigenmode $U_n(x)$, with $n \in \mathbb{N}^+$ and defined on a domain $\Omega$, satisfies the eigenvalue problem: find $(U_n, \lambda_n) \in \mathcal{V} \times \mathbb{R}^+$ such that 
\begin{equation}
    \left(\mathcal{K} - \lambda_n \mathcal{M}\right) \, \eigenvec_n(x) \; = \;  0 \, , \quad x \in \Omega \, . 
\label{gep} \end{equation}
Here, $\mathcal{M}$ and $\mathcal{K}$ are the mass and stiffness operators, $\lambda_n = \omega_n^2$ is the $n^{\text{th}}$ eigenvalue equal to 
the square of the $n^{\text{th}}$ eigenfrequency $\omega_n$, and $\mathcal{V}$ is the space of functions with sufficient regularity that allows the differential operators in $\mathcal{M}$ and $\mathcal{K}$ to be applied.

The strong form of the generalized eigenvalue problem \eqref{gep} can be transferred into a variational form via the standard Galerkin method and subsequently discretized with $N$ finite element basis functions $B_i (x)$. The resulting discrete eigenvalue problem can be expressed via the following matrix equations: find $(U^h_n, \lambda^h_n) \in \mathcal{V}^h \times \mathbb{R}^+$ such that
\begin{equation} \label{dgep}
    \left( \mat{K} - \lambda_n^h \, \mat{M} \right) \, \mat{\eigenvec}_n^h \; = \; 0 \, , \qquad n = 1,2,\ldots,N \, ,
\end{equation}
where $\mat{\eigenvec}_n^h$ denotes the vector of unknown coefficients, such that the $n^{\text{th}}$ discrete eigenmode is
\begin{align}
  \eigenvec_n^h(x) = \begin{bmatrix}
        B_1 (x) \; \ldots \; B_N (x) 
    \end{bmatrix} \, \mat{\eigenvec}_n^h \, , \qquad \eigenvec_n^h(x) \in \mathcal{V}^h \subset \mathcal{V} \, . \nonumber
\end{align}
Here, $\mat{K}$ and $\mat{M}$ denote the stiffness and consistent mass matrix, $\lambda^h_n = (\omega_n^h)^2$ is the $n^{\text{th}}$ discrete eigenvalue equal to 
the square of the $n^{\text{th}} $ discrete eigenfrequency $\omega_n^h$, and $\mathcal{V}^h$ is the space of finite element basis functions with sufficient regularity. 
The discrete eigenmodes $\eigenvec^h_n$ are orthogonal in the $L^2$ norm and thus form a basis for the solution of any boundary value problem defined with the same model. 

We note that in many applications, the use of a lumped mass matrix instead of the consistent mass matrix is common. Lumping schemes, however, often significantly affect the accuracy of the discrete spectrum, see e.g. \cite{Cottrell2006}, and are therefore not considered in this study.

\subsection{Ordering of eigenvalues, rank sufficiency}

We recall the following important properties due to their relevance in the remainder of the paper. The  eigenvalues $\lambda^h_n$ can be  sorted in ascending order, 
where the corresponding eigenmodes can be ordered arbitrarily. 
Under the condition that \eqref{dgep} is derived from a homogeneous Neumann eigenvalue problem, that is, no Dirichlet boundary conditions are specified, the $N \times N$ stiffness matrix is symmetric positive semi-definite and the $N \times N$ consistent mass matrix is symmetric positive definite for linear structural systems. As a consequence, all eigenvalues are nonnegative real numbers ordered as
\begin{align}
0 \leq \lambda^h_1 \leq \lambda^h_2 \leq \ldots \leq \lambda^h_n \leq \ldots \leq \lambda^h_N \, .
\label{ordering}
\end{align}
A stable finite element scheme satisfies the notion of rank sufficiency based on the following three requirements \cite{Hughes2000, Schillinger:14.2}:
\begin{enumerate}
\item The number of zero eigenvalues corresponds exactly to the number of rigid body modes, given by the specific structural system under consideration. The proper imposition of Dirichlet boundary conditions removes all rigid body modes and corresponding zero eigenvalues. 
\item All eigenvalues are real, and the smallest non-zero eigenvalue converges to a finite value larger than zero. This ensures that no further zero eigenvalues occur, since the set of eigenvalues is bounded from below due to \eqref{ordering}.
\item The set of eigenvalues is bounded from above, i.e.\ the largest eigenvalue is finite. 
\end{enumerate}

\subsection{Error measures in spectral analysis}

In this paper, we will investigate the error globally across the complete spectrum of eigenvalues and eigenmodes. To this end, we first define the following two error measures:
\begin{align}
    & \left|\frac{\lambda_n^h - \lambda_n}{\lambda_n}\right|, \qquad \text{relative eigenvalue error}, \label{eve}\\
    & \frac{\|\eigenvec_n^h - \eigenvec_n \|}{\| \eigenvec_n \|}, \qquad \text{relative mode error in the } L^2\text{-norm}, \label{eme}
\end{align}
which we will use extensively throughout this paper to quantify locking effects. We recall that for every mode $n$, the relative errors in the corresponding eigenvalue and mode, \eqref{eve} and \eqref{eme}, sum to the mode error in the energy norm \cite[Section~6.3, p. 233]{Strang2008}:
\begin{equation}\label{eq:pythagorean_error_theorem}
    \frac{\lambda_n^h - \lambda_n}{\lambda_n} + \frac{\|\eigenvec_n^h - \eigenvec_n \|^2}{\| \eigenvec_n \|^2} = \frac{\|\eigenvec_n^h - \eigenvec_n \|^2_E}{\| \eigenvec_n \|^2_E}, \quad \forall \, n = 1,2, \ldots, N \, ,
\end{equation}
provided that $\|\eigenvec_n^h \|_{L^2} = \|\eigenvec_n \|_{L^2}$. 
This relationship, denoted as the \textit{Pythagorean eigenvalue error theorem}, is used extensively in \cite{Hughes2014} to evaluate both standard finite element and isogeometric approximations of eigenvalue, boundary-value, and initial-value problems. We refer to \cite{Hughes2014} for an in-depth discussion of error measures used in eigenvalue problems.

\subsection{The role of the lowest eigenvalues and eigenmodes}
\label{sec:lowmode}

It is important to note that for the approximation power of a finite element scheme, the accuracy of the lower part of the discrete spectrum is particularly crucial. To illustrate this key statement, we consider the discrete form $\mat{K} \mat{x} = \mat{f}$ that results from a finite element discretization of an elliptic boundary value problem, where $\mat{K}$ and $\mat{f}$ denote the stiffness matrix and the force vector, and $\mat{x}$ is the vector of unknowns. As the discrete eigenmodes $\eigenvec^h_n$ form a basis for the solution, we can expand the solution coefficients of the finite element basis in terms of the coefficients of the eigenmodes as
\begin{equation}\label{eq:em_ext}
    \mat{x} = \sum_n \mat{\eigenvec}_n^h \, c_n \, .
\end{equation}
Using basic algebra and the orthogonality properties of the eigenmodes with respect to the stiffness and mass matrices, one can derive a closed-form expression for  each unknown $c_n$ of the eigenmode expansion \eqref{eq:em_ext} that reads 
\begin{equation}\label{eq:expansion}
     c_n = \frac{1}{\lambda_n^h} \; \frac{(\mat{\eigenvec}^h_n)^T}{(\mat{\eigenvec}^h_n)^T \, \mat{M} \; \mat{\eigenvec}^h_n} \; \mat{f} \, .
\end{equation}
For the intermediate steps of this derivation, we refer to \cite{Schillinger:14.2,clough1993dynamics}. 
Each coefficient $c_n$ is inversely proportional to the size of the corresponding eigenvalue $\lambda_n^h$.  
Due to the ordering \eqref{ordering}, the magnitude  
of the eigenvalues monotonically increases with mode number $n$. Therefore, for discretized elliptic boundary value problems, the contribution of higher eigenmodes with $n \gg 1$ will typically be significantly smaller than the contribution of the lowest eigenmodes $n=1,2,3,4,\ldots$ For practical meshes with more than a few basis functions, we can even discard the contribution of the high modes completely, as this tendency becomes more pronounced, when the number of degrees of freedom and hence the number of eigenvalues is increased.

\section{Free vibration of the Euler-Bernoulli circular ring} \label{sec:formulations1}

\begin{figure}[h!]
\begin{center}
	\def\svgwidth{0.7\textwidth}
	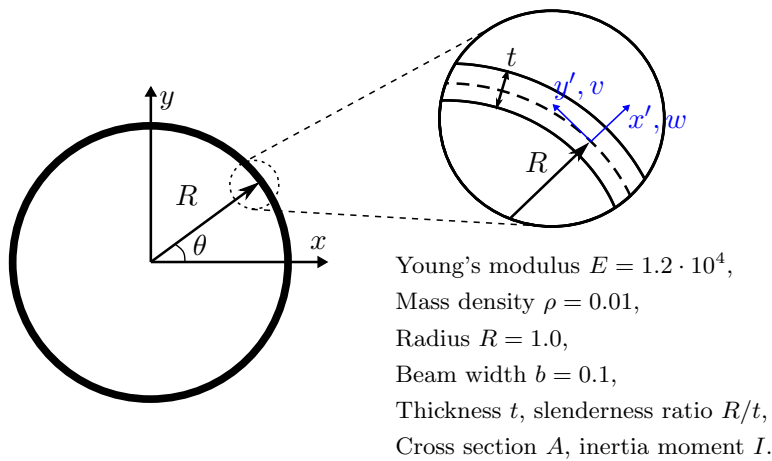
\end{center}
\caption{Closed circular ring modeled as a curved Euler-Bernoulli beam.}
\label{fig:ring_geometry}
\end{figure}

To illustrate our idea to apply spectral analysis for assessing locking phenomena, we consider the free vibration response of a thin circular ring that we will model as a curved Euler-Bernoulli beam in two dimensions and numerically solve with different finite element formulations, both locking and locking-free. A basic illustration of the Euler-Bernoulli ring, which we assume to be unconstrained and undamped, is given in Fig.~\ref{fig:ring_geometry}. Focusing our attention on a single benchmark entails the following advantages. On the one hand, the resulting discrete model is a representative example for membrane locking, but not susceptible to any other form of locking. Therefore, we can a priori exclude any interaction of different locking phenomena. In addition, the continuous problem still allows for an analytical solution, so that the error measures \eqref{eve} and \eqref{eme} can be evaluated. On the other hand, beam formulations can be written up in concise format, including their different locking free variants, which facilitates comparison and avoids deviation from our main focus on the assessment of membrane locking through spectral analysis.

\subsection{Strong form of the eigenvalue problem in mixed format}
\label{sec:variational_form_mixed}

We briefly review the derivation of the generalized eigenvalue problem that governs the free vibration response of the unconstrained Euler-Bernoulli ring from the system of equations of motion.
For a circular ring, the radius of curvature $R$ is constant and the  
arc-length coordinate $s=R\theta$ can be expressed in terms of the angle $\theta \in [0, 2\pi]$ \cite{Soedel2004}. To facilitate the application to some of the locking-free formulations later on, we state the equations of motion in mixed form, where the kinematic relations are added as individual constraint equations: 
\begin{subequations} 
\begin{align}
	& \frac{EI}{R^2} \kappa_{, \theta} + \frac{EA}{R} \epsilon_{, \theta} = \rho A \ddot{v} \, ,  \label{eq1} \\
	& \frac{EI}{R^2} \kappa_{, \theta \theta} - \frac{EA}{R} \epsilon = \rho A \ddot{w} \, , \label{eq2}  \\
	& \epsilon = \frac{1}{R} v_{,\theta} + \frac{1}{R} w  \, , \label{eq3} \\ 
	& \kappa = \frac{1}{R^2} v_{,\theta} - \frac{1}{R^2} w_{,\theta \theta} \, . \label{eq4}
\end{align}
\end{subequations}
The field variables are the circumferential and transverse displacement components $v$ and $w$, respectively, the membrane strain $\epsilon$, and the change of curvature $\kappa$, which, at fixed radius $R$, are functions of the angle $\theta$ and time $t$. Young's modulus $E$, cross-section area $A$, moment of inertia $I$, and mass density $\rho$ are assumed constant. 
The double dot operator indicates second derivatives with respect to time $t$, and $(\cdot)_{,\theta}$ and $(\cdot)_{,\theta \theta}$ indicate first and second derivatives with respect to the angular coordinate $\theta$. 

We now assume that the field solutions of the free vibration problem are composed of a set of spatial solutions that depend on $\theta$, multiplied by a modulation $T_n$ that depends on time $t$: 
\begin{align}
\begin{split}
	& v(\theta, t) = \sum_n \hat{v}_n (\theta) T_n(t), \quad w(\theta, t) = \sum_n \hat{w}_n (\theta) T_n(t) \, ,  \\
	& \epsilon(\theta, t) = \sum_n \hat{\epsilon}_n (\theta) T_n(t) \, ,  \quad 
	 \kappa(\theta, t) = \sum_n \hat{\kappa}_n (\theta) T_n(t) \, , \label{sep}
	 \end{split}
\end{align}
where $n = 1, 2, \ldots, \infty$. We note that the time modulation is of the form $T=\exp(j \, \omega_n t)$, where $\omega_n$ is an eigenfrequency of the ring and $j$ is the imaginary unit. When we insert relations \eqref{sep} into the equations of motion \eqref{eq1} and \eqref{eq2}, we find for each component $n$ of the solution
\begin{align}
	& \left[\frac{EI}{R^2} \hat{\kappa}_{n, \theta} + \frac{EA}{R} \hat{\epsilon}_{n, \theta}\right] \, T_n - \left[\rho A \hat{v}_n\right] \, \ddot{T}_n = 0 \, , \label{eq5} \\
	& \left[\frac{EI}{R^2} \hat{\kappa}_{n, \theta \theta} - \frac{EA}{R} \hat{\epsilon}_n \right] \, T_n - \left[ \rho A \hat{w}_n\right] \, \ddot{T}_n = 0 \, \label{eq6} .
\end{align}
For each $n$, we can now separate the field variables that depend on space and time by dividing \eqref{eq5} and \eqref{eq6} by $T_n \left(\rho A \hat{v}_n \right)$ and $T_n \left(\rho A \hat{w}_n \right)$, respectively. After this operation, the first term in \eqref{eq5} and \eqref{eq6} only depends on $\theta$ and the second term only on $t$. For the equations of motion \eqref{eq5} and \eqref{eq6}, separation of variables thus allows us to write
\begin{align} \label{eq7}
	\frac{\frac{EI}{R^2} \hat{\kappa}_{n, \theta} + \frac{EA}{R} \hat{\epsilon}_{n, \theta}}{\rho A \hat{v}_n} \; = \;\frac{\frac{EI}{R^2} \hat{\kappa}_{n, \theta \theta} - \frac{EA}{R} \hat{\epsilon}_n}{\rho A \hat{w}_n} \; = \;\frac{\ddot{T}_n}{T_n} \; =\;  - \lambda_{in} \, .
\end{align}
As the angular coordinate $\theta$ and the time coordinate $t$ can be varied independently, all terms in \eqref{eq7} must remain equal to a constant, denoted here as $- \lambda_{in}$. As a result of the system of two equations of motion, one can show that for each $n$, two different constants exist \cite{Soedel2004}, which we refer to with the additional index $i=1, 2$.

Combining \eqref{eq7} and the kinematic relations \eqref{eq3} and \eqref{eq4} that are obviously true for each $n$ and arbitrary mode coefficients $T_n(t)$ results in the generalized eigenvalue problem for the unconstrained circular Euler-Bernoulli ring: find $( \{ \hat{v}_n, \hat{w}_n, \hat{\epsilon}_n, \hat{\kappa}_n \}, \lambda_{in}) \in \mathcal{V} \times \mathbb{R}^+$ such that
\begin{subequations}
\begin{align}
	& \frac{EI}{R^2} \hat{\kappa}_{n, \theta} + \frac{EA}{R} \hat{\epsilon}_{n, \theta} + \lambda_{in} \, \rho A \hat{v}_n = 0 \, , \label{eq91} \\
	& \frac{EI}{R^2} \hat{\kappa}_{n, \theta \theta} - \frac{EA}{R} \hat{\epsilon}_n + \lambda_{in} \, \rho A \hat{w}_n = 0 \, , \label{eq92} \\
	& \hat{\epsilon}_n = \frac{1}{R} \hat{v}_{n,\theta} + \frac{1}{R} \hat{w}_n \, , \label{eq93} \\ 
	& \hat{\kappa}_n = \frac{1}{R^2} \hat{v}_{n,\theta} - \frac{1}{R^2} \hat{w}_{n,\theta \theta} \, . \label{eq94}
\end{align}
\end{subequations}
where $n = 1, 2, \ldots, \infty$, $i=1, 2$, and $\mathcal{V}$ is a set of four spaces of continuous functions with sufficient regularity. 
Equations \eqref{eq91} and \eqref{eq92} can be identified as a generalized eigenvalue problem of the form \eqref{gep}, accompanied by two additional kinematic constraints \eqref{eq93} and \eqref{eq94}. The constants $\lambda_{in}$ form the $n^{\text{th}}$ eigenvalue pair, the square of the $n^{\text{th}}$ eigenfrequency pair $\omega_{in}$ of the ring.

\captionsetup[subfigure]{labelformat=empty}
\begin{figure}[!t]
	\centering
	\subfloat[$n = 0$]{\includegraphics[width=0.205\textwidth]{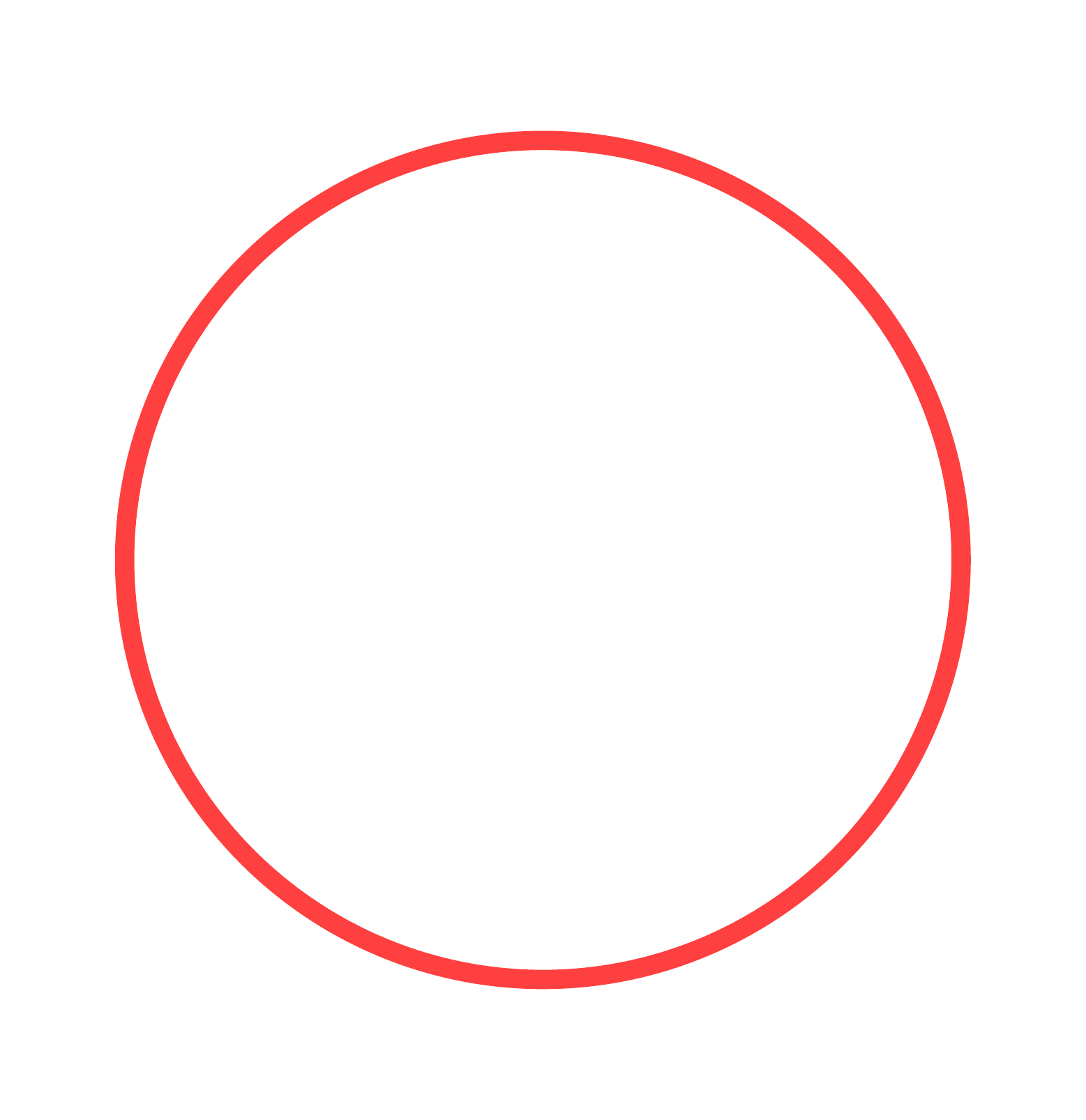}}
	\subfloat[$n = 1$]{\includegraphics[width=0.205\textwidth]{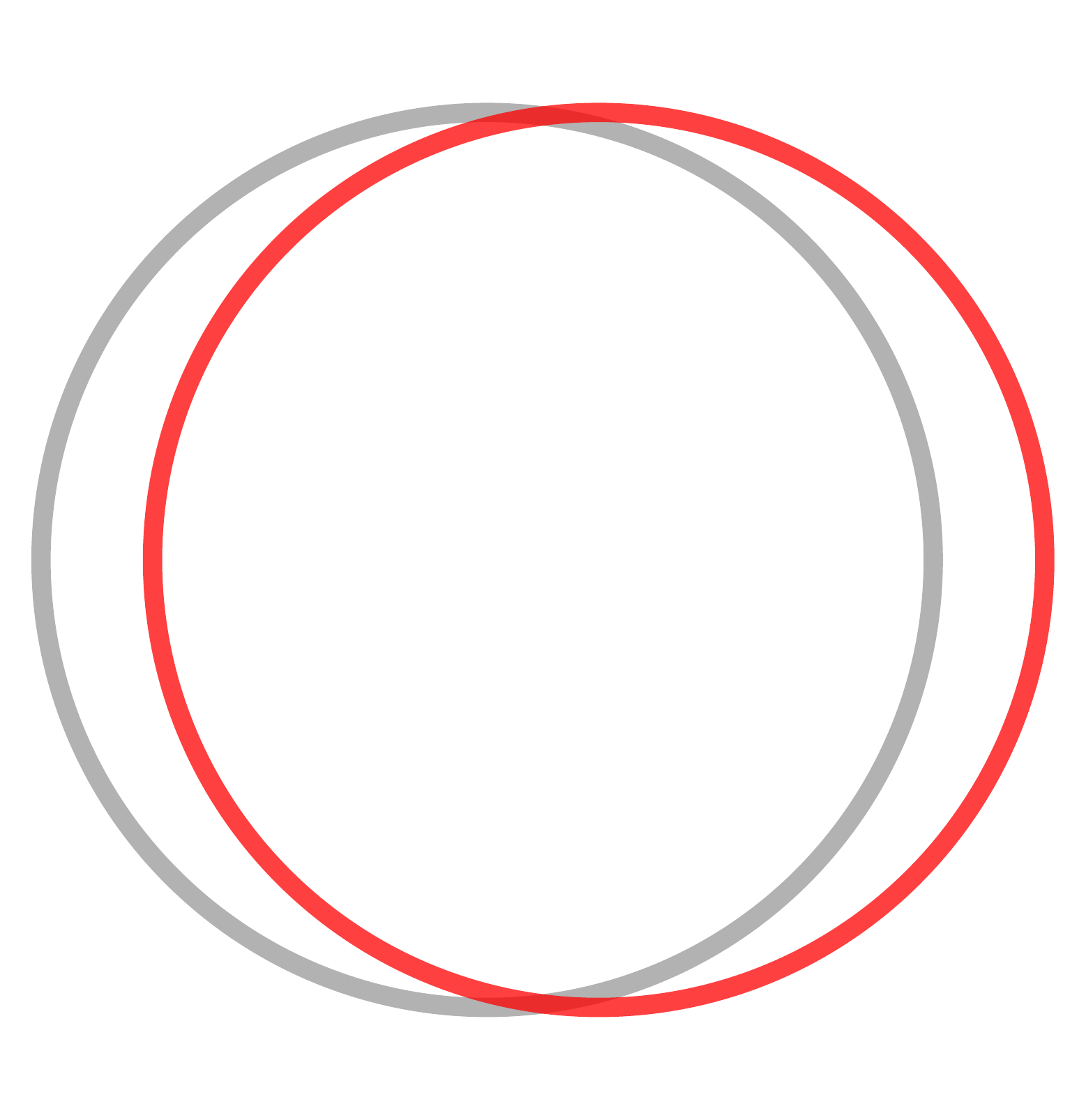}}\hspace{0.1cm}
	\subfloat[$n = 2$]{\includegraphics[width=0.205\textwidth]{sections/figures/ring_standard_mode1_n2.pdf}}\hspace{0.1cm}
	\subfloat[$n = 3$]{\includegraphics[width=0.205\textwidth]{sections/figures/ring_standard_mode1_n3.pdf}}

	\subfloat[$n = 4$]{\includegraphics[width=0.205\textwidth]{sections/figures/ring_standard_mode1_n4.pdf}}\hspace{0.1cm}
	\subfloat[$n = 5$]{\includegraphics[width=0.205\textwidth]{sections/figures/ring_standard_mode1_n5.pdf}}\hspace{0.1cm}
	\subfloat[$n = 6$]{\includegraphics[width=0.205\textwidth]{sections/figures/ring_standard_mode1_n6.pdf}}\hspace{0.1cm}
	\subfloat[$n = 7$]{\includegraphics[width=0.205\textwidth]{sections/figures/ring_standard_mode1_n7.pdf}}
	\caption{Analytical transverse eigenmodes (corresponding to $\lambda_{1n}$) of the circular ring with a slenderness ratio of $R/t = 1\cdot10^3$, where $R$ is the radius and $t$ the thickness of the ring. The plotted modes represent displacements in Cartesian coordinates, ($\eigenvec_x, \eigenvec_y$), obtained via the transformation \eqref{eq:modes_rotation}.
	\label{fig:ring_first12modes1}} 

	\vspace{0.8cm}

	\subfloat[$n = 0$]{\includegraphics[width=0.205\textwidth]{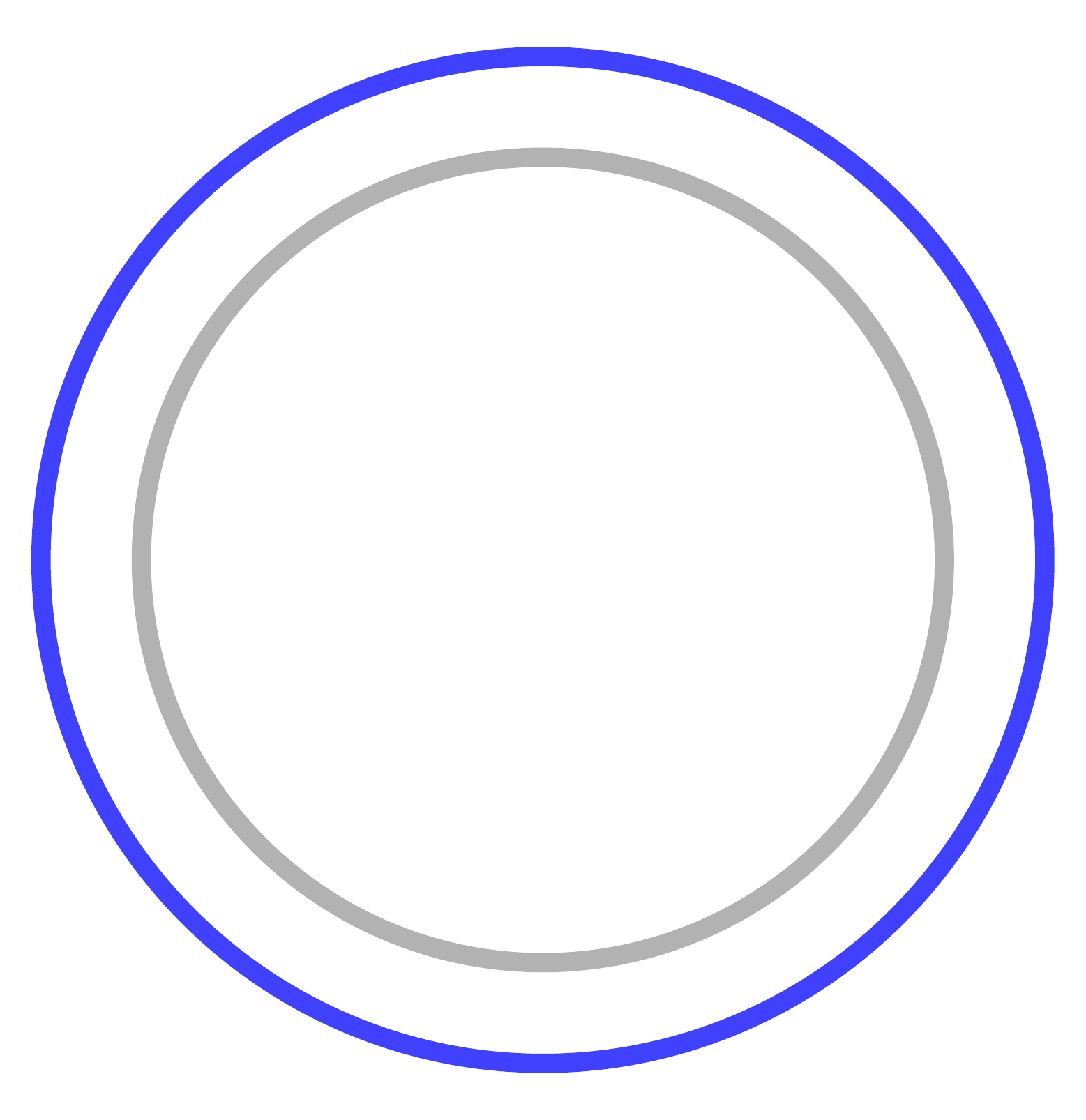}}\hspace{0.1cm}
	\subfloat[$n = 1$]{\includegraphics[width=0.205\textwidth]{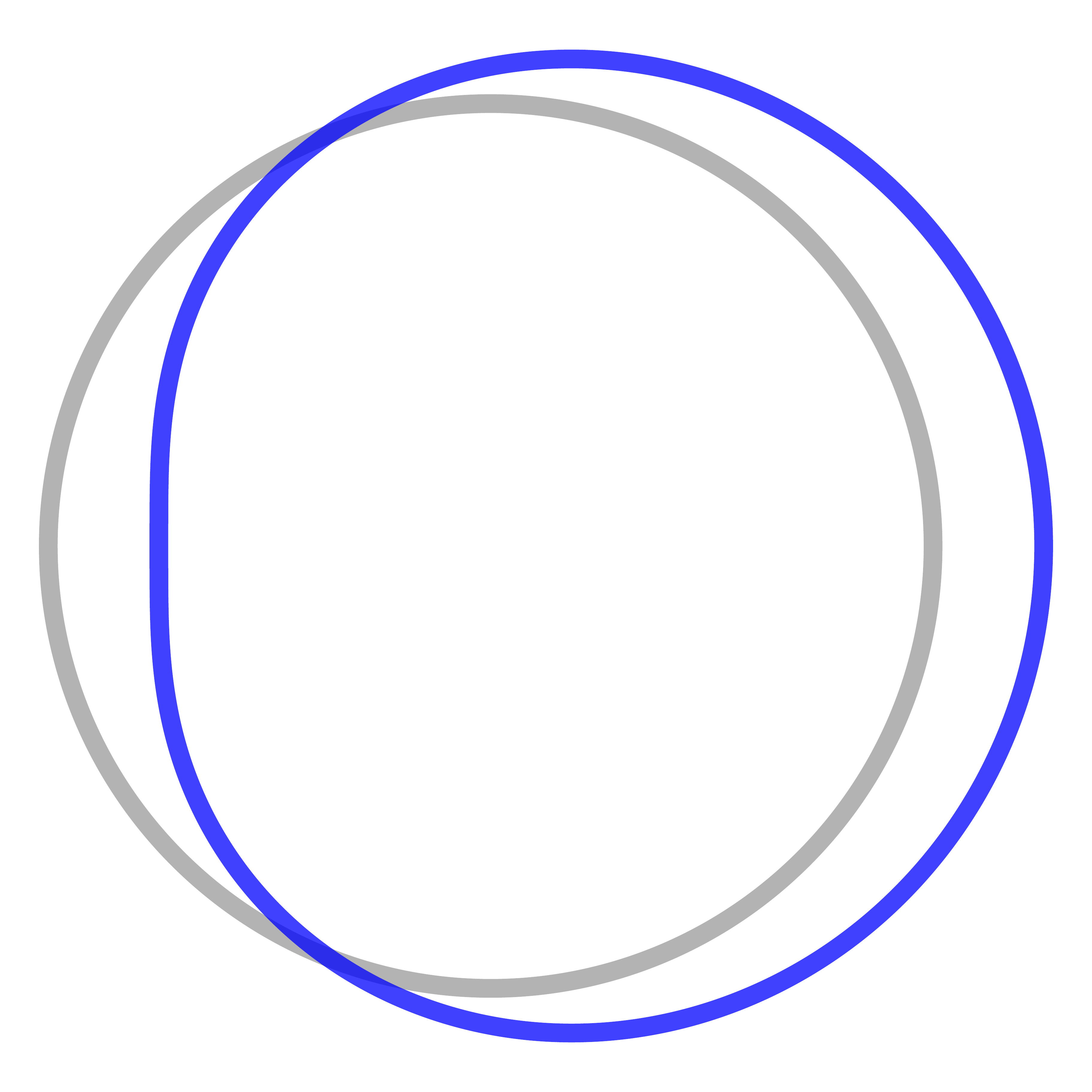}}\hspace{0.1cm}
	\subfloat[$n = 2$]{\includegraphics[width=0.205\textwidth]{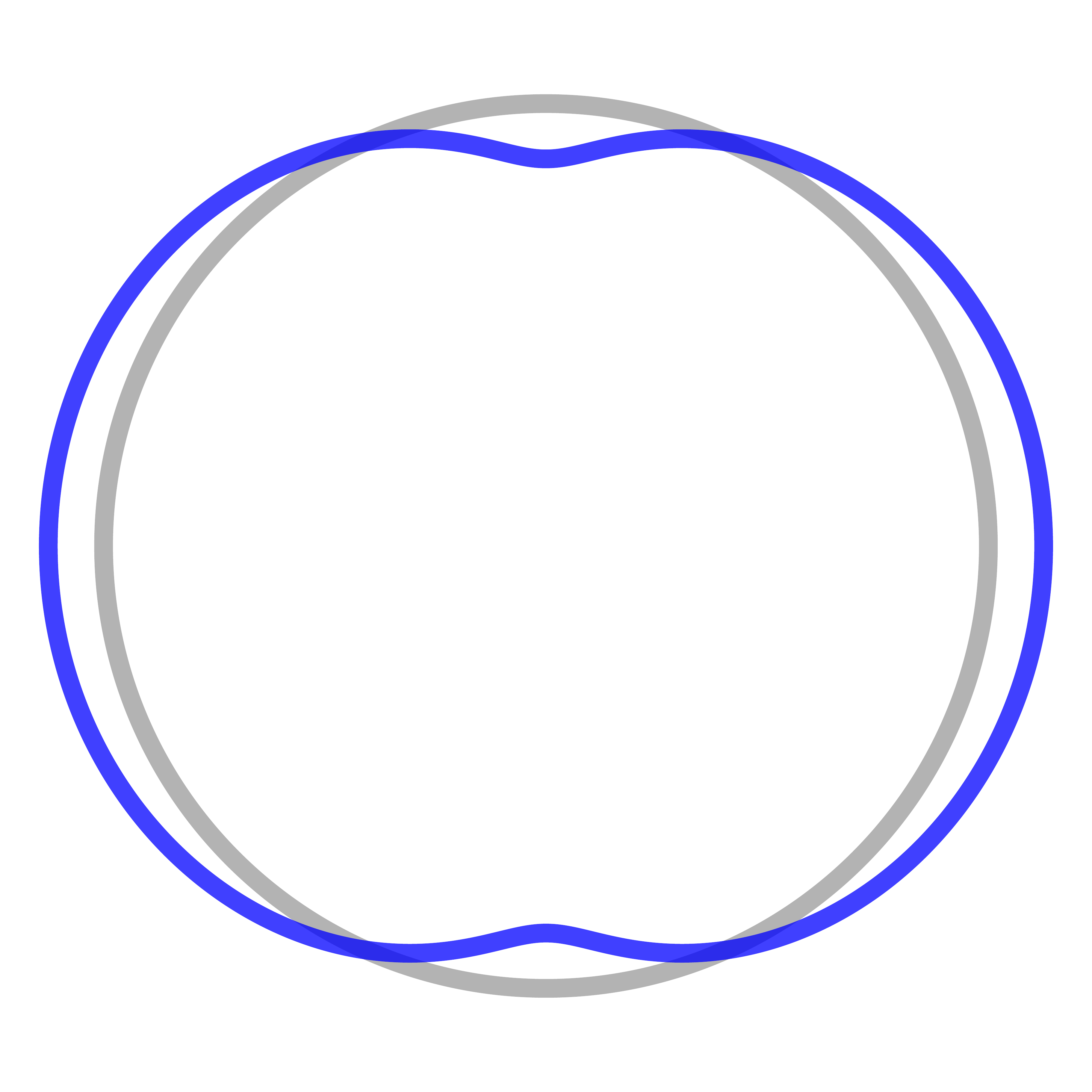}}\hspace{0.1cm}
	\subfloat[$n = 3$]{\includegraphics[width=0.205\textwidth]{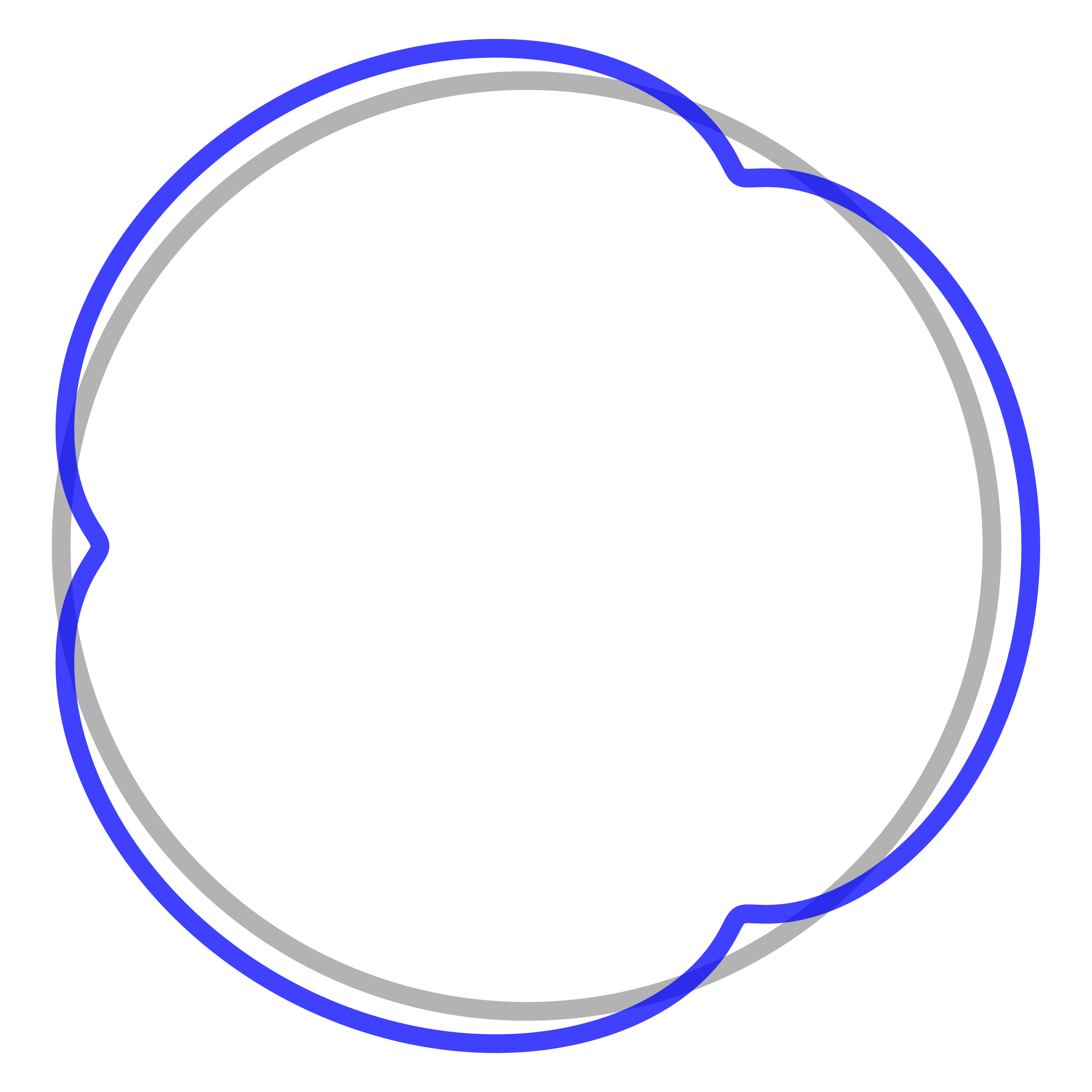}}

	\subfloat[$n = 4$]{\includegraphics[width=0.205\textwidth]{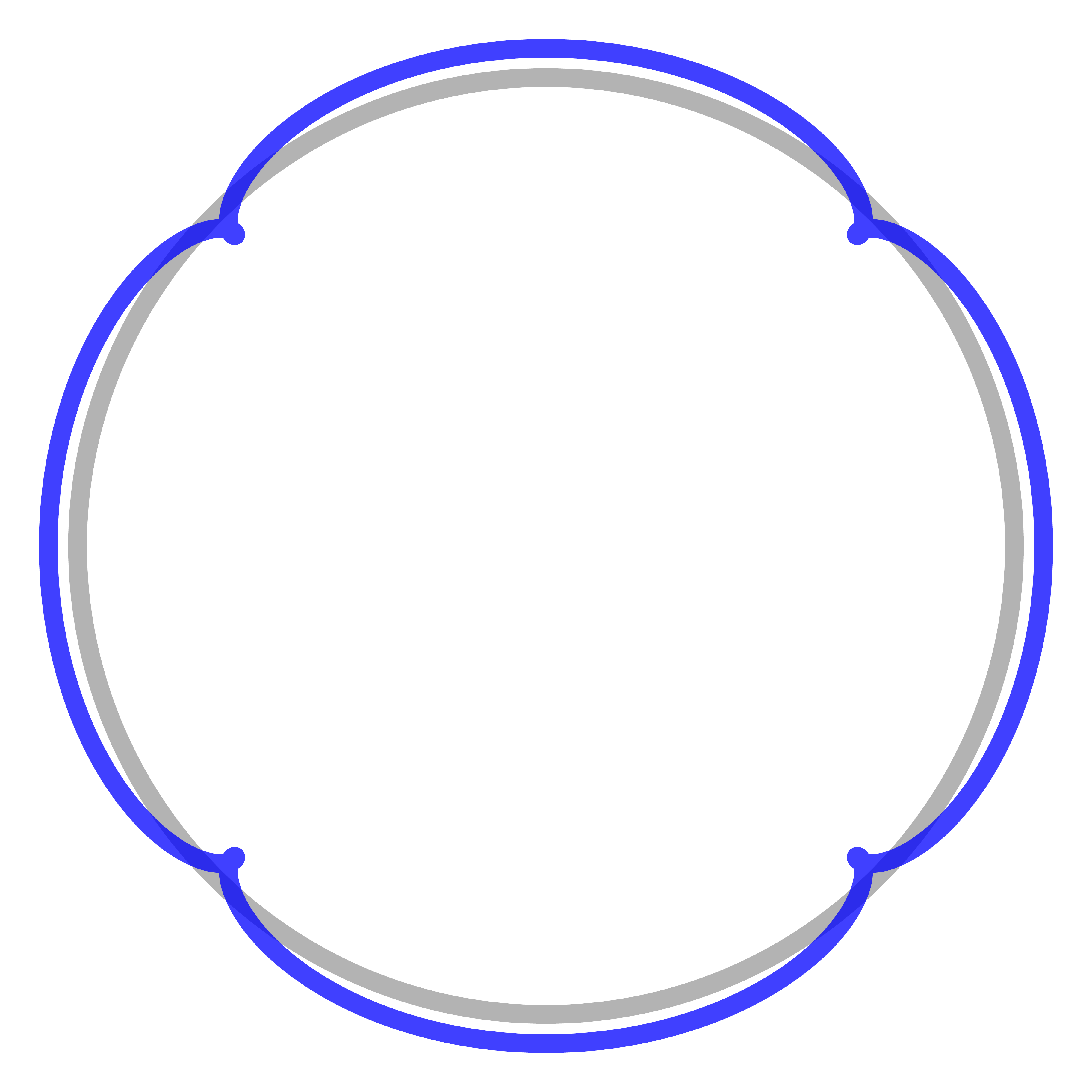}}\hspace{0.1cm}
	\subfloat[$n = 5$]{\includegraphics[width=0.205\textwidth]{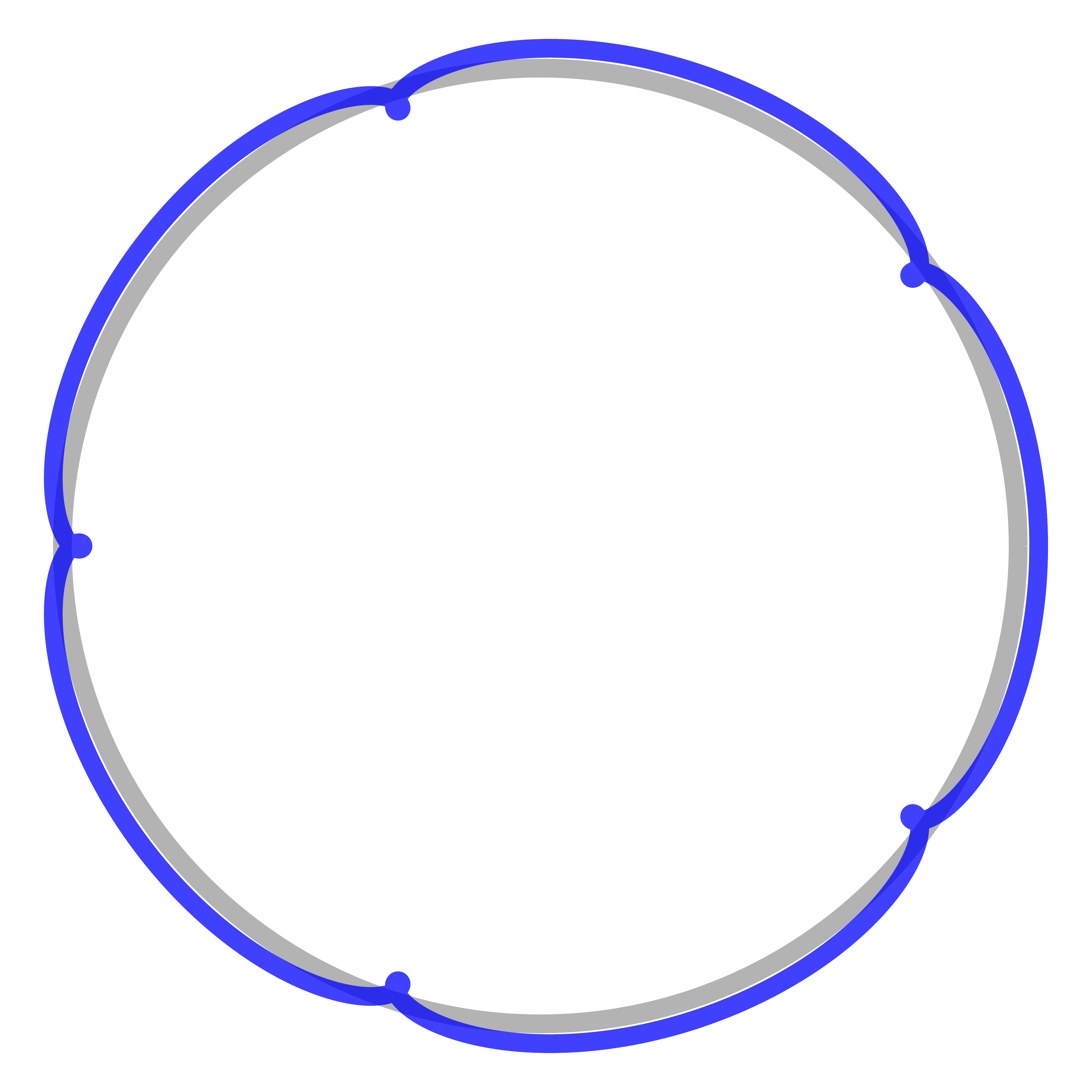}}\hspace{0.1cm}
	\subfloat[$n = 6$]{\includegraphics[width=0.205\textwidth]{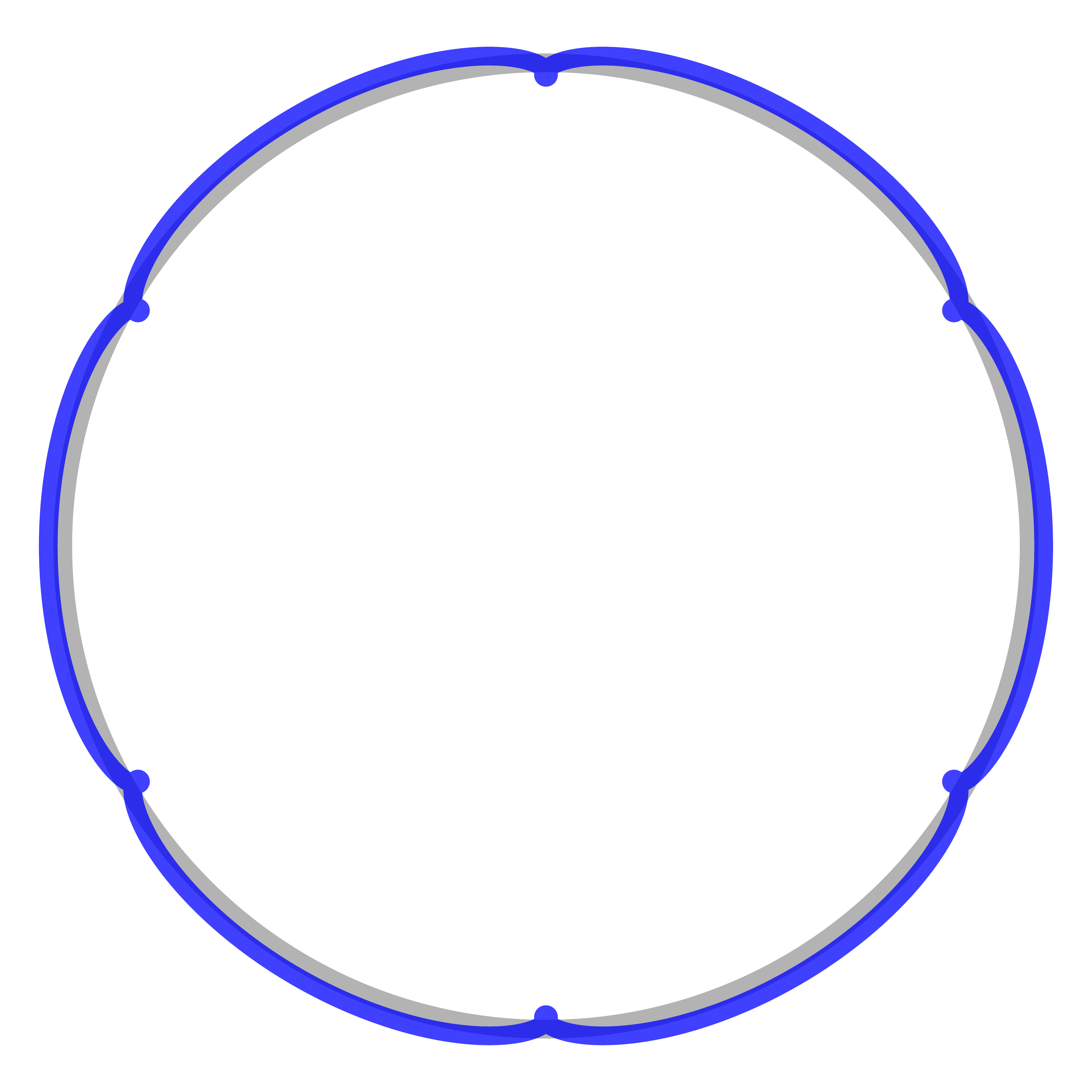}}\hspace{0.1cm}
	\subfloat[$n = 7$]{\includegraphics[width=0.205\textwidth]{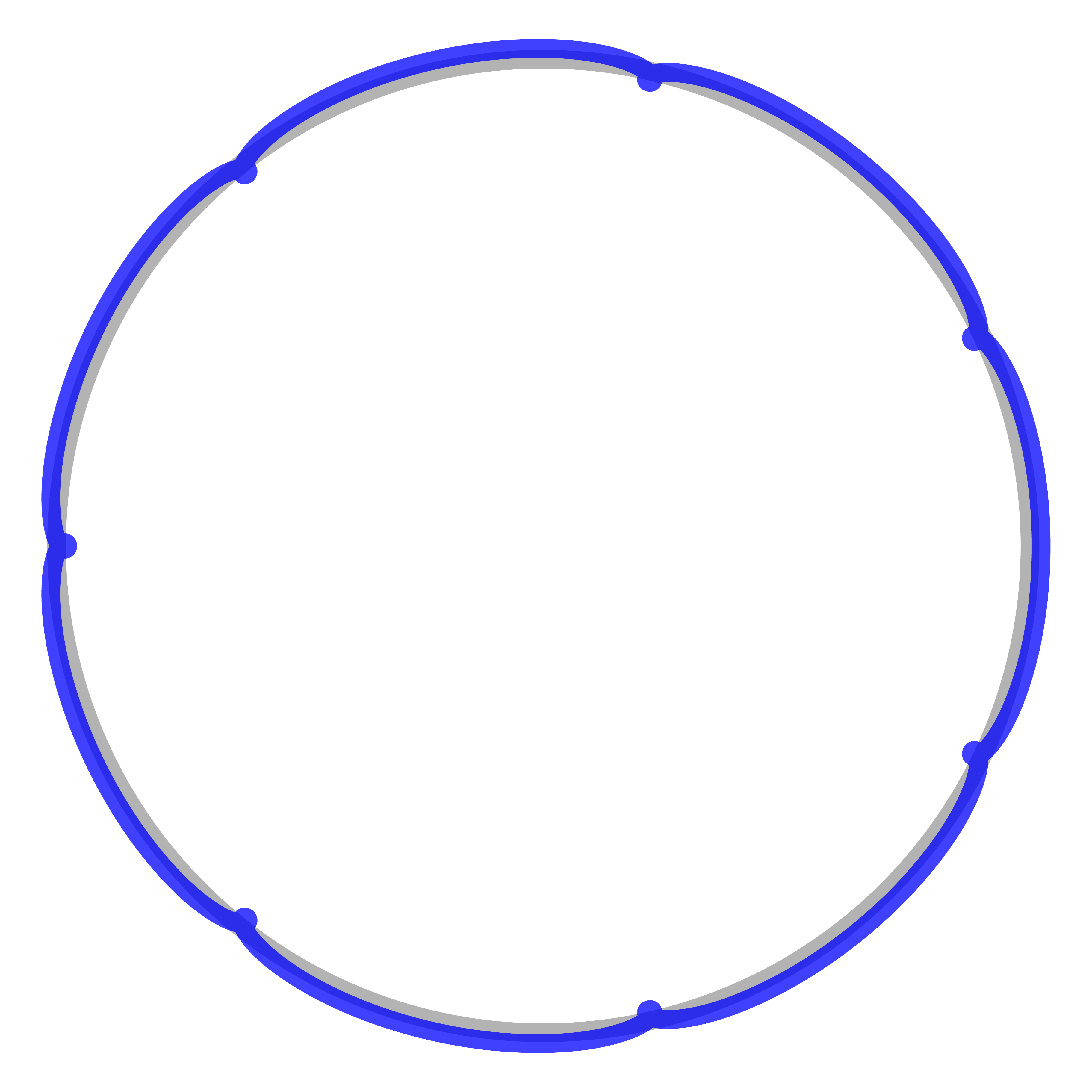}}
	\caption{Analytical circumferential eigenmodes (corresponding to $\lambda_{2n}$) of the circular ring with a slenderness ratio of $R/t = 1\cdot10^3$, where $R$ is the radius and $t$ the thickness of the ring. The plotted modes represent displacements in Cartesian coordinates, ($\eigenvec_x, \eigenvec_y$), obtained via the transformation \eqref{eq:modes_rotation}.
	\label{fig:ring_first12modes2}}
\end{figure}

An analytical solution of the eigenvalue problem of the circular Euler-Bernoulli ring is due to Soedel \cite{Soedel2004}, which is summarized in \ref{sec:analytic_eigen_ring} for completeness. As a consequence of the eigenvalue pairs, the solution naturally splits into two types of eigenmodes, which can be associated with transverse and circumferential displacement behavior. 
In this paper, we refer to the transverse and circumferential deflection dominating modes of \cite{Soedel2004} as transverse and circumferential modes. 
Figures~\ref{fig:ring_first12modes1} and \ref{fig:ring_first12modes2} illustrate the analytical shapes of some of the transverse and circumferential modes, plotted for the geometric and material parameters given in Fig.~\ref{fig:ring_geometry}, and a slenderness $R/t = 1\cdot10^3$, where $t$ is the thickness of the ring.

\begin{remark}
	The first two transverse modes in Figure~\ref{fig:ring_first12modes1} denote a rigid body rotation ($n=0$) and translation in x-direction ($n=1$), respectively. The remaining rigid body translation in y-direction is part of another solution set,  \eqref{analytical_mode2a} - \eqref{analytical_mode2b}, with a $\pi/2$ phase-shift, see \cite{Soedel2004} and \ref{sec:analytic_eigen_ring}. The first circumferential mode in Figure~\ref{fig:ring_first12modes2} is called ``breathing mode'' \cite{Soedel2004}. For our numerical studies in Section \ref{sec:spectra_beam}, we exclude the rigid body modes and the corresponding zero eigenvalues in both the numerical and analytical solutions.
\end{remark}

\begin{remark} 
The ``kinks'' that appear in the circumferential modes shown in Fig.~\ref{fig:ring_first12modes2} do not represent singularities, but are an artifact from plotting the modes for a finite displacement increment.
\end{remark}

\subsection{Variational formulation}

For ease of notation, we omit the hat on all displacement and strain field variables that we introduced in \eqref{sep}, with the understanding that from now on all displacement and strain fields will only depend on a spatial variable.
We also define the following strain-displacement operators for membrane strain and change of curvature that act on circumferential and transverse displacements $v$ and $w$:
\begin{subequations}	
	\begin{align}
		& L_\epsilon (v,w) = \frac{1}{R} v_{,\theta} + \frac{1}{R} w \, ,  \label{eq19}\\ 
		& L_\kappa ( v,w) = \frac{1}{R^2} v_{,\theta} - \frac{1}{R^2} w_{,\theta \theta} \, , \label{eq20}
	\end{align}
\end{subequations}
where we assume sufficient regularity, such that all derivatives with respect to $\theta$ are well defined. 

\subsubsection{Curvilinear displacements}

To transfer the strong form of the generalized eigenvalue problem \eqref{eq91} and \eqref{eq92} with kinematic constraints \eqref{eq93} and \eqref{eq94} into a variational format, we can apply the weighted residual method \cite{Hughes2000}. To this end, we bring each equation in residual form, multiply with a suitable test function and integrate over the ring domain:
\begin{subequations}
\begin{align} 
	& \int_0^{2\pi}\left[ \frac{EI}{R^2} {\kappa}_{n, \theta} + \frac{EA}{R} {\epsilon}_{n, \theta} + \lambda_{in} \, \rho A \, {v}_n \right] \, \delta {v} \, R \, \mathrm{d} \theta = 0 \, , \label{eq10} \\
	& \int_0^{2\pi} \left[ \frac{EI}{R^2} {\kappa}_{n, \theta \theta} - \frac{EA}{R} {\epsilon}_n + \lambda_{in} \, \rho A \, {w}_n \right] \, \delta {w} \, R \, \mathrm{d} \theta = 0 \, , \label{eq11} \\
	& \int_0^{2\pi} \left[{L}_{\epsilon}({{v}}_n, {w}_n) - {\epsilon}_n \right] \, \delta  {N} \; R \, \mathrm{d} \theta = 0 \, , \label{eq12} \\
	& \int_0^{2\pi} \left[ {L}_{\kappa}({{v}}_n, {w}_n) -{\kappa}_n\right] \, \delta  {M} \; R \, \mathrm{d} \theta = 0 \, . \label{eq13} 
\end{align}
\end{subequations}
From an energetic consistency viewpoint, we can identify the test functions as the virtual displacements $\delta{v}$ and $\delta{w}$ in circumferential and transverse direction, and the virtual membrane force and bending moment $\delta {N}$ and $\delta {M}$, respectively. 

Assuming sufficient regularity, we now integrate \eqref{eq10} and \eqref{eq11} by parts to shift all derivatives from the strain field variables to the virtual displacements:
\begin{align}
 \label{eq14_1} 
\begin{split}
	& \int_0^{2\pi} \left[ -\frac{EI}{R^2} {\kappa}_n \, \delta {v}_{,\theta} - \frac{EA}{R} {\epsilon}_n \, \delta {v}_{,\theta} + \lambda_{in} \, \rho A {v}_n \, \delta {v} \right] \, R \, \mathrm{d} \theta = 0 \, , \\
	& \int_0^{2\pi} \left[ \frac{EI}{R^2} {\kappa}_n \, \delta {w}_{,\theta \theta} - \frac{EA}{R} {\epsilon}_n \, \delta {w} + \lambda_{in} \, \rho A {w}_n \, \delta {w} \right] \, R \, \mathrm{d} \theta = 0 \, .
\end{split}
\end{align}
We note that due to the periodic nature of the ring, the integration by parts procedure does not produce any boundary terms.

In \eqref{eq12} and \eqref{eq13}, we use the constitutive relations for the membrane force and the bending moment, $\delta {N} =EA \, \delta {\epsilon}$ and $\delta {M} =EI \, \delta {\kappa}$, to obtain expressions based on corresponding virtual membrane strain and change of curvature functions:
\begin{align}
\label{eq14_2} 
\begin{split}
		& \int_0^{2\pi} \left[ EA \; {L_\epsilon}({{v}}_n, {w}_n) \, \delta {\epsilon} -  EA \, {\epsilon}_n \, \delta {\epsilon} \right] \, R \, \mathrm{d} \theta = 0 \, , \\
		& \int_0^{2\pi} \left[   EI \, {L_\kappa}({{v}}_n,{w}_n) \, \delta{\kappa} - EI \, {\kappa}_n \, \delta{\kappa} \right] \, R \, \mathrm{d} \theta = 0 \, . 
		\end{split}
\end{align}

Summing up the two equations in \eqref{eq14_1} and the two equations in \eqref{eq14_2}, we can write the variational mixed formulation of the circular Euler-Bernoulli ring in curvilinear coordinates in concise format: find $( \{ {v}_n, {w}_n \} , \{{\epsilon}_n, {\kappa}_n\}, \lambda_{in}) \in \mathcal{W} \times \mathcal{S} \times \mathbb{R}^+$ such that   
\begin{align}
\label{eq:weak_form_mixedHR_curved}
\int_0^{2\pi} \left[ EA \, {\epsilon}_n \, {L}_{\epsilon} (\delta{{v}},\delta{{w}}) + EI {\kappa}_n \,{L}_{\kappa} (\delta {{u}},\delta {{w}}) \right] \, R \, \mathrm{d} \theta  \;\;\; - \qquad \qquad & \nonumber  \\ 
\lambda_{in} \int_0^{2\pi} \rho A  \left[ {{v}}_n \, \delta{{v}} + {{w}}_n \, \delta{{w}} \right] \, R \, \mathrm{d} \theta \;= \;0  & \qquad  \forall \; \{ \delta{v}, \delta{w}\}  \in  \mathcal{W}  \, , \\
\int_0^{2\pi} \left[EA \, {L}_{\epsilon}({{v}}_n, {w}_n) \, \delta {\epsilon} + EI \, {L}_{\kappa}({{v}}_n,{w}_n) \, \delta{\kappa}  \right] \, R \, \mathrm{d} \theta  \;\;\; - \qquad \qquad & \nonumber  \\
\int_0^{2\pi} \left[EA \, {\epsilon}_n \, \delta {\epsilon} + EI \, {\kappa}_n \, \delta{\kappa} \right] \, R \, \mathrm{d} \theta \;= \;0 & \qquad \forall \; \{\delta{\epsilon},\delta{\kappa}\} \in  \mathcal{S}  \, ,
\end{align}
where $\mathcal{W}= ( H^1 \times H^2 ) $ and $\mathcal{S} = ( L^2 \times L^2) $ are the Sobolev spaces of periodic functions, all defined on the ring domain $[0, 2\pi]$.

\subsubsection{Cartesian displacements}\label{sec:variational_form_cartesian}

At each point of the ring parametrized by the angular coordinate $\theta$, we can express circumferential and transverse displacements $v$ and $w$ in terms of Cartesian displacements $u_x$ and $u_y$ that refer to a fixed \textcolor{black}{global} coordinate system. 
(see Fig.~\ref{fig:ring_geometry}). The corresponding transformation rule is:
\begin{align}\label{eq:modes_rotation}
	& \begin{bmatrix}
		w \\ v
	\end{bmatrix} = \begin{bmatrix}
		\cos(\theta) & \sin(\theta) \\
		-\sin(\theta) & \cos(\theta)
	\end{bmatrix} \begin{bmatrix}
		u_x \\ u_y
	\end{bmatrix}.
\end{align}
Substituting this transformation in \eqref{eq19} and \eqref{eq20}, we obtain the corresponding strain-displacement operators,
\begin{subequations}
	\begin{align}
		& L_{\epsilon}(u_x, u_y) = \frac{1}{R} \left(- u_{x,\theta} \sin(\theta) + u_{y,\theta} \cos(\theta) \right) \label{eq21} \\
		& L_{\kappa}(u_x, u_y) = \frac{1}{R^2} \left(- u_{x,\theta \theta} \cos(\theta) + u_{x,\theta} \sin(\theta) - u_{y,\theta \theta} \sin(\theta) - u_{y,\theta} \cos(\theta) \right). \label{eq22}
	\end{align}
\end{subequations}
that act on Cartesian displacements. From \eqref{eq:weak_form_mixedHR_curved}, we can derive the variational formulation of the generalized eigenvalue problem with respect to Cartesian displacements by replacing all circumferential and transverse displacements via \eqref{eq:modes_rotation}, \eqref{eq21} and \eqref{eq22}. 

The result is: find $( \{u_{x,n}, u_{y,n} \}, \{{\epsilon}_n, {\kappa}_n\}, \lambda_{in}) \in \mathcal{U} \times \mathcal{S} \times \mathbb{R}^+$ such that   
\begin{align}
\label{eq:weak_form_mixedHR_cart1}
 \int_0^{2\pi} \left[ EA \, {\epsilon}_n \, {L}_{\epsilon} (\delta{{u_x}},\delta{{u_y}}) + EI \, {\kappa}_n \,{L}_{\kappa} (\delta {{u_x}},\delta {{u_y}}) \right] \, R \, \mathrm{d} \theta  \;\;\; - \qquad \qquad & \nonumber  \\ 
\lambda_{in} \int_0^{2\pi} \rho A  \left[ u_{x,n} \, \delta{{u_x}} +  u_{y,n} \, \delta{{u_y}} \right] \, R \, \mathrm{d} \theta \;= \;0  & \qquad  \forall \; \{\delta{v}, \delta{w}\}  \in  \mathcal{U} \, , \\
\int_0^{2\pi} \left[EA \, {L}_{\epsilon}(u_{x,n}, u_{y,n}) \, \delta {\epsilon} + EI \, {L}_{\kappa}(u_{x,n}, u_{y,n}) \, \delta{\kappa} \right] \, R \, \mathrm{d} \theta \;\;\; - \qquad \qquad & \nonumber  \\ 
\int_0^{2\pi} \left[ EA \, {\epsilon}_n \, \delta {\epsilon} + EI \, {\kappa}_n \, \delta{\kappa} \right] \, R \, \mathrm{d} \theta \;= \;0 & \qquad \forall \; \{\delta{\epsilon},\delta{\kappa}\} \in  \mathcal{S}  \, , \label{eq:weak_form_mixedHR_cart2}
\end{align}
where $\mathcal{U}= ( H^2 \times H^2 ) $ and $\mathcal{S} = ( L^2 \times L^2) $ are the Sobolev spaces of periodic functions, defined on the ring domain $[0, 2\pi]$. In the remainder of this work, we will apply the variational formulation \eqref{eq:weak_form_mixedHR_cart1} and \eqref{eq:weak_form_mixedHR_cart2} as well as the strain-displacement relations \eqref{eq21} and \eqref{eq22} as the basis for understanding different finite element discretization schemes.

\subsection{Standard isogeometric finite element discretization} \label{sec:standardIGA}

In this paper, we employ splines as basis functions, which are widely used today in the context of isogeometric analysis \cite{Hughes2005,Cottrell2009}. For further details, we refer to the recent reviews \cite{Haberleitner:17.1,Hughes:17.1,Schillinger:18.1} and the references therein. 

\subsubsection{Uniform periodic B-splines on a circular ring}

\begin{figure}[h!]
\begin{center}
	\def\svgwidth{0.92\textwidth}
	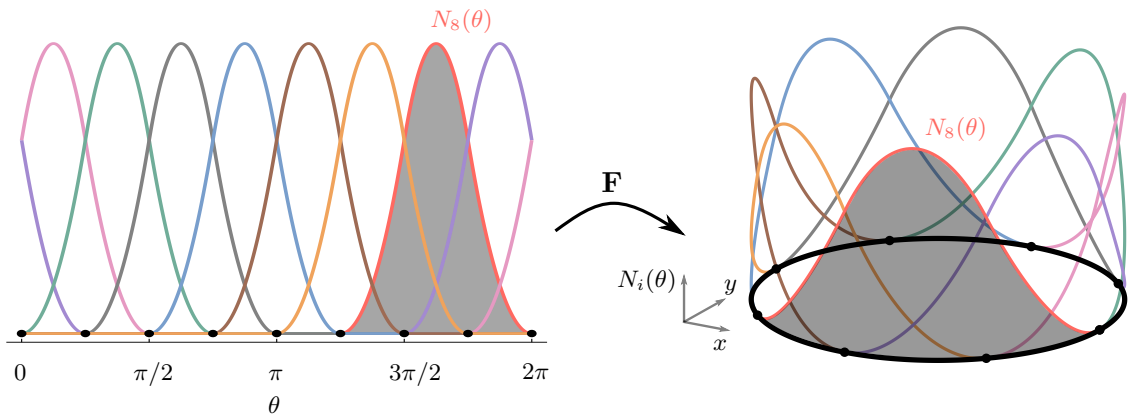
\end{center}
\caption{The space $\mathcal{V}_8^{2,1}(0, 2 \pi)$, consisting of $\hat{n}_b=8$ quadratic periodic B-splines on the circle.}
\label{fig:periodic_splines_circle}
\end{figure}

A spline is a piecewise polynomial, characterized by the polynomial degree $p$ of its segments and the prescribed smoothness at the segment interfaces. In the following, we employ smooth B-splines with maximum continuity $C^{p-1}$ with $p \geq 2$, defined on a uniform partition of the circle. We construct periodic spline discretizations of dimension $\hat{n}_b$ by taking $\hat{n}_b+p$ sequential B-splines and applying suitable end conditions to the last $p$ B-splines. We denote the resulting discrete space spanned by the $\hat{n}_b$ periodic B-spline basis functions of polynomial degree $p$ \textcolor{black}{with continuity $C^{p-1}$ as $\mathcal{V}^{p,p-1}_{\hat{n}_b}$}. 
For the circular ring, we apply an exact geometric mapping, $\mathbf{F}$, using trigonometric functions:
\begin{align}
(x,y) = \mathbf{F}(\theta) = (R \cos(\theta), R \sin(\theta))
\end{align}
Figure~\ref{fig:periodic_splines_circle} shows a graphical illustration. 
Construction, differentiation and integration of B-splines can be performed using standard spline formulae, see \cite{Schumaker2007,Piegl1996}.

\begin{remark}
Spline discretization with repeated knots lead to the appearance of outlier eigenvalues and eigenmodes at the end of the spectrum \cite{Cottrell2006,Hughes2008,Hiemstra:21.1}. The use of uniform periodic B-spline discretizations of the circular ring eliminates this issue entirely, highlighting another advantage of our choice of benchmark problem. 
\end{remark}

\subsubsection{Displacement-based stiffness and mass matrices}

Standard displacement-based finite element methods presume that the strain-displacement relations are exactly satisfied by the finite element approximation. On the one hand, the strain field variables can therefore be eliminated from the variational formulation by replacing them by the displacement-based expressions incorporated in the strain-displacement operators \eqref{eq21} and \eqref{eq22}. It is thus sufficient to discretize the Cartesian displacements and virtual displacements by a finite sum of $\hat{n}_b$ B-spline basis functions $N_i(\theta)$ multiplied by unknown coefficients:
\begin{align} 
u^h_x(\theta) & = \sum_{i=1}^{\hat{n}_{\text{b}}} N_i(\theta) \, \eigenvec_x^i \, , & u^h_y(\theta) = \sum_{i=1}^{\hat{n}_{\text{b}}} N_i(\theta) \, \eigenvec_y^i \, , \label{eq23}\\
\delta u^h_x(\theta) & = \sum_{i=1}^{\hat{n}_{\text{b}}} N_i(\theta) \, \delta \eigenvec_x^i \, , & \delta u^h_y(\theta) = \sum_{i=1}^{\hat{n}_{\text{b}}} N_i(\theta) \, \delta \eigenvec_y^i \, . \label{eq24}
\end{align}
On the other hand, the weak form of the kinematic constraints \eqref{eq:weak_form_mixedHR_cart2} is a priori satisfied strongly, and can thus be removed from the variational formulation. 

Inserting \eqref{eq23} and \eqref{eq24} in the remaining weak form \eqref{eq:weak_form_mixedHR_cart1} yields the standard finite element formulation of the generalized eigenvalue problem: find $( \{u^h_{x,n}, u^h_{y,n} \}, \lambda^h_{in}) \in \mathcal{V} \times \mathbb{R}^+$ such that   
\begin{align}
\label{eq:weak_form_standard}
\int_0^{2\pi} \left[ EA \, {L}_{\epsilon}(u^h_{x,n},u^h_{y,n}) \, {L}_{\epsilon} (\delta{{u^h_x}},\delta{{u^h_y}}) + EI \, {L}_{\kappa}(u^h_{x,n},u^h_{y,n}) \,{L}_{\kappa} (\delta {{u^h_x}},\delta {{u^h_y}}) \right] & \, R \, \mathrm{d} \theta  \;\;\; - \nonumber  \\ 
\lambda^h_{in} \int_0^{2\pi} \rho A  \left[ u^h_{x,n} \, \delta{{u^h_x}} +  u^h_{y,n} \, \delta{{u^h_y}} \right] \, R \, \mathrm{d} \theta \;= \;0  \qquad  \forall \; & \{\delta{v}, \delta{w}\}  \in  \mathcal{V}^{p,p-1}_{\hat{n}_b} \, , 
\end{align}
with $n=1,2, \ldots, \hat{n}_b$ and $i=1,2$. The space $\mathcal{V} = (\mathcal{V}^{p,p-1}_{\hat{n}_b} \times \mathcal{V}^{p,p-1}_{\hat{n}_b})$, where $p\geq2$, consists of $\hat{n}_b$ periodic \textcolor{black}{$C^{p-1}$} B-spline functions of at least quadratic polynomial degree defined on the ring domain $[0, 2\pi]$. It entails displacements and virtual displacements in the sense of the Galerkin method.

From \eqref{eq:weak_form_standard}, it is straightforward to retrieve the standard form of a discrete generalized eigenvalue problem \eqref{dgep} \cite{Hughes2000}. For the circular Euler-Bernoulli ring, the stiffness matrix of the standard displacement-based finite element method is 
\begin{align}\label{eq:stiffness_matrix_ring}
\mat{K} = \underbrace{EA \int_0^{2\pi} \mat{B}_m^T \, \mat{B}_m \, R \,\mathrm d \theta}_{\mat{K}_m} + \underbrace{EI \int_0^{2\pi} \mat{B}_b^T \, \mat{B}_b \, R \, \mathrm d \theta}_{\mat{K}_b} \, ,
\end{align}
that can be divided into a membrane part, $\mat{K}_m$, and a bending part, $\mat{K}_b$. The discrete strain-displacement matrices for the membrane strain and the change of curvature, $\mat{B}_m$ and $\mat{B}_b$, are: 
\begin{align}\label{eq:B_matrices}
	& \mat{B}_m = \frac{1}{R} \, \begin{bmatrix}
		-N_{1,\theta} (\theta) \sin(\theta) \; \ldots \; -N_{\hat{n}_b,\theta} (\theta) \sin(\theta) \;\; N_{1,\theta} (\theta) \cos(\theta) \; \ldots \; N_{\hat{n}_b,\theta} (\theta) \cos(\theta)
	\end{bmatrix} \, , \\ 
	& \mat{B}_b = \frac{1}{R^2} \, \begin{bmatrix}
		-N_{1,\theta \theta} (\theta) \cos(\theta) + N_{1,\theta} (\theta) \sin(\theta) \; \ldots \; 
		-N_{1,\theta \theta} (\theta) \sin(\theta) - N_{1,\theta} (\theta) \cos(\theta) \; \ldots \; 
	\end{bmatrix} \, .
\end{align}
The consistent mass matrix is a $2 \times 2$ block diagonal matrix of the following form
	\begin{align}\label{eq:mass_matrix_ring}
		 \mat{M} = \begin{bmatrix}
			\mmat{M} & \\ & \mmat{M}
		\end{bmatrix}, \;\; \text{with} \;
		 M_{ij} = \rho A \int_0^{2\pi} N_i (\theta) \, N_j (\theta) \, R \, \mathrm d \theta \, , \; \{i,j\} = 1, 2, \ldots, \hat{n}_b \, .
	\end{align}

\subsubsection{Full versus reduced integration}

The standard finite element formulation \eqref{eq:weak_form_standard} of the Euler-Bernoulli circular ring uses Gauss quadrature with $(p+1)$ quadrature points to numerically integrate the entries of the stiffness and mass matrices \eqref{eq:stiffness_matrix_ring} and \eqref{eq:mass_matrix_ring}, commonly denoted as full integration.
It is well-known, however, that the standard formulation with full integration suffers from severe membrane locking.

In the case of the Euler-Bernoulli beam, selective reduced integration performs numerical integration of the membrane part of the stiffness matrix in \eqref{eq:stiffness_matrix_ring} with a quadrature rule that accurately integrates only a subset of all polynomials within an element. 
On the one hand, reduced selective integration is simple to implement and operates with the same displacement-based standard variational formulation \eqref{eq:weak_form_standard}. On the other hand, 
reduced integration can imply unstable solution behavior due to the appearance of spurious zero-energy modes \cite{Flanagan1981,Belytschko1984,Liu1984}. 
For other reduced quadrature schemes suitable for IGA, we refer the reader to \cite{Zou_quadrature2021,Adam2015b,Hiemstra2017a} and the references therein. 
In this study, we will employ a reduced quadrature scheme based on Gaussian quadrature that uses $p$ quadrature points per spline segment for the integration of the membrane stiffness matrix $\mat{K}_m$ in \eqref{eq:stiffness_matrix_ring}, and $p+1$ quadrature points for the integration of the bending stiffness and mass matrices $\mat{K}_b$ in \eqref{eq:stiffness_matrix_ring} and $\mat{M}$ in \eqref{eq:mass_matrix_ring}. One can show that this choice
preserves full accuracy and still avoids spurious modes \cite{Schillinger:14.2}.

\section{Three membrane locking-free finite element formulations} \label{sec:formulations2}

In the following, we briefly review and compare the key concepts of three well-established methods that are widely used to mitigate membrane locking. These are B-bar strain projection, the discrete strain gap method, and a mixed formulation based on the Hellinger-Reissner principle. We note that their use is not limited to membrane locking in the Euler-Bernoulli beam formulation, but all three have been successfully employed for mitigating a variety of locking phenomena in different structure and material models. For the sake of conciseness, we state their formulation directly for the Euler-Bernoulli ring and refer to the literature for a more general presentation. 


\begin{remark}
The assumed natural strain (ANS) method is equivalent to the B-bar method, since the spaces of assumed strain fields and projection spaces are equivalent to appropriate projection operators \cite{Antolin2019}. The ANS method can thus be expected to provide similar results to the B-bar method. The enhanced assumed strain (EAS) method is in some sense equivalent to the Hellinger-Reissner formulation \cite{Bischoff1999}, and can thus be expected to provide similar results. We therefore do not include the ANS and EAS methods in the following study.
\end{remark}

\subsection{B-bar strain projection}\label{sec:Bbar_method}

The B-bar strain projection method was initially developed to treat volumetric locking \cite{Hughes1977, Hughes1980} and then extended to isogeometric analysis and other types of locking such as transverse shear and membrane locking \cite{Elguedj2007, Antolin2019, Bouclier2013}. The basic idea is to project the strain components associated with locking onto a basis of lower dimension so that the locking effect is alleviated. 
A common choice for the definition of a projector is the minimization of the $L^2$ norm. 
The B-bar strain projection method can be applied for any type of locking phenomenon and with any polynomial basis of arbitrary degree and spatial dimension. The projected strain fields result in a modified strain tensor and modified strain-displacement matrix, $\mat{B}$, hence the name. There are no additional point sets to evaluate and no additional stiffness terms or variables are required.

For our study, we adapt the Timoshenko beam formulation presented in \cite{Bouclier2012} to the Euler-Bernoulli case, resulting in a modification of the membrane stiffness matrix $\mat{K}_m$ in \eqref{eq:stiffness_matrix_ring}. We choose a projection using the $L^2$ norm for the membrane strain and a basis of degree $p-1$ \textcolor{black}{with continuity $C^{p-2}$} and $p$ \textcolor{black}{with continuity $C^{p-1}$, where $p \geq 2$,} for the projected membrane strain and the displacement field, respectively. The modified membrane stiffness matrix is then: 
\begin{align}
\bar{\mathbf{K}}_m = EA \;\; \bar{\mat{B}}_m^T \, \bar{\mat{M}}^{-1} \, \bar{\mat{B}}_m \, .
\end{align}
For our example of the circular Euler-Bernoulli ring discretized with a periodic B-spline basis, the matrices resulting from the projection procedure are defined as		
\begin{align}	
	& \bar{\mat{B}}_m = \left[\bar{\mat{B}}_{1} \; \bar{\mat{B}}_{2} \right], \;\; \text{with} \;\;
	\bar{B}_{1,ij} = - \int_0^{2\pi} \sin(\theta) \, \bar{N}_i \, N_{j,\theta} \, \mathrm d \theta \, , \;\; 
	\bar{B}_{2,ij} = \int_0^{2\pi} \cos(\theta) \, \bar{N}_i \, N_{j,\theta} \, \mathrm d \theta, \nonumber \\
	& \text{and} \;\; \bar{M}_{ij} = \int_0^{2\pi} \bar{N}_i \, \bar{N}_j \, R \, \mathrm d \theta \, , \quad \text{where} \;\;  \{i,j\} = 1,2, \ldots, \hat{n}_b \, ,  \;\; (N, \bar{N} )\in \mathcal{V}^{p,p-1}_{\hat{n}_b} \times \tilde{\mathcal{V}}^{p-1,p-2}_{\hat{n}_b} \, . \nonumber
\end{align}
The entries in $\bar{\mat{B}}$ and $\bar{\mat{M}}$, corresponding to the projected strain in a basis of degree $p-1$, are evaluated with $p$ quadrature points in each \Bezier element. The bending stiffness matrix, $\mathbf{K}_b$, and the consistent mass matrix, $\mathbf{M}$, remain unaffected by the projection.

\subsection{Discrete strain gap method}\label{sec:DSG_method}

The discrete strain gap (DSG) method was originally developed to alleviate transverse shear locking in plates and shells \cite{Bletzinger2000,Bischoff2001}, and then extended to membrane locking \cite{Koschnick2005} and isogeometric analysis \cite{Bouclier2012, Echter2010, Echter2013}. Its main idea is to enable the strain fields associated with locking to represent zero strains by modifying the interpolation of these strain fields. The DSG method can be classified as a B-bar method since it results in a modified strain-displacement matrix. For details on the procedure to modify the strain interpolation and obtain the modified stiffness matrix, we refer to \cite{Bouclier2012, Echter2010}. 
For our example of the circular Euler-Bernoulli ring, we adapt the procedure described for the Timoshenko curved beam in \cite{Bouclier2012} and obtain the modified strain-displacement matrix, $\bar{\mat{B}}_m$, in the following form:
\begin{align}
\bar{\mathbf{B}}_m = \frac{1}{R} \, \left[\tilde{N}_{1,\theta} (\theta) \text{ } \tilde{N}_{2,\theta} (\theta) \text{ } \ldots \tilde{N}_{N,\theta} (\theta) \right] \mat{A}^{-1} \mat{C} \; \mat{D}  \, .
\end{align}
The matrices are defined as
\begin{align}
A_{ij} = \tilde{N}_j (\theta_i) \, , \quad C_{ij} = - \frac{1}{R} \int_0^{\theta_i} \sin(\theta) \, N_{j,\theta} (\theta) \, R \, \mathrm{d} \theta \, , \quad D_{ij} = \frac{1}{R} \int_0^{\theta_i} \cos(\theta) \, N_{j,\theta} (\theta) \, R \, \mathrm{d} \theta, \nonumber
\end{align}
where $N_i (\theta)$ and $\tilde{N}_i (\theta)$ denote the basis functions interpolating the displacement fields and the modified membrane contribution, respectively, and $\theta_i$ is the angular coordinate corresponding to the $i^{\text{th}}$ collocation point. Substituting the modified matrix, $\bar{\mat{B}}_m$, in \eqref{eq:stiffness_matrix_ring} results in the modified membrane stiffness matrix, following the DSG method, which is equivalent to what is described in \cite[eq.~55]{Bouclier2012}. 
The evaluation of the matrices, $\mat{A}, \mat{C}, \mat{D}$, depends only on the set of collocation points and does not require any assembly routine. The basis functions $\tilde{N}_{i,\theta} (\theta)$ in the first term of $\bar{\mathbf{B}}_m$ (the first row vector) are evaluated in an assembly routine at quadrature points in each element.
	
In \cite{Bouclier2012, Echter2010}, the authors apply NURBS to describe the geometry and to interpolate all variable fields including the modified strain contribution (i.e. NURBS basis functions as $\tilde{N}_i$). In our computations in section \ref{sec:spectra_beam}, we choose a space of uniform B-splines that is of the same degree and defined on the same uniform open knot vector as the space of the uniform periodic \textcolor{black}{$C^{p-1}$} B-splines, $\mathcal{V}^h_p$, to interpolate the modified membrane contribution. Thus:
\begin{align}
	(N, \tilde{N} )\in \mathcal{V}^{p,p-1}_{\hat{n}_b} \times \mathcal{S}^{p,p-1}_{\tilde{n}_b} \, , \nonumber 
\end{align}	
where $\mathcal{V}^{p,p-1}_{\hat{n}_b}$ denotes the space of uniform periodic B-splines of degree $p$ with continuity $C^{p-1}$, and $\mathcal{S}^{p,p-1}_{\tilde{n}_b}$ denotes the space of uniform B-splines of degree $p$ with continuity $C^{p-1}$, defined on an open knot vector. 
Our choice of collocation points is the Greville abscissa corresponding to the uniform B-splines of the modified membrane contribution \cite{Bouclier2012}.

\subsection{Hellinger-Reissner principle}\label{sec:mixedHR_method}

The third locking-free formulation that we consider is the mixed formulation that follows from the Hellinger-Reissner principle \cite{Bischoff1999, Echter2013, Taylor2011, Stolarski1986}. This is a two-field formulation of displacement and either stress or strain fields. When we choose appropriate approximation spaces for these fields, we can eliminate locking. The Hellinger-Reissner mixed formulation can be derived from the general three-field mixed formulation (Hu-Washizu principle) by satisfying the  constitutive relation strongly \cite{Bischoff1999}. 	
	
The mixed variational formulation based on the Hellinger-Reissner principle of the eigenvalue problem for an Euler-Bernoulli circular ring in Cartesian coordinates was derived in section \ref{sec:variational_form_cartesian}. Discretizing the independent variable fields, $\left( \mat{u}^h, \boldsymbol{\epsilon}^h \right) = \left( [u_{x}, u_{y} ]^T, [{\epsilon}, {\kappa}]^T \right) \in \mathcal{V}^{p,p-1}_{\hat{n}_b} \times \tilde{\mathcal{V}}^{p-1,p-2}_{\hat{n}_b}$, and inserting those in the variational formulation, \eqref{eq:weak_form_mixedHR_cart1} and \eqref{eq:weak_form_mixedHR_cart2}, yields the following matrix equation of the eigenvalue problem based on the Hellinger-Reissner principle for the Euler-Bernoulli circular ring:
\begin{align}\label{eq:mixed_form_HR_matrix}
		\begin{bmatrix}
		\mathbf{K}_{11} & \mathbf{K}_{12} \\
		\mathbf{K}_{12}^T & 0
	\end{bmatrix} \begin{bmatrix} \boldsymbol{\epsilon} \\ \mat{\eigenvec} \end{bmatrix} = \begin{bmatrix}
		0 \\ \boldsymbol{\lambda} \, \mat{M} \mat{\eigenvec}
	\end{bmatrix}, \quad \text{with} \;\;\;
		\mathbf{K}_{11} = \begin{bmatrix}
		\mathbf{k}_{11} & 0 \\ 0 & \mathbf{k}_{22}
	\end{bmatrix}, \quad \mathbf{K}_{12} = \begin{bmatrix}
		\mathbf{k}_{13} & \mathbf{k}_{14} \\ \mathbf{k}_{23} & \mathbf{k}_{24}
	\end{bmatrix} \, , \nonumber
\end{align}
where the entries of the different blocks of the stiffness matrix are defined as
\begin{align}
	& k_{11,ij} = -EA \int_0^{2\pi} \bar{N}_i \bar{N}_j \, R \, \mathrm d \theta, \quad k_{22,ij} = -EI \int_0^{2\pi} \bar{N}_i \bar{N}_j \, R \, \mathrm d \theta, \\
	& k_{13,ij} = EA \int_0^{2\pi} \bar{N}_i \left(-\frac{1}{R} \sin(\theta) N_{j,\theta} \right) \, R \, \mathrm d \theta \\
	& k_{14,ij} = EA \int_0^{2\pi} \bar{N}_i \left(\frac{1}{R} \cos(\theta) N_{j,\theta} \right) \, R \, \mathrm d \theta \\
	& k_{23,ij} = EI \int_0^{2\pi} \bar{N}_i \left(- \frac{1}{R^2} \cos(\theta) N_{j,\theta \theta} + \frac{1}{R^2} \sin(\theta) N_{j,\theta} \right) \, R \, \mathrm d \theta \\
	& k_{24,ij} = EI \int_0^{2\pi} \bar{N}_i \left(- \frac{1}{R^2} \sin(\theta) N_{j,\theta \theta} - \frac{1}{R^2} \cos(\theta) N_{j,\theta} \right) \, R \, \mathrm d \theta \, , \\
	& \text{where} \;\;  \{i,j\} = 1,2, \ldots, \hat{n}_b \, ,  \;\;  (N, \bar{N} )\in \mathcal{V}^{p,p-1}_{\hat{n}_b} \times \tilde{\mathcal{V}}^{p-1,p-2}_{\hat{n}_b} \, . \nonumber 
\end{align}	
To eliminate the secondary field, we can apply static condensation which leads to the final stiffness matrix of the eigenvalue problem based on the Hellinger-Reissner mixed formulation:
\begin{align}
	\mat{K} = - \mathbf{K}_{12}^T \mathbf{K}_{11}^{-1} \mathbf{K}_{12} \; .
\end{align}
	
We note that the mass matrix $\mat{M}$ remains unchanged to what is defined in \eqref{eq:mass_matrix_ring}.

\section{Assessing membrane locking and unlocking via spectral analysis}\label{sec:spectra_beam}

In this section, we demonstrate for a slender Euler-Bernoulli circular ring that spectral analysis can be an effective tool to assess locking and unlocking in finite element formulations. To this end, we compare the errors of the eigenvalues and eigenmodes across the spectrum for the standard finite element formulations with full and reduced integration as well as the three locking-free formulations that we reviewed above. In particular, we use spectral analysis to assess their sensitivity to locking on a coarse mesh, under mesh refinement, and for $p$-refinement (i.e. increasing polynomial degree as well as smoothness). We show that the spectral approximation properties of each formulation can be directly related to its locking deficiency or unlocking capability. 

\subsection{Locking indicator based on spectral analysis}
In the following, we measure locking and unlocking from a spectral analysis viewpoint based on the following criterion:

\vspace{0.36cm}

\noindent\begin{tabular}{| p{16cm} |}
\hline
\rowcolor{black!5}

\vspace{-0.2cm}
\noindent\textbf{Locking indicator:} A method is locking-free if the normalized spectra obtained on coarse meshes are ``close'' to asymptotic refinement curves obtained from ``overkill'' discretizations, that is, the normalized spectra do not significantly change with mesh refinement. Accordingly, a method is locking-prone if its normalized spectra differ significantly from on asymptotic refinement curves obtained from ``overkill'' discretizations, that is, the normalized spectra significantly change with mesh refinement.
\vspace{0.1cm}
\\ \hline
\end{tabular}
\vspace{0.5cm}

\begin{figure}[h!]
	\centering
	\includegraphics[width=0.9\textwidth]{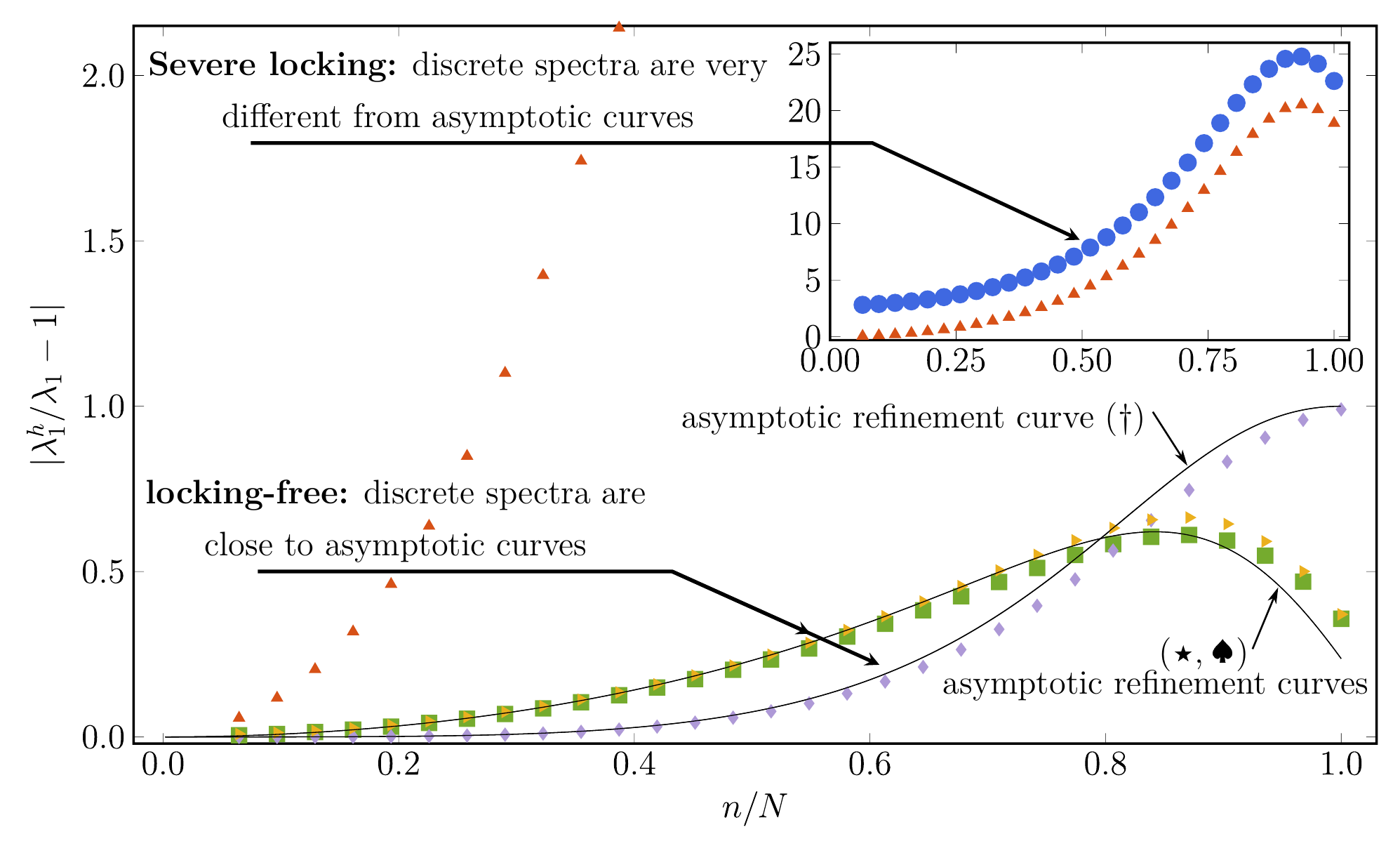}
	\def\svgwidth{0.7\textwidth}
	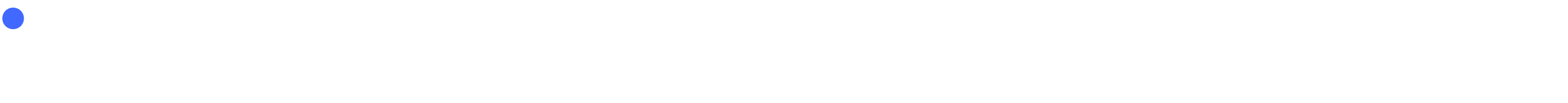
	\caption{Our locking indicator provides a visual tool to understand locking and unlocking in discrete spectra.}
	\label{fig:locking_indicator}
\end{figure}

The above statement is best described by means of an illustration. Figure \ref{fig:locking_indicator} depicts the normalized error in the eigenvalues $\lambda_1^h$ computed with quadratic B-splines ($p=2$) on a mesh of 64 elements ($N=32$), for ``large'' slenderness ratio $R/t = 2000/3$. We observe from the plot that the spectra obtained with full integration and reduced integration differ significantly from the asymptotic refinement curves, and hence the corresponding formulations severely lock over the entire range of the spectrum. In contrast, the discrete spectra of the B-bar, Hellinger-Reissner and DSG methods are close their asymptotic refinement curves, and are thus locking-free.

\vspace{0.36cm}

From a practical standpoint, we can now proceed as follows. For the finite element formulation in question, eigenvalue and mode errors are computed on a coarse discretization and corresponding asymptotic eigenvalue and mode errors are computed with an ``overkill'' discretization. The eigenvalue and mode errors from the coarse mesh are related to the corresponding asymptotic eigenvalue and mode errors by plotting both sets with respect to the normalized mode number $n/N$, where $N$ denotes the total number of modes in each discretization. This relation is based on the notion that - given the underlying solution behavior is sufficiently resolved - all spectral error curves plotted over their normalized mode numbers must be identical, irrespective of the mesh size and the associated number of degrees of freedom. As a consequence, the finite element formulation is locking-free, if the corresponding spectral error curves are matching up irrespective of the mesh size, and the finite element formulation is locking-prone, if the corresponding spectral error curves are different, implying that the spectral error curve changes with mesh refinement.

\subsection{In-depth comparison for quadratic splines}\label{sec:results_methods_thin_beam}
\label{Sec51}

\begin{figure}[h!]
	\centering
	\begin{center}
		\def\svgwidth{0.85\textwidth}
\begingroup%
  \makeatletter%
  \providecommand\color[2][]{%
    \errmessage{(Inkscape) Color is used for the text in Inkscape, but the package 'color.sty' is not loaded}%
    \renewcommand\color[2][]{}%
  }%
  \providecommand\transparent[1]{%
    \errmessage{(Inkscape) Transparency is used (non-zero) for the text in Inkscape, but the package 'transparent.sty' is not loaded}%
    \renewcommand\transparent[1]{}%
  }%
  \providecommand\rotatebox[2]{#2}%
  \newcommand*\fsize{\dimexpr\f@size pt\relax}%
  \newcommand*\lineheight[1]{\fontsize{\fsize}{#1\fsize}\selectfont}%
  \ifx\svgwidth\undefined%
    \setlength{\unitlength}{574.77697754bp}%
    \ifx\svgscale\undefined%
      \relax%
    \else%
      \setlength{\unitlength}{\unitlength * \real{\svgscale}}%
    \fi%
  \else%
    \setlength{\unitlength}{\svgwidth}%
  \fi%
  \global\let\svgwidth\undefined%
  \global\let\svgscale\undefined%
  \makeatother%
  \begin{picture}(1,0.59406518)%
    \lineheight{1}%
    \setlength\tabcolsep{0pt}%
    \put(0,0){\includegraphics[width=\unitlength,page=1]{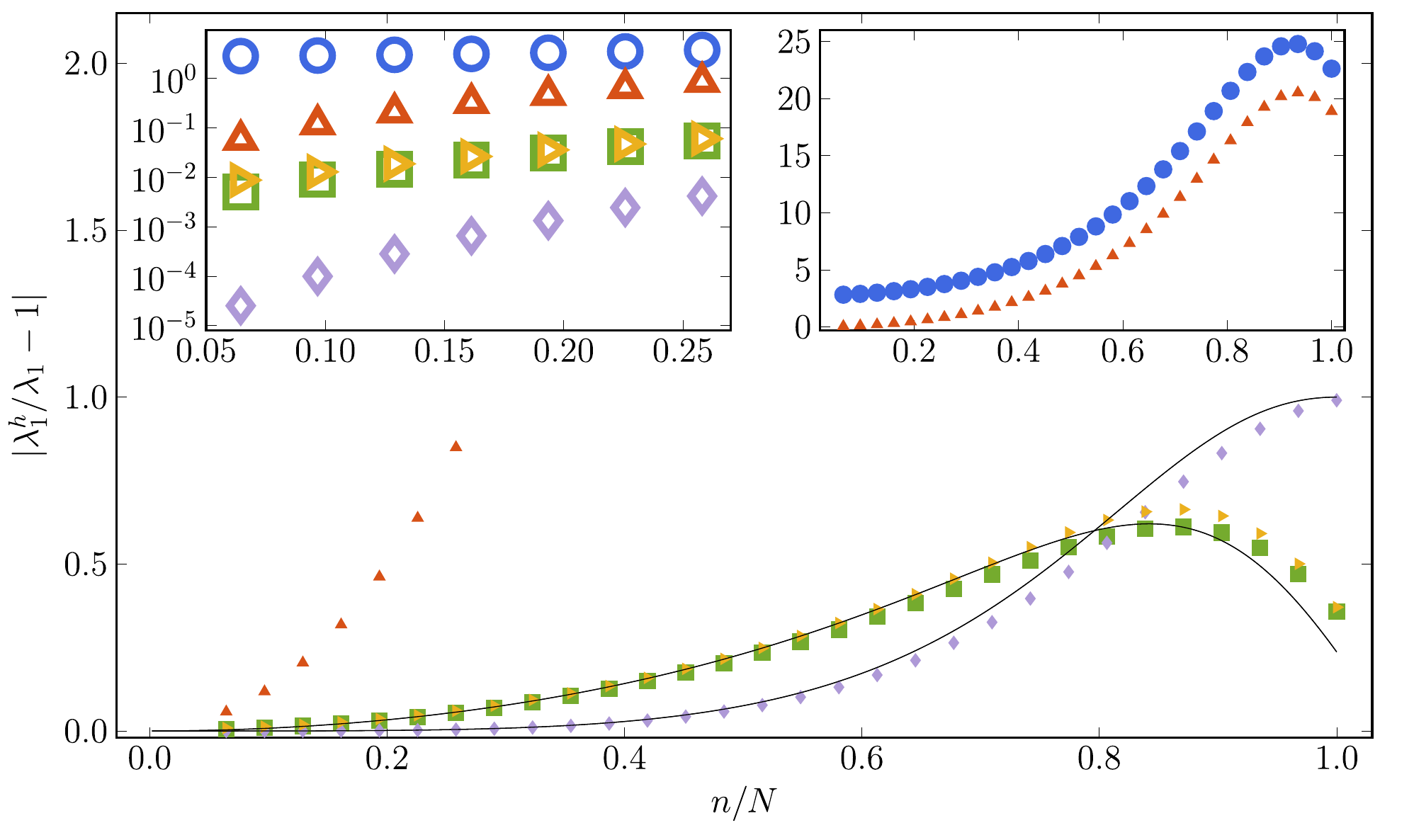}}%
    \put(0.78085676,0.29830717){\color[rgb]{0,0,0}\makebox(0,0)[lt]{\lineheight{1.25}\smash{\begin{tabular}[t]{l}$(\dagger)$\end{tabular}}}}%
    \put(0,0){\includegraphics[width=\unitlength,page=2]{eigen_ring_methods_spectra1_t0.0015_p2_64ele.pdf}}%
    \put(0.84239302,0.12021465){\color[rgb]{0,0,0}\makebox(0,0)[lt]{\lineheight{1.25}\smash{\begin{tabular}[t]{l}$(\star, \spadesuit)$\end{tabular}}}}%
    \put(0,0){\includegraphics[width=\unitlength,page=3]{eigen_ring_methods_spectra1_t0.0015_p2_64ele.pdf}}%
  \end{picture}%
\endgroup%
 \\
		\footnotesize{(a) Normalized error in eigenvalues $\lambda_1^h$} (associated with transverse modes $\eigenvec_1^h$) \\
	\end{center}
	\begin{center}
		\def\svgwidth{0.85\textwidth}
\begingroup%
  \makeatletter%
  \providecommand\color[2][]{%
    \errmessage{(Inkscape) Color is used for the text in Inkscape, but the package 'color.sty' is not loaded}%
    \renewcommand\color[2][]{}%
  }%
  \providecommand\transparent[1]{%
    \errmessage{(Inkscape) Transparency is used (non-zero) for the text in Inkscape, but the package 'transparent.sty' is not loaded}%
    \renewcommand\transparent[1]{}%
  }%
  \providecommand\rotatebox[2]{#2}%
  \newcommand*\fsize{\dimexpr\f@size pt\relax}%
  \newcommand*\lineheight[1]{\fontsize{\fsize}{#1\fsize}\selectfont}%
  \ifx\svgwidth\undefined%
    \setlength{\unitlength}{574.77697754bp}%
    \ifx\svgscale\undefined%
      \relax%
    \else%
      \setlength{\unitlength}{\unitlength * \real{\svgscale}}%
    \fi%
  \else%
    \setlength{\unitlength}{\svgwidth}%
  \fi%
  \global\let\svgwidth\undefined%
  \global\let\svgscale\undefined%
  \makeatother%
  \begin{picture}(1,0.59406518)%
    \lineheight{1}%
    \setlength\tabcolsep{0pt}%
    \put(0,0){\includegraphics[width=\unitlength,page=1]{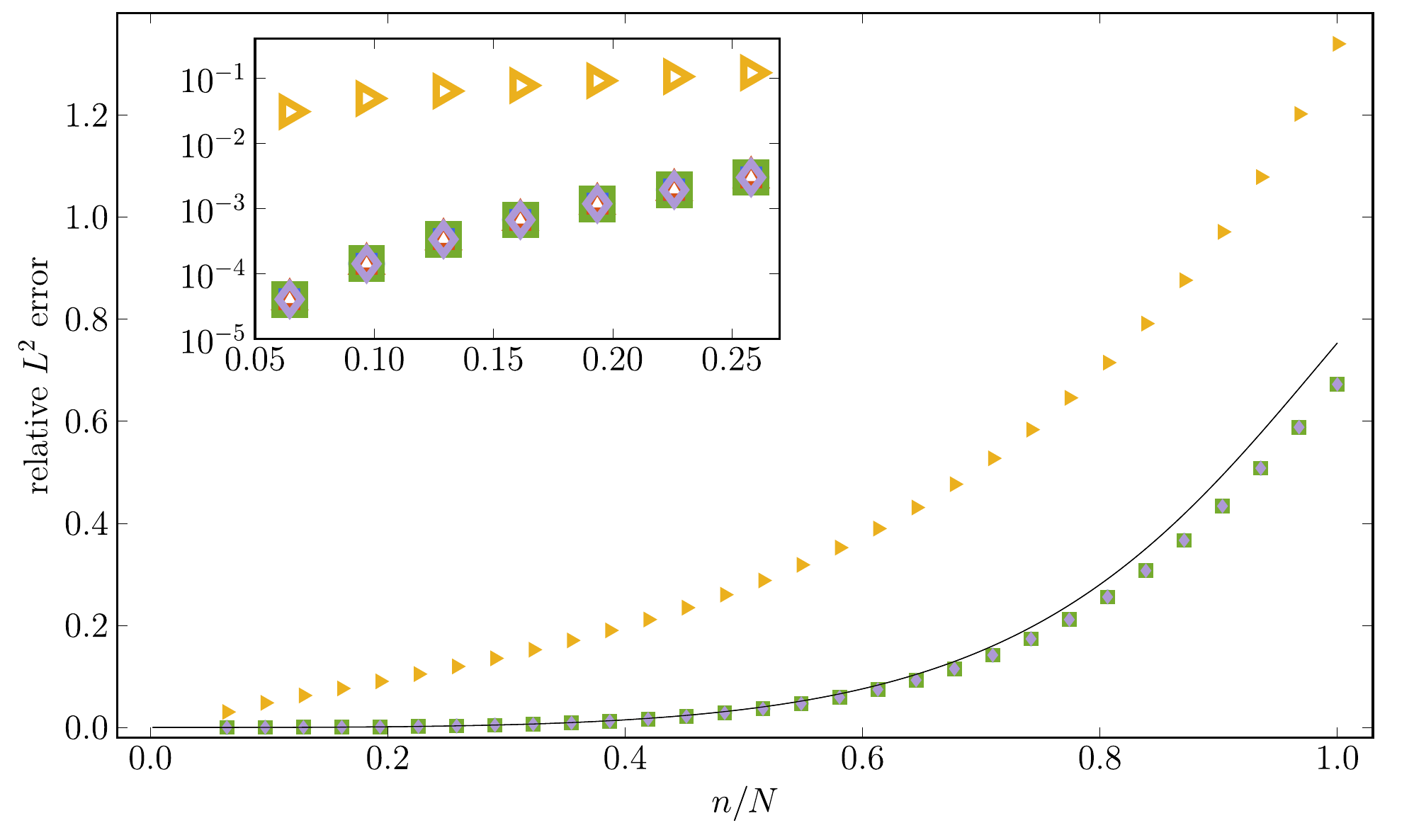}}%
    \put(0.86946889,0.18194544){\color[rgb]{0,0,0}\makebox(0,0)[lt]{\lineheight{1.25}\smash{\begin{tabular}[t]{l}$(\star, \dagger, \spadesuit)$\end{tabular}}}}%
    \put(0,0){\includegraphics[width=\unitlength,page=2]{eigen_ring_methods_L2mode1_t0.0015_p2_64ele.pdf}}%
    \put(0.56412378,0.5218114){\color[rgb]{0,0,0}\makebox(0,0)[lt]{\lineheight{1.25}\smash{\begin{tabular}[t]{l}\footnotesize{Overlapping}\end{tabular}}}}%
    \put(0,0){\includegraphics[width=\unitlength,page=3]{eigen_ring_methods_L2mode1_t0.0015_p2_64ele.pdf}}%
  \end{picture}%
\endgroup%
 \\
		\footnotesize{(b) Normalized $L^2$-norm error in transverse mode shapes $\eigenvec_1^h$}
	\end{center}
	\begin{center}
		\def\svgwidth{0.7\textwidth}
		\input{sections/figures/svg/legend_methods.pdf_tex}
	\end{center}
	\caption{Normalized errors in eigenvalues $\lambda_1^h$ and \textbf{transverse} mode shapes $\eigenvec_1^h$ computed with \textbf{quadratic B-splines} ($p=2$) on a mesh of \textbf{64 elements} ($N=32$), for ``large'' slenderness ratio $R/t = 2000/3$.}
	\label{fig:spectra1_compare_formulations}
\end{figure}

\begin{figure}[h!]
	\centering
	\begin{center}
		\def\svgwidth{0.85\textwidth}
\begingroup%
  \makeatletter%
  \providecommand\color[2][]{%
    \errmessage{(Inkscape) Color is used for the text in Inkscape, but the package 'color.sty' is not loaded}%
    \renewcommand\color[2][]{}%
  }%
  \providecommand\transparent[1]{%
    \errmessage{(Inkscape) Transparency is used (non-zero) for the text in Inkscape, but the package 'transparent.sty' is not loaded}%
    \renewcommand\transparent[1]{}%
  }%
  \providecommand\rotatebox[2]{#2}%
  \newcommand*\fsize{\dimexpr\f@size pt\relax}%
  \newcommand*\lineheight[1]{\fontsize{\fsize}{#1\fsize}\selectfont}%
  \ifx\svgwidth\undefined%
    \setlength{\unitlength}{574.77697754bp}%
    \ifx\svgscale\undefined%
      \relax%
    \else%
      \setlength{\unitlength}{\unitlength * \real{\svgscale}}%
    \fi%
  \else%
    \setlength{\unitlength}{\svgwidth}%
  \fi%
  \global\let\svgwidth\undefined%
  \global\let\svgscale\undefined%
  \makeatother%
  \begin{picture}(1,0.59928459)%
    \lineheight{1}%
    \setlength\tabcolsep{0pt}%
    \put(0,0){\includegraphics[width=\unitlength,page=1]{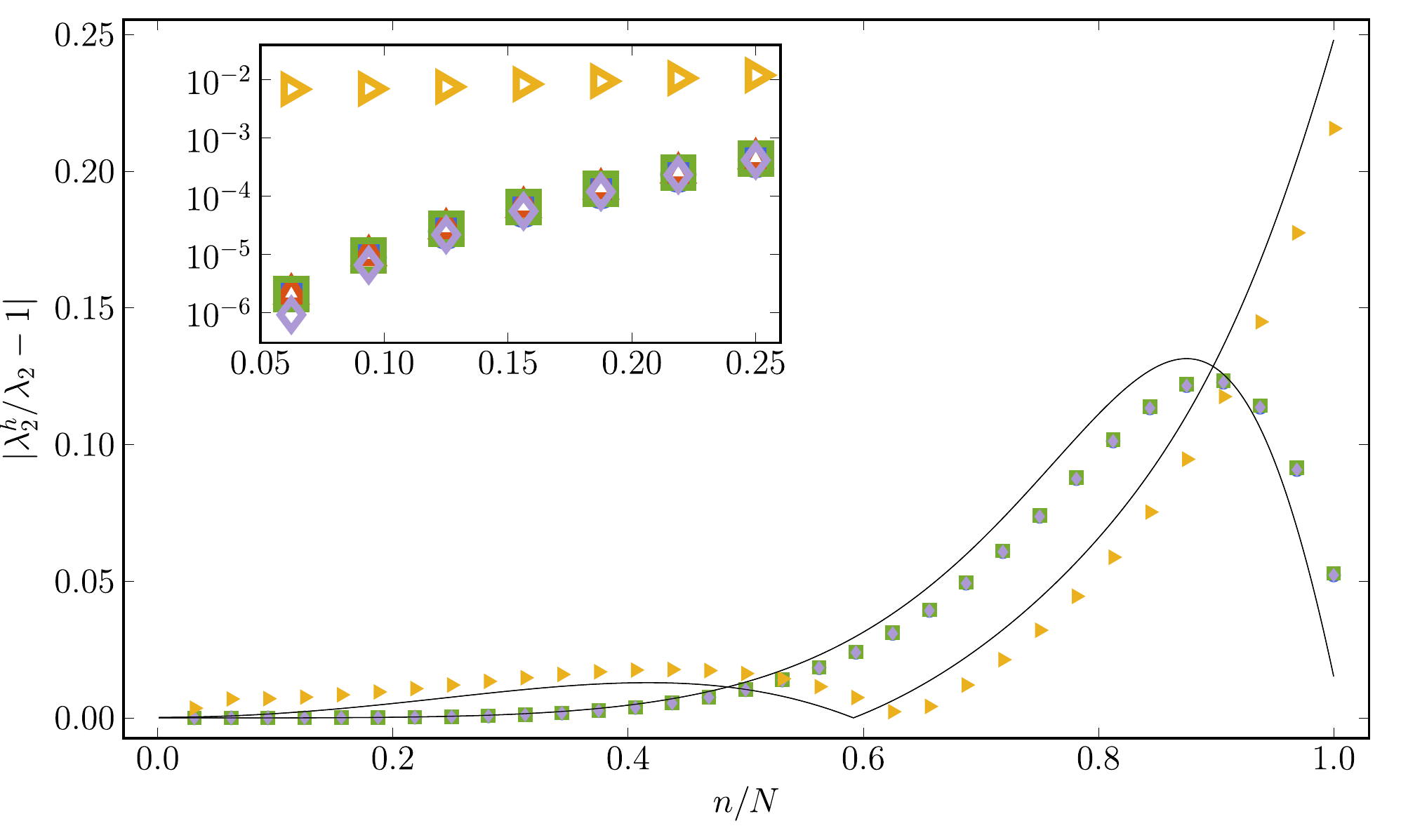}}%
    \put(0.78211908,0.40550574){\color[rgb]{0,0,0}\makebox(0,0)[lt]{\lineheight{1.25}\smash{\begin{tabular}[t]{l}$(\spadesuit)$\end{tabular}}}}%
    \put(0,0){\includegraphics[width=\unitlength,page=2]{eigen_ring_methods_spectra2_t0.0015_p2_64ele.pdf}}%
    \put(0.7140887,0.33491665){\color[rgb]{0,0,0}\makebox(0,0)[lt]{\lineheight{1.25}\smash{\begin{tabular}[t]{l}$(\star, \dagger)$\end{tabular}}}}%
    \put(0,0){\includegraphics[width=\unitlength,page=3]{eigen_ring_methods_spectra2_t0.0015_p2_64ele.pdf}}%
    \put(0.57002074,0.52550239){\color[rgb]{0,0,0}\makebox(0,0)[lt]{\lineheight{1.25}\smash{\begin{tabular}[t]{l}\footnotesize{Overlapping}\end{tabular}}}}%
    \put(0,0){\includegraphics[width=\unitlength,page=4]{eigen_ring_methods_spectra2_t0.0015_p2_64ele.pdf}}%
  \end{picture}%
\endgroup%
 \\
		\footnotesize{(a) Normalized error in eigenvalues $\lambda_2^h$ (associated with circumferential modes  $\eigenvec_2^h$)}
	\end{center}
	\begin{center}
		\def\svgwidth{0.85\textwidth}
\begingroup%
  \makeatletter%
  \providecommand\color[2][]{%
    \errmessage{(Inkscape) Color is used for the text in Inkscape, but the package 'color.sty' is not loaded}%
    \renewcommand\color[2][]{}%
  }%
  \providecommand\transparent[1]{%
    \errmessage{(Inkscape) Transparency is used (non-zero) for the text in Inkscape, but the package 'transparent.sty' is not loaded}%
    \renewcommand\transparent[1]{}%
  }%
  \providecommand\rotatebox[2]{#2}%
  \newcommand*\fsize{\dimexpr\f@size pt\relax}%
  \newcommand*\lineheight[1]{\fontsize{\fsize}{#1\fsize}\selectfont}%
  \ifx\svgwidth\undefined%
    \setlength{\unitlength}{574.77697754bp}%
    \ifx\svgscale\undefined%
      \relax%
    \else%
      \setlength{\unitlength}{\unitlength * \real{\svgscale}}%
    \fi%
  \else%
    \setlength{\unitlength}{\svgwidth}%
  \fi%
  \global\let\svgwidth\undefined%
  \global\let\svgscale\undefined%
  \makeatother%
  \begin{picture}(1,0.59928459)%
    \lineheight{1}%
    \setlength\tabcolsep{0pt}%
    \put(0,0){\includegraphics[width=\unitlength,page=1]{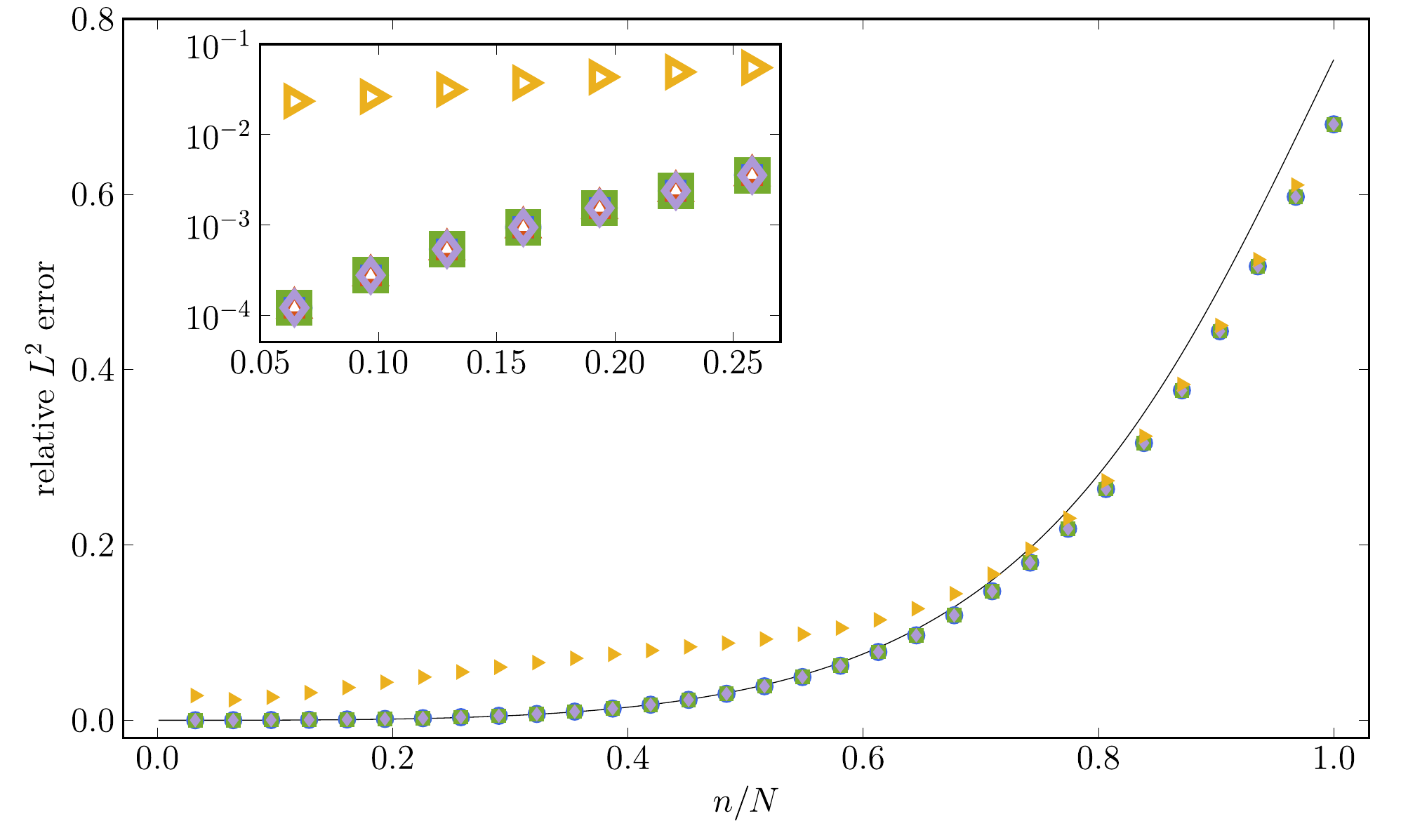}}%
    \put(0.76369494,0.40214788){\color[rgb]{0,0,0}\makebox(0,0)[lt]{\lineheight{1.25}\smash{\begin{tabular}[t]{l}$(\star, \dagger, \spadesuit)$\end{tabular}}}}%
    \put(0,0){\includegraphics[width=\unitlength,page=2]{eigen_ring_methods_L2mode2_t0.0015_p2_64ele.pdf}}%
    \put(0.56960146,0.52580356){\color[rgb]{0,0,0}\makebox(0,0)[lt]{\lineheight{1.25}\smash{\begin{tabular}[t]{l}\footnotesize{Overlapping}\end{tabular}}}}%
    \put(0,0){\includegraphics[width=\unitlength,page=3]{eigen_ring_methods_L2mode2_t0.0015_p2_64ele.pdf}}%
  \end{picture}%
\endgroup%
 \\
		\footnotesize{(b) Normalized $L^2$-norm error in circumferential mode shapes $\eigenvec_2^h$}
	\end{center}
	\begin{center}
		\def\svgwidth{0.7\textwidth}
		\input{sections/figures/svg/legend_methods.pdf_tex}
	\end{center}
	\caption{Normalized errors in eigenvalues $\lambda_2^h$ and \textbf{circumferential} mode shapes $\eigenvec_2^h$ computed with \textbf{quadratic B-splines} ($p=2$) on a mesh of \textbf{64 elements} ($N=32$), for ``large'' slenderness ratio $R/t = 2000/3$.}
	\label{fig:spectra2_compare_formulations}
\end{figure}

In the first step, we study the effect of membrane locking on the spectral approximation properties for ``large'' ring slenderness $R/t=2000/3$. For this case, we can expect severe membrane locking to occur, as demonstrated in our initial cantilever example in the introduction (see Fig.~\ref{fig:cantilever_convergence}). For each finite element formulation, we compute the discrete eigenvalues and modes using periodic uniform B-splines of polynomial degree two, defined on 64 B\'ezier elements. Figures~\ref{fig:spectra1_compare_formulations} and \ref{fig:spectra2_compare_formulations} plot the relative eigenvalue errors \eqref{eve} and the relative $L^2$-norm mode errors \eqref{eme} across the normalized spectrum for the transverse eigenmodes $\eigenvec_1^h$ and the associated eigenvalues $\lambda_1^h$ and the circumferential eigenmodes $\eigenvec_2^h$ and the associated eigenvalues $\lambda_2^h$, respectively. The eigenvalue and mode errors are obtained with respect to the analytical solutions given in \ref{sec:analytic_eigen_ring}. Furthermore, we compute asymptotic spectral error curves numerically for each finite element formulation with an ``overkill'' discretization of 2048 elements.

\begin{remark}
Readers interested in the technical details for identifying transverse and circumferential modes and ordering them correctly to comply with the analytical ordering are referred to \ref{sec:postprocessing}.
\end{remark}

\subsubsection{Standard formulation with full and reduced integration}

We start by considering the spectral analysis results obtained with the standard formulation with full integration, plotted in Figs.~\ref{fig:spectra1_compare_formulations} and \ref{fig:spectra2_compare_formulations} with blue circles. We would like to identify the impact of membrane locking on the accuracy of the spectrum. Firstly, we focus on the eigenvalue error corresponding to the transverse modes, plotted in Fig.~\ref{fig:spectra1_compare_formulations}a. We observe that the transverse eigenvalues obtained with the standard finite element formulation show significant error levels, with the error curves being far away from the asymptotic reference curve plotted in black. We attribute this increase in error level to the effect of membrane locking, which is further supported by a look at the locking-free formulations that do not show a similar increase, producing transverse eigenvalue errors that match the asymptotic reference curve. 
As the accuracy of the transverse eigenvalues is heavily affected over the complete spectrum, we conclude that beam computations with the standard formulation at this mesh size will yield completely inaccurate results. This is confirmed by the convergence plots in Fig.~\ref{fig:cantilever_convergence_methods}a computed for our initial cantilever example, where the standard formulation does not converge for practical mesh sizes.

Secondly, we focus on the remaining spectral error quantities, that is, the mode errors for the transverse and circumferential mode shapes and the eigenvalue error corresponding to the circumferential modes, plotted in Fig.~\ref{fig:spectra1_compare_formulations}b, Fig.~\ref{fig:spectra2_compare_formulations}a and Fig.~\ref{fig:spectra2_compare_formulations}b. We observe that they show exactly the same error as the locking-free formulations, with the error curves being practically identical to the asymptotic reference curves. We conclude that for the case of the Euler-Bernoulli ring, membrane locking only influences the accuracy of the eigenvalues of the transverse modes, while the transverse mode shapes and both the eigenvalues and mode shapes of the circumferential modes do not lock.

Thirdly, we consider the spectral analysis results obtained with the standard formulation with selective reduced integration, which  in Figs.~\ref{fig:spectra1_compare_formulations} and \ref{fig:spectra2_compare_formulations} are plotted with red triangles. On the one hand, we observe in Fig.~\ref{fig:spectra1_compare_formulations}a that compared to full integration, selective reduced integration is able to improve the spectral accuracy of the lowest eigenvalues of the transverse modes. On the other hand, the spectral accuracy degenerates very quickly with increasing mode number. As a consequence, accurate finite element approximations of beam solutions on coarse meshes, where most of the spectrum is required to actively contribute, are not possible. Our conclusion is supported by Fig.~\ref{fig:cantilever_convergence_methods}a for the cantilever example, where convergence at the best possible accuracy level can only be achieved for finer mesh sizes.

\subsubsection{B-bar method}

We then move forward to the locking-free formulations. We first consider the results obtained with the B-bar formulation, which in Figs.~\ref{fig:spectra1_compare_formulations} and \ref{fig:spectra2_compare_formulations} are plotted with green squares.
We observe that the B-bar method eliminates the effect of membrane locking in the entire spectrum of the eigenvalues corresponding to the transverse modes plotted in Fig.~\ref{fig:spectra1_compare_formulations}a. In addition, all error curves, both for the transverse and circumferential mode shapes and associated eigenvalues, correspond well to the asymptotic error curves, irrespective of the mesh size. This indicates that the finite element formulation with the B-bar method already achieves full spectral accuracy on the current coarser mesh. On the one hand, we conclude that the B-bar formulation successfully mitigates membrane locking. On the other hand, we can preclude any negative effect from the B-bar method on the convergence properties of the finite element formulation. Our conclusions therefore confirm that the B-bar method is an effective locking-free finite element formulation for the Euler-Bernoulli beam. They are supported by our initial cantilever example whose convergence results in Fig.~\ref{fig:cantilever_convergence_methods}a show the best possible accuracy on coarse and fine meshes that can be achieved for $p=2$ in a purely displacement based formulation.

\subsubsection{DSG method}

We then consider the results obtained with the DSG formulation, which in Figs.~\ref{fig:spectra1_compare_formulations} and \ref{fig:spectra2_compare_formulations} are plotted with yellow triangles. On the one hand, Figure~\ref{fig:spectra1_compare_formulations}a shows that the DSG method completely eliminates the effect of locking on the locking-prone eigenvalues of the transverse modes, with the error curve closely matching the asymptotic error curve. On the other hand, the DSG method exhibits an increased level of error across large parts of the spectrum for the transverse mode shapes, the circumferential mode shapes and the eigenvalue error of the circumferential modes, which can be observed in Fig.~\ref{fig:spectra1_compare_formulations}b, Figs.~\ref{fig:spectra2_compare_formulations}a and \ref{fig:spectra2_compare_formulations}b. These observations are remarkable as we concluded in the discussion above that these three quantities are not affected by membrane locking for the current Euler-Bernoulli ring problem. 
In particular, we can see in the inset plots that the low modes that are important for the approximation power of the basis (see Section~\ref{sec:lowmode}) are significantly less accurate compared to the other formulations. We therefore conclude that the DSG formulations itself is responsible for the increase in error levels, and thus deteriorates the accuracy of the original standard finite element formulation with respect to part of the spectrum. We note that in Fig.~\ref{fig:cantilever_convergence_methods}a, the DSG method still achieves the best possible accuracy on coarse and fine meshes, as in our cantilever example, the transverse mode behavior dominates the overall accuracy of the analysis.

\begin{remark}
The kink in the circumferential eigenvalue spectrum obtained with the DSG method is due to the fact that the ratio $\lambda_2^h / \lambda_2$ changes from positive to negative at that location.
\end{remark}

\begin{figure}[h!]
	\centering
	\begin{center}
		\def\svgwidth{0.85\textwidth}
\begingroup%
  \makeatletter%
  \providecommand\color[2][]{%
    \errmessage{(Inkscape) Color is used for the text in Inkscape, but the package 'color.sty' is not loaded}%
    \renewcommand\color[2][]{}%
  }%
  \providecommand\transparent[1]{%
    \errmessage{(Inkscape) Transparency is used (non-zero) for the text in Inkscape, but the package 'transparent.sty' is not loaded}%
    \renewcommand\transparent[1]{}%
  }%
  \providecommand\rotatebox[2]{#2}%
  \newcommand*\fsize{\dimexpr\f@size pt\relax}%
  \newcommand*\lineheight[1]{\fontsize{\fsize}{#1\fsize}\selectfont}%
  \ifx\svgwidth\undefined%
    \setlength{\unitlength}{574.77697754bp}%
    \ifx\svgscale\undefined%
      \relax%
    \else%
      \setlength{\unitlength}{\unitlength * \real{\svgscale}}%
    \fi%
  \else%
    \setlength{\unitlength}{\svgwidth}%
  \fi%
  \global\let\svgwidth\undefined%
  \global\let\svgscale\undefined%
  \makeatother%
  \begin{picture}(1,0.59928459)%
    \lineheight{1}%
    \setlength\tabcolsep{0pt}%
    \put(0,0){\includegraphics[width=\unitlength,page=1]{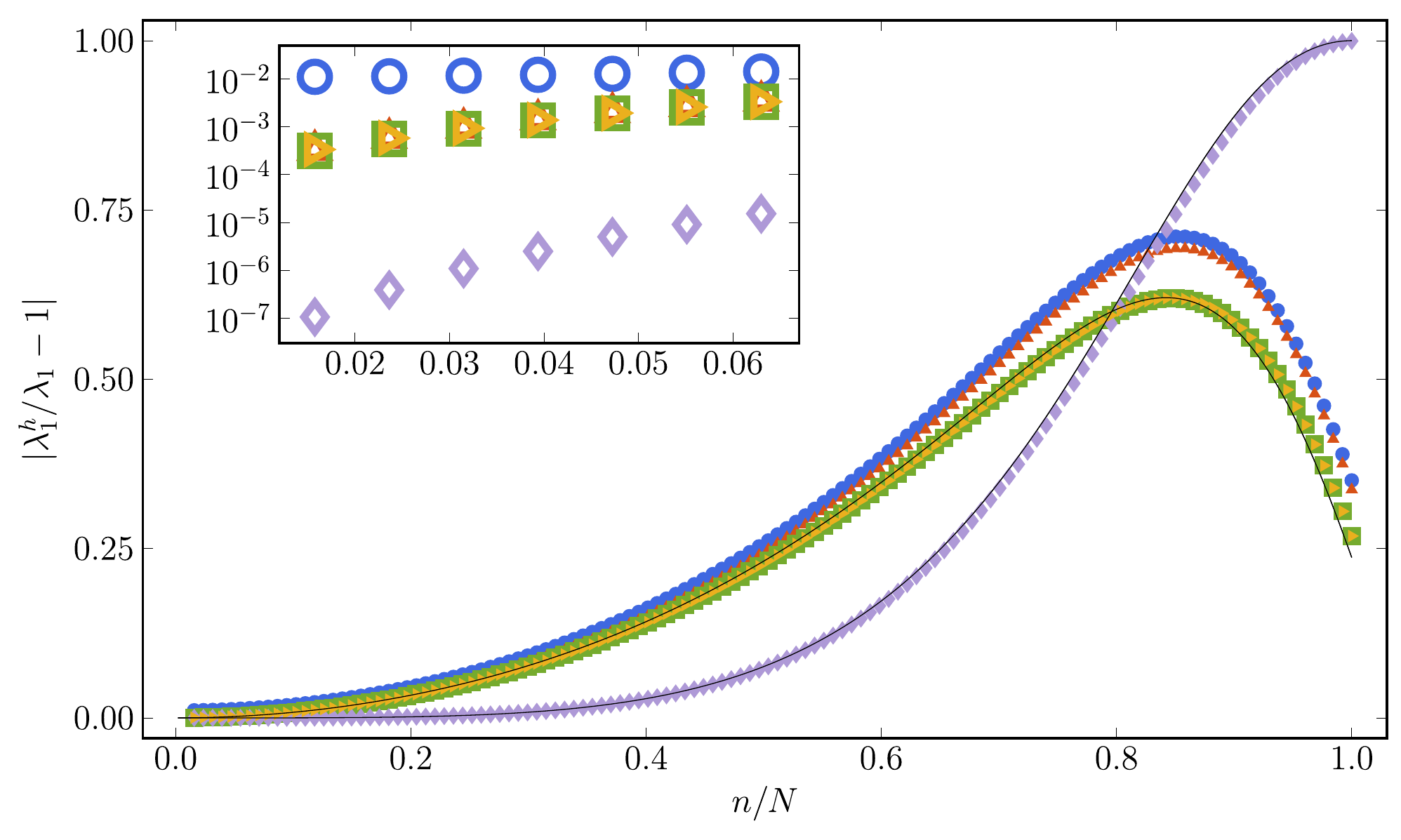}}%
    \put(0.83608545,0.55298184){\color[rgb]{0,0,0}\makebox(0,0)[lt]{\lineheight{1.25}\smash{\begin{tabular}[t]{l}$(\dagger)$\end{tabular}}}}%
    \put(0,0){\includegraphics[width=\unitlength,page=2]{eigen_ring_methods_spectra1_t0.0015_p2_256ele.pdf}}%
    \put(0.91318376,0.17646236){\color[rgb]{0,0,0}\makebox(0,0)[lt]{\lineheight{1.25}\smash{\begin{tabular}[t]{l}$(\star, \spadesuit)$\end{tabular}}}}%
    \put(0,0){\includegraphics[width=\unitlength,page=3]{eigen_ring_methods_spectra1_t0.0015_p2_256ele.pdf}}%
    \put(0.58309687,0.48635676){\color[rgb]{0,0,0}\makebox(0,0)[lt]{\lineheight{1.25}\smash{\begin{tabular}[t]{l}\footnotesize{Overlapping}\end{tabular}}}}%
    \put(0,0){\includegraphics[width=\unitlength,page=4]{eigen_ring_methods_spectra1_t0.0015_p2_256ele.pdf}}%
  \end{picture}%
\endgroup%
 \\
		\footnotesize{(a) Normalized error in eigenvalues $\lambda_1^h$ (associated with transverse modes  $\eigenvec_1^h$)}
	\end{center}
	\begin{center}
		\def\svgwidth{0.85\textwidth}
\begingroup%
  \makeatletter%
  \providecommand\color[2][]{%
    \errmessage{(Inkscape) Color is used for the text in Inkscape, but the package 'color.sty' is not loaded}%
    \renewcommand\color[2][]{}%
  }%
  \providecommand\transparent[1]{%
    \errmessage{(Inkscape) Transparency is used (non-zero) for the text in Inkscape, but the package 'transparent.sty' is not loaded}%
    \renewcommand\transparent[1]{}%
  }%
  \providecommand\rotatebox[2]{#2}%
  \newcommand*\fsize{\dimexpr\f@size pt\relax}%
  \newcommand*\lineheight[1]{\fontsize{\fsize}{#1\fsize}\selectfont}%
  \ifx\svgwidth\undefined%
    \setlength{\unitlength}{574.77697754bp}%
    \ifx\svgscale\undefined%
      \relax%
    \else%
      \setlength{\unitlength}{\unitlength * \real{\svgscale}}%
    \fi%
  \else%
    \setlength{\unitlength}{\svgwidth}%
  \fi%
  \global\let\svgwidth\undefined%
  \global\let\svgscale\undefined%
  \makeatother%
  \begin{picture}(1,0.59928459)%
    \lineheight{1}%
    \setlength\tabcolsep{0pt}%
    \put(0,0){\includegraphics[width=\unitlength,page=1]{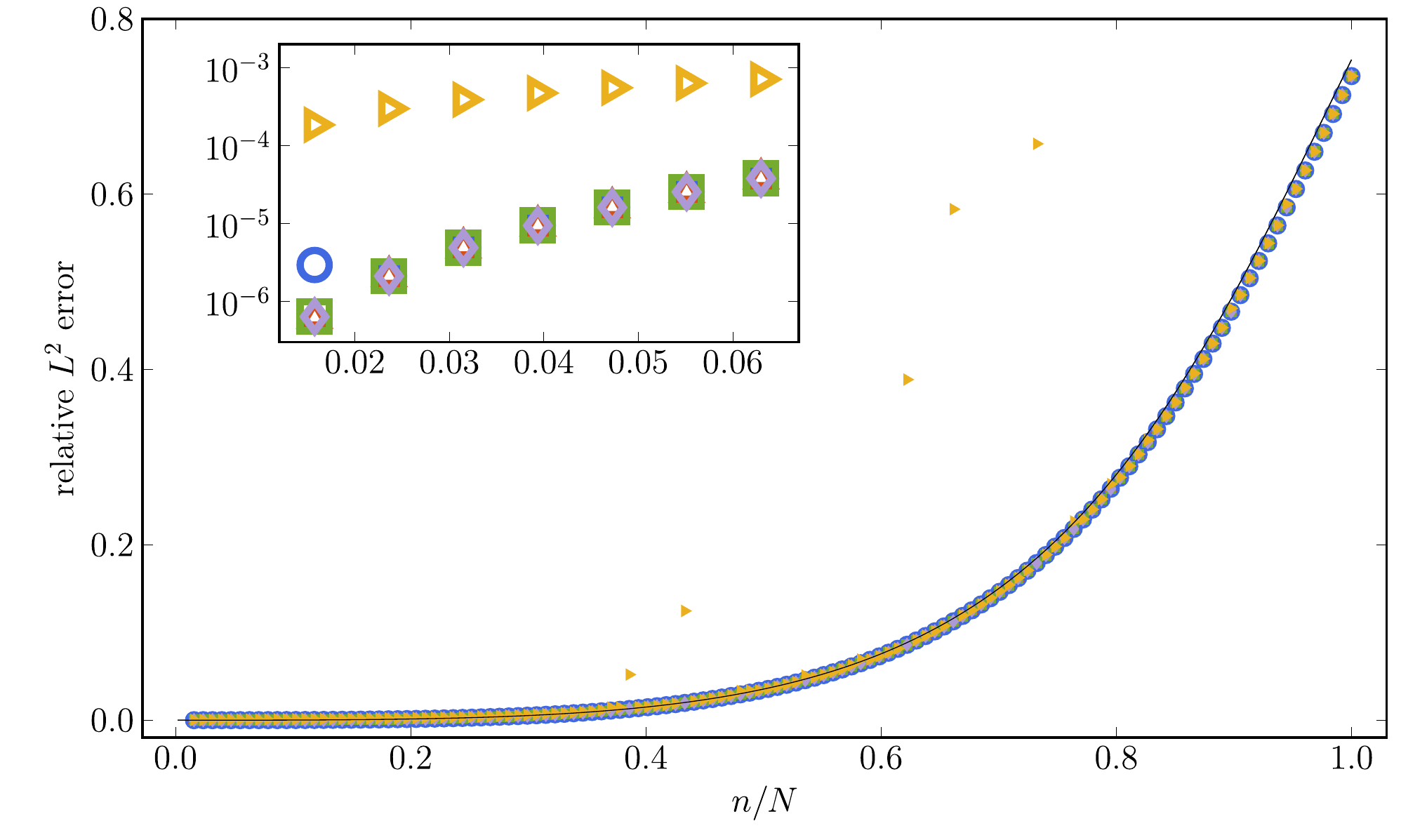}}%
    \put(0.81611481,0.47924378){\color[rgb]{0,0,0}\makebox(0,0)[lt]{\lineheight{1.25}\smash{\begin{tabular}[t]{l}$(\star, \dagger, \spadesuit)$\end{tabular}}}}%
    \put(0,0){\includegraphics[width=\unitlength,page=2]{eigen_ring_methods_L2mode1_t0.0015_p2_256ele.pdf}}%
    \put(0.59875519,0.54113295){\color[rgb]{0,0,0}\makebox(0,0)[lt]{\lineheight{1.25}\smash{\begin{tabular}[t]{l}\footnotesize{Overlapping}\end{tabular}}}}%
    \put(0,0){\includegraphics[width=\unitlength,page=3]{eigen_ring_methods_L2mode1_t0.0015_p2_256ele.pdf}}%
  \end{picture}%
\endgroup%
 \\
		\footnotesize{(b) Normalized $L^2$-norm error in transverse mode shapes $\eigenvec_1^h$}
	\end{center}
	\begin{center}
		\def\svgwidth{0.7\textwidth}
		\input{sections/figures/svg/legend_methods.pdf_tex}
	\end{center}
	\caption{Normalized errors in eigenvalues $\lambda_1^h$ and \textbf{transverse} mode shapes $\eigenvec_1^h$ computed with \textbf{quadratic B-splines} ($p=2$) on a mesh of \textbf{256 elements} ($N=128$), for ``large'' slenderness ratio $R/t = 2000/3$.}
	\label{fig:spectra1_compare_formulations_h}
\end{figure}

\begin{figure}[h!]
	\centering
	\begin{center}
		\def\svgwidth{0.85\textwidth}
\begingroup%
  \makeatletter%
  \providecommand\color[2][]{%
    \errmessage{(Inkscape) Color is used for the text in Inkscape, but the package 'color.sty' is not loaded}%
    \renewcommand\color[2][]{}%
  }%
  \providecommand\transparent[1]{%
    \errmessage{(Inkscape) Transparency is used (non-zero) for the text in Inkscape, but the package 'transparent.sty' is not loaded}%
    \renewcommand\transparent[1]{}%
  }%
  \providecommand\rotatebox[2]{#2}%
  \newcommand*\fsize{\dimexpr\f@size pt\relax}%
  \newcommand*\lineheight[1]{\fontsize{\fsize}{#1\fsize}\selectfont}%
  \ifx\svgwidth\undefined%
    \setlength{\unitlength}{574.77697754bp}%
    \ifx\svgscale\undefined%
      \relax%
    \else%
      \setlength{\unitlength}{\unitlength * \real{\svgscale}}%
    \fi%
  \else%
    \setlength{\unitlength}{\svgwidth}%
  \fi%
  \global\let\svgwidth\undefined%
  \global\let\svgscale\undefined%
  \makeatother%
  \begin{picture}(1,0.59406518)%
    \lineheight{1}%
    \setlength\tabcolsep{0pt}%
    \put(0,0){\includegraphics[width=\unitlength,page=1]{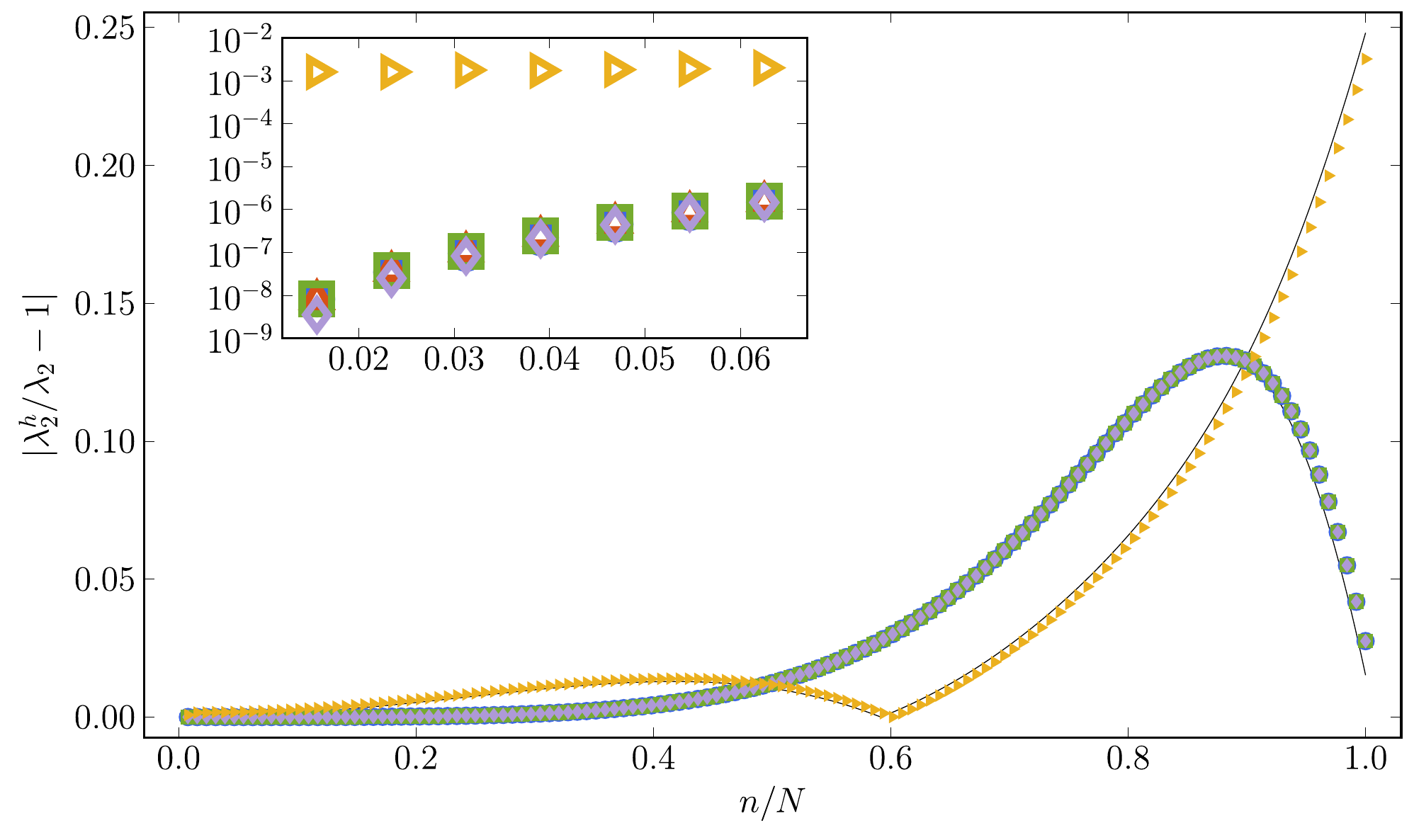}}%
    \put(0.87848649,0.12823943){\color[rgb]{0,0,0}\makebox(0,0)[lt]{\lineheight{1.25}\smash{\begin{tabular}[t]{l}$(\star, \dagger)$\end{tabular}}}}%
    \put(0,0){\includegraphics[width=\unitlength,page=2]{eigen_ring_methods_spectra2_t0.0015_p2_256ele.pdf}}%
    \put(0.81770178,0.39961588){\color[rgb]{0,0,0}\makebox(0,0)[lt]{\lineheight{1.25}\smash{\begin{tabular}[t]{l}$(\spadesuit)$\end{tabular}}}}%
    \put(0,0){\includegraphics[width=\unitlength,page=3]{eigen_ring_methods_spectra2_t0.0015_p2_256ele.pdf}}%
    \put(0.58437408,0.5085393){\color[rgb]{0,0,0}\makebox(0,0)[lt]{\lineheight{1.25}\smash{\begin{tabular}[t]{l}\footnotesize{Overlapping}\end{tabular}}}}%
    \put(0,0){\includegraphics[width=\unitlength,page=4]{eigen_ring_methods_spectra2_t0.0015_p2_256ele.pdf}}%
  \end{picture}%
\endgroup%
 \\
		\footnotesize{(a) Normalized error in eigenvalues $\lambda_2^h$ (associated with circumferential modes  $\eigenvec_2^h$)}
	\end{center}
	\begin{center}
		\def\svgwidth{0.85\textwidth}
\begingroup%
  \makeatletter%
  \providecommand\color[2][]{%
    \errmessage{(Inkscape) Color is used for the text in Inkscape, but the package 'color.sty' is not loaded}%
    \renewcommand\color[2][]{}%
  }%
  \providecommand\transparent[1]{%
    \errmessage{(Inkscape) Transparency is used (non-zero) for the text in Inkscape, but the package 'transparent.sty' is not loaded}%
    \renewcommand\transparent[1]{}%
  }%
  \providecommand\rotatebox[2]{#2}%
  \newcommand*\fsize{\dimexpr\f@size pt\relax}%
  \newcommand*\lineheight[1]{\fontsize{\fsize}{#1\fsize}\selectfont}%
  \ifx\svgwidth\undefined%
    \setlength{\unitlength}{574.77697754bp}%
    \ifx\svgscale\undefined%
      \relax%
    \else%
      \setlength{\unitlength}{\unitlength * \real{\svgscale}}%
    \fi%
  \else%
    \setlength{\unitlength}{\svgwidth}%
  \fi%
  \global\let\svgwidth\undefined%
  \global\let\svgscale\undefined%
  \makeatother%
  \begin{picture}(1,0.59928459)%
    \lineheight{1}%
    \setlength\tabcolsep{0pt}%
    \put(0,0){\includegraphics[width=\unitlength,page=1]{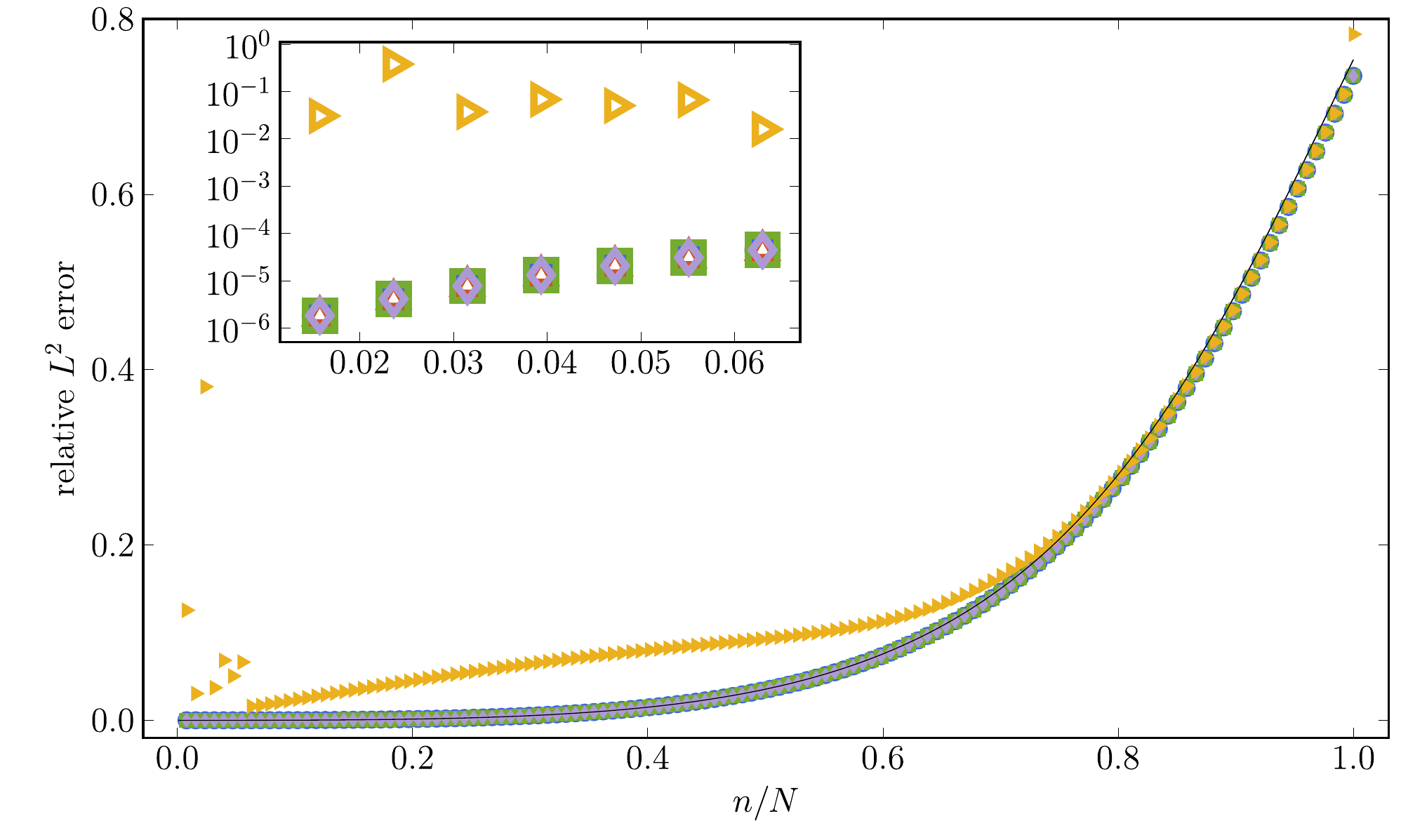}}%
    \put(0.73363693,0.40038703){\color[rgb]{0,0,0}\makebox(0,0)[lt]{\lineheight{1.25}\smash{\begin{tabular}[t]{l}$(\star, \dagger, \spadesuit)$\end{tabular}}}}%
    \put(0,0){\includegraphics[width=\unitlength,page=2]{eigen_ring_methods_L2mode2_t0.0015_p2_256ele.pdf}}%
    \put(0.58330673,0.48226923){\color[rgb]{0,0,0}\makebox(0,0)[lt]{\lineheight{1.25}\smash{\begin{tabular}[t]{l}\footnotesize{Overlapping}\end{tabular}}}}%
    \put(0,0){\includegraphics[width=\unitlength,page=3]{eigen_ring_methods_L2mode2_t0.0015_p2_256ele.pdf}}%
  \end{picture}%
\endgroup%
 \\
		\footnotesize{(b) Normalized $L^2$-norm error in circumferential mode shapes $\eigenvec_2^h$}
	\end{center}
	\begin{center}
		\def\svgwidth{0.7\textwidth}
		\input{sections/figures/svg/legend_methods.pdf_tex}
	\end{center}
	\caption{Normalized errors in eigenvalues $\lambda_2^h$ and \textbf{circumferential} mode shapes $\eigenvec_2^h$ computed with \textbf{quadratic B-splines} ($p=2$) on a mesh of \textbf{256 elements} ($N=128$), for ``large'' slenderness ratio $R/t = 2000/3$.}
	\label{fig:spectra2_compare_formulations_h}
\end{figure}

\subsubsection{Hellinger-Reissner formulation}

We finally consider the results obtained with the Hellinger-Reissner formulation, which in Figs.~\ref{fig:spectra1_compare_formulations} and \ref{fig:spectra2_compare_formulations} are plotted with purple diamonds. 
We observe that the effect of membrane locking is eliminated in the entire spectrum of the eigenvalues of the transverse modes. The corresponding error curve closely matches the asymptotic error curve, irrespective of the mesh size. We note that the Hellinger-Reissner formulation has a different asymptotic error curve, which is different from all other formulations considered here. We observe in the inset plot of Fig.~\ref{fig:spectra1_compare_formulations}a that the Hellinger-Reissner formulation achieves the best accuracy of the lowest eigenvalues of the locking-prone transverse modes, which are of particular importance for the approximation power of the basis, see Section~\ref{sec:lowmode}. This accuracy advantage is maintained over 80\% of the normalized spectrum.

The Hellinger-Reissner formulation is a mixed method, which requires the discretization of both displacement and strain fields. To render it comparable to the other methods that rely only on displacement variables, the Hellinger-Reissner formulation requires additional computational effort for the static condensation of the strain variables. In addition, for the Euler-Bernoulli beam model, a mixed formulation does not require basis functions that are in the space $H^3$ to achieve optimal rates of convergence \cite{engel2002continuous,Tagliabue2014}. As a consequence, for quadratic basis functions that are only in $H^2$, the displacements converge with $\mathcal{O}(3)$ in the $L^2$ norm in the Hellinger-Reissner formulation, while displacement-based formulations achieve only $\mathcal{O}(2)$. This advantage of the Hellinger-Reissner formulation, however, is expected to disappear, when we consider basis functions of polynomial degree $p\ge3$ that are in $H^3$, for which all methods achieve the same optimal rates $\mathcal{O}(p+1)$.

We therefore conclude that in terms of the effective prevention of membrane locking, the Hellinger-Reissner formulation hits a sweet spot for $p=2$ and therefore seems to be the most effective choice for quadratic spline discretizations. This conclusion is confirmed by our initial cantilever example, where the accuracy gap between the Hellinger-Reissner formulation on the one hand and the B-bar and DSG methods on the other hand is clearly demonstrated by Fig.~\ref{fig:cantilever_convergence_methods}a.

\subsection{Sensitivity with respect to mesh refinement}\label{sec:results_mesh_refine}
\label{Sec52}

Mesh refinement will eventually remove most locking phenomena. This comes, however, at the price of a significantly increased computational cost that is always uneconomical and often prohibitive with respect to the available computing resources. Therefore, from a practical viewpoint, mesh refinement is not a viable option to mitigate locking. We use spectral analysis to illustrate the lacking efficiency of mesh refinement. To this end, we compute the discrete eigenvalues and modes for each finite element formulation at hand, using quadratic B-splines defined on 256 B\'ezier elements. Figures~\ref{fig:spectra1_compare_formulations_h} and \ref{fig:spectra2_compare_formulations_h} plot the relative eigenvalue errors \eqref{eve} and the relative $L^2$-norm mode errors \eqref{eme} across the normalized spectrum for the transverse and circumferential modes, respectively.

\subsubsection{Inefficiency of the standard formulation}

We first focus on the spectrum of eigenvalues $\lambda_1^h$ associated with the transverse modes plotted in Fig.~\ref{fig:spectra1_compare_formulations_h}a, which is the quantity affected by membrane locking. We observe that the results obtained with the standard finite element formulation with full and reduced integration both improve significantly, with their spectral error now being in range of the error of the locking-free asymptotic solution. For selective reduced integration, we achieve the same accuracy in the lowest modes as for the locking-free B-bar and DSG methods. For full integration, however, the error of the lowest modes is still two orders of magnitude larger than the error level of the locking-free formulations, and therefore still prevents a high-fidelity solution despite the prohibitively fine mesh size. The effect on the accuracy in analysis that corresponds to these observations is illustrated in Fig.~\ref{fig:cantilever_convergence_methods}a for our initial cantilever example. While the convergence curve obtained with selective reduced integration catches up with the locking-free solutions for finer mesh sizes, the curve obtained with full integration still lags significantly behind.

\begin{figure}[h!]
	\centering
	\begin{center}
		\def\svgwidth{0.84\textwidth}
\begingroup%
  \makeatletter%
  \providecommand\color[2][]{%
    \errmessage{(Inkscape) Color is used for the text in Inkscape, but the package 'color.sty' is not loaded}%
    \renewcommand\color[2][]{}%
  }%
  \providecommand\transparent[1]{%
    \errmessage{(Inkscape) Transparency is used (non-zero) for the text in Inkscape, but the package 'transparent.sty' is not loaded}%
    \renewcommand\transparent[1]{}%
  }%
  \providecommand\rotatebox[2]{#2}%
  \newcommand*\fsize{\dimexpr\f@size pt\relax}%
  \newcommand*\lineheight[1]{\fontsize{\fsize}{#1\fsize}\selectfont}%
  \ifx\svgwidth\undefined%
    \setlength{\unitlength}{574.77697754bp}%
    \ifx\svgscale\undefined%
      \relax%
    \else%
      \setlength{\unitlength}{\unitlength * \real{\svgscale}}%
    \fi%
  \else%
    \setlength{\unitlength}{\svgwidth}%
  \fi%
  \global\let\svgwidth\undefined%
  \global\let\svgscale\undefined%
  \makeatother%
  \begin{picture}(1,0.59406518)%
    \lineheight{1}%
    \setlength\tabcolsep{0pt}%
    \put(0,0){\includegraphics[width=\unitlength,page=1]{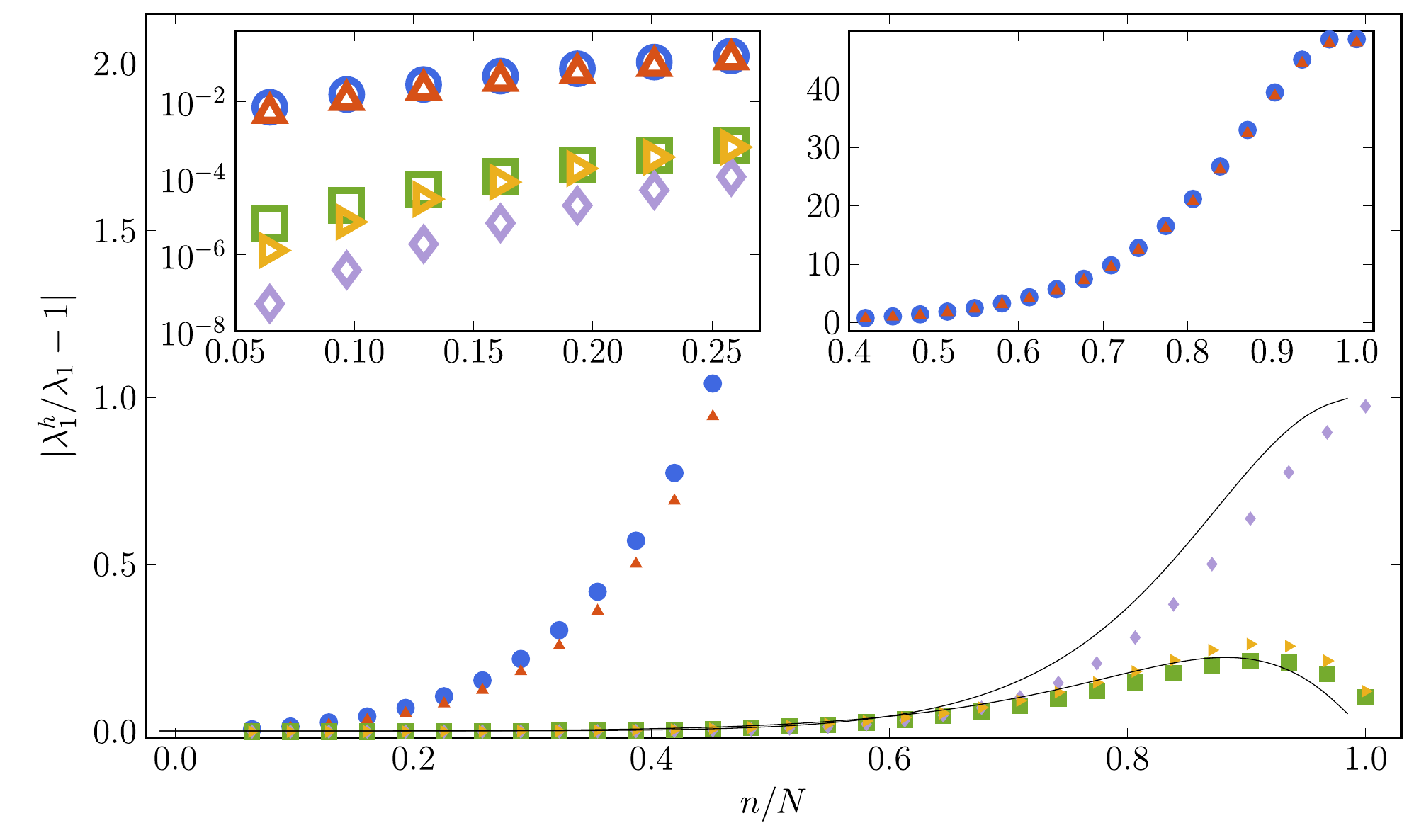}}%
    \put(0.78095265,0.23267782){\color[rgb]{0,0,0}\makebox(0,0)[lt]{\lineheight{1.25}\smash{\begin{tabular}[t]{l}$(\dagger)$\end{tabular}}}}%
    \put(0,0){\includegraphics[width=\unitlength,page=2]{eigen_ring_methods_spectra1_t0.0015_p3_64ele.pdf}}%
    \put(0.86943508,0.17497365){\color[rgb]{0,0,0}\makebox(0,0)[lt]{\lineheight{1.25}\smash{\begin{tabular}[t]{l}$(\star, \spadesuit)$\end{tabular}}}}%
    \put(0,0){\includegraphics[width=\unitlength,page=3]{eigen_ring_methods_spectra1_t0.0015_p3_64ele.pdf}}%
  \end{picture}%
\endgroup%
 \\
		\footnotesize{(a) Normalized error in eigenvalues $\lambda_1^h$ (associated with transverse modes  $\eigenvec_1^h$)}
	\end{center}
	\begin{center}
		\def\svgwidth{0.84\textwidth}
\begingroup%
  \makeatletter%
  \providecommand\color[2][]{%
    \errmessage{(Inkscape) Color is used for the text in Inkscape, but the package 'color.sty' is not loaded}%
    \renewcommand\color[2][]{}%
  }%
  \providecommand\transparent[1]{%
    \errmessage{(Inkscape) Transparency is used (non-zero) for the text in Inkscape, but the package 'transparent.sty' is not loaded}%
    \renewcommand\transparent[1]{}%
  }%
  \providecommand\rotatebox[2]{#2}%
  \newcommand*\fsize{\dimexpr\f@size pt\relax}%
  \newcommand*\lineheight[1]{\fontsize{\fsize}{#1\fsize}\selectfont}%
  \ifx\svgwidth\undefined%
    \setlength{\unitlength}{574.77697754bp}%
    \ifx\svgscale\undefined%
      \relax%
    \else%
      \setlength{\unitlength}{\unitlength * \real{\svgscale}}%
    \fi%
  \else%
    \setlength{\unitlength}{\svgwidth}%
  \fi%
  \global\let\svgwidth\undefined%
  \global\let\svgscale\undefined%
  \makeatother%
  \begin{picture}(1,0.59406518)%
    \lineheight{1}%
    \setlength\tabcolsep{0pt}%
    \put(0,0){\includegraphics[width=\unitlength,page=1]{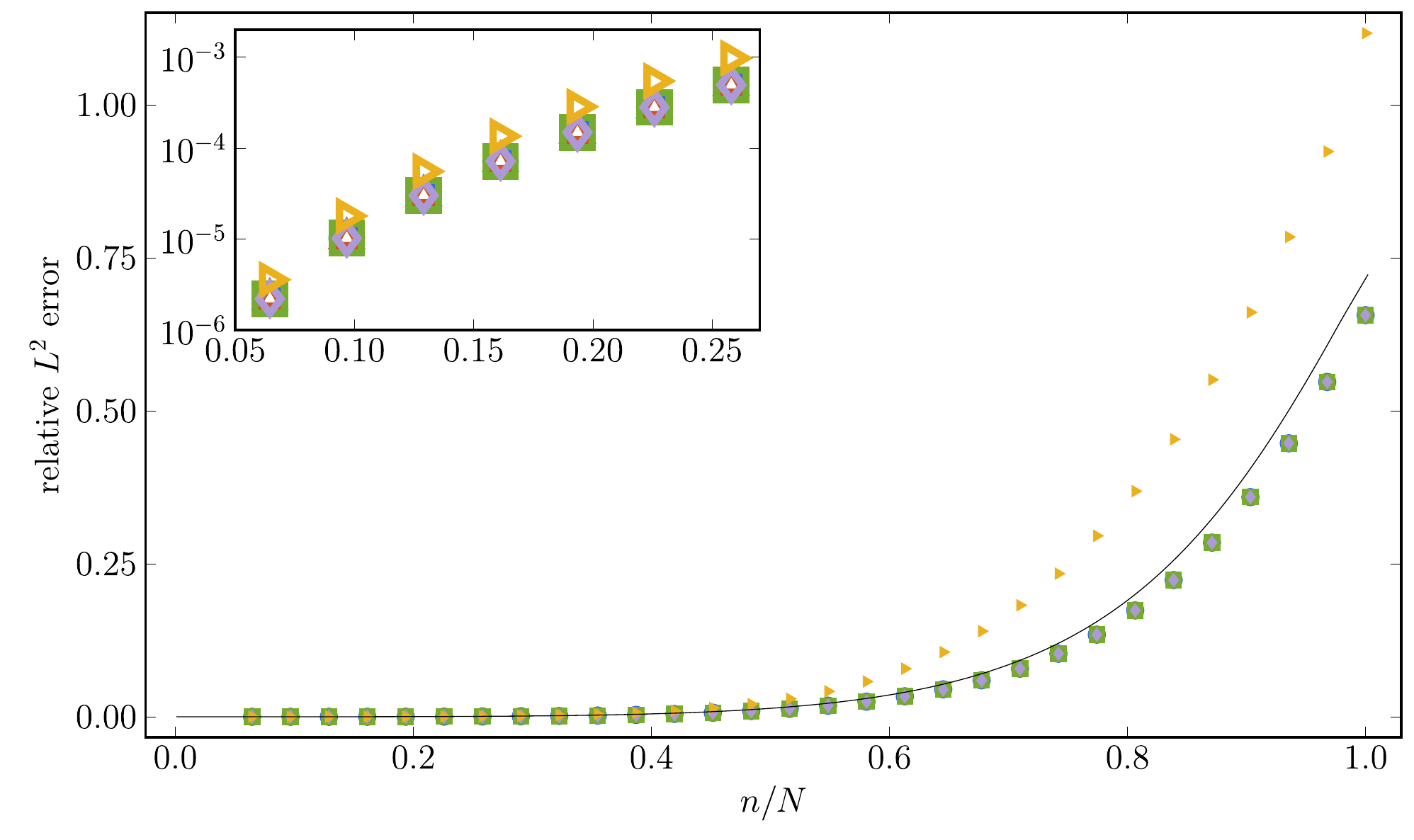}}%
    \put(0.88800465,0.18565145){\color[rgb]{0,0,0}\makebox(0,0)[lt]{\lineheight{1.25}\smash{\begin{tabular}[t]{l}$(\star, \dagger, \spadesuit)$\end{tabular}}}}%
    \put(0,0){\includegraphics[width=\unitlength,page=2]{eigen_ring_methods_L2mode1_t0.0015_p3_64ele.pdf}}%
    \put(0.55044787,0.4432966){\color[rgb]{0,0,0}\makebox(0,0)[lt]{\lineheight{1.25}\smash{\begin{tabular}[t]{l}\footnotesize{Overlapping}\end{tabular}}}}%
    \put(0,0){\includegraphics[width=\unitlength,page=3]{eigen_ring_methods_L2mode1_t0.0015_p3_64ele.pdf}}%
  \end{picture}%
\endgroup%
 \\
		\footnotesize{(b) Normalized $L^2$-norm error in transverse mode shapes $\eigenvec_1^h$}
	\end{center}
	\begin{center}
		\def\svgwidth{0.7\textwidth}
		\input{sections/figures/svg/legend_methods.pdf_tex}
	\end{center}
	\caption{Normalized errors in eigenvalues $\lambda_1^h$ and \textbf{transverse} mode shapes $\eigenvec_1^h$ computed with \textbf{cubic B-splines} ($p=3$) on a mesh of \textbf{64 elements} ($N=32$), for ``large'' slenderness ratio $R/t = 2000/3$.}
	\label{fig:spectra1_compare_formulations_p3}
\end{figure}

\begin{figure}[h!]
	\centering
	\begin{center}
		\def\svgwidth{0.84\textwidth}
\begingroup%
  \makeatletter%
  \providecommand\color[2][]{%
    \errmessage{(Inkscape) Color is used for the text in Inkscape, but the package 'color.sty' is not loaded}%
    \renewcommand\color[2][]{}%
  }%
  \providecommand\transparent[1]{%
    \errmessage{(Inkscape) Transparency is used (non-zero) for the text in Inkscape, but the package 'transparent.sty' is not loaded}%
    \renewcommand\transparent[1]{}%
  }%
  \providecommand\rotatebox[2]{#2}%
  \newcommand*\fsize{\dimexpr\f@size pt\relax}%
  \newcommand*\lineheight[1]{\fontsize{\fsize}{#1\fsize}\selectfont}%
  \ifx\svgwidth\undefined%
    \setlength{\unitlength}{574.77697754bp}%
    \ifx\svgscale\undefined%
      \relax%
    \else%
      \setlength{\unitlength}{\unitlength * \real{\svgscale}}%
    \fi%
  \else%
    \setlength{\unitlength}{\svgwidth}%
  \fi%
  \global\let\svgwidth\undefined%
  \global\let\svgscale\undefined%
  \makeatother%
  \begin{picture}(1,0.59406518)%
    \lineheight{1}%
    \setlength\tabcolsep{0pt}%
    \put(0,0){\includegraphics[width=\unitlength,page=1]{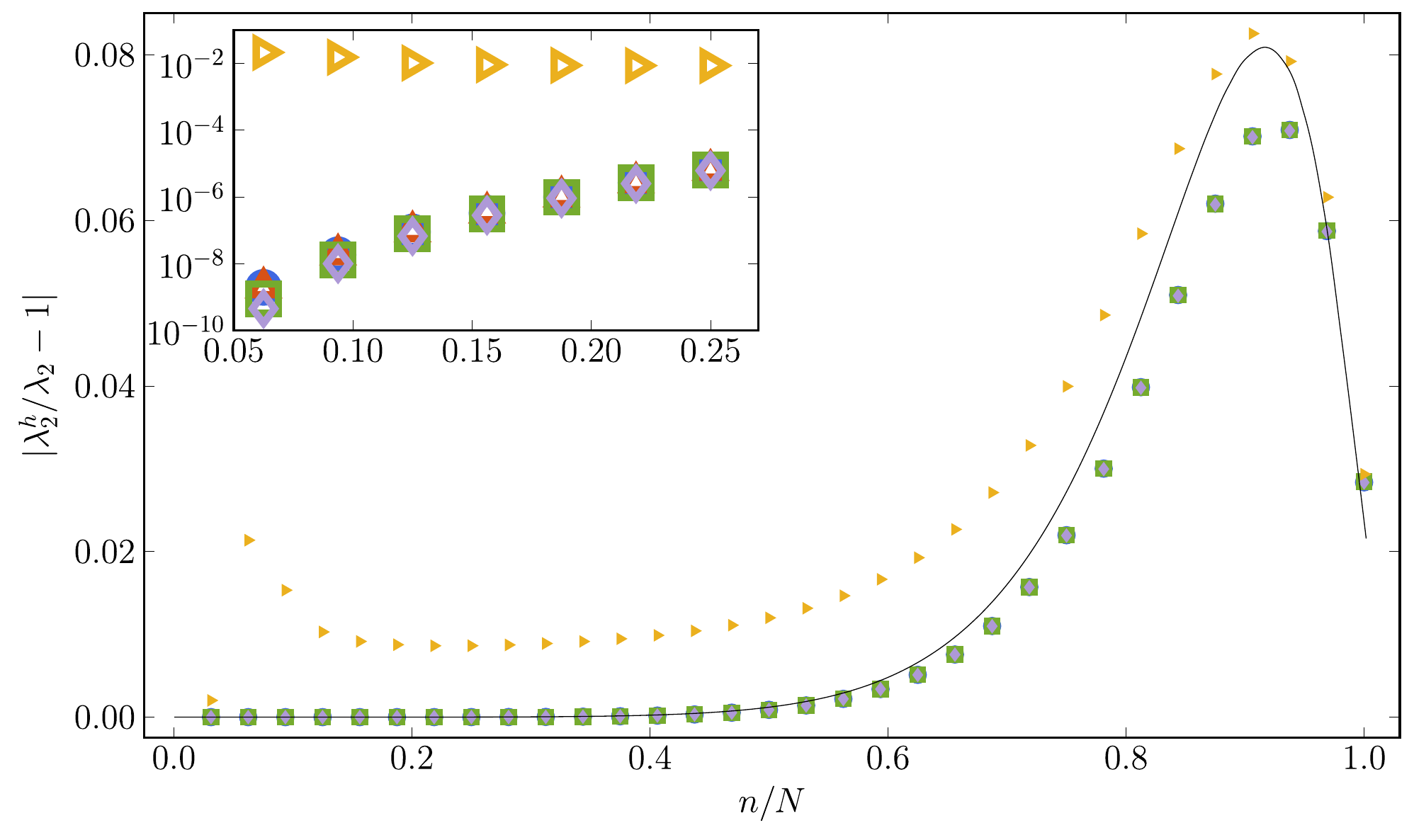}}%
    \put(0.86295142,0.21587355){\color[rgb]{0,0,0}\makebox(0,0)[lt]{\lineheight{1.25}\smash{\begin{tabular}[t]{l}$(\star, \dagger, \spadesuit)$\end{tabular}}}}%
    \put(0,0){\includegraphics[width=\unitlength,page=2]{eigen_ring_methods_spectra2_t0.0015_p3_64ele.pdf}}%
    \put(0.54878097,0.51489279){\color[rgb]{0,0,0}\makebox(0,0)[lt]{\lineheight{1.25}\smash{\begin{tabular}[t]{l}\footnotesize{Overlapping}\end{tabular}}}}%
    \put(0,0){\includegraphics[width=\unitlength,page=3]{eigen_ring_methods_spectra2_t0.0015_p3_64ele.pdf}}%
  \end{picture}%
\endgroup%
 \\
		\footnotesize{(a) Normalized error in eigenvalues $\lambda_2^h$ (associated with circumferential modes  $\eigenvec_2^h$)}
	\end{center}
	\begin{center}
		\def\svgwidth{0.84\textwidth}
\begingroup%
  \makeatletter%
  \providecommand\color[2][]{%
    \errmessage{(Inkscape) Color is used for the text in Inkscape, but the package 'color.sty' is not loaded}%
    \renewcommand\color[2][]{}%
  }%
  \providecommand\transparent[1]{%
    \errmessage{(Inkscape) Transparency is used (non-zero) for the text in Inkscape, but the package 'transparent.sty' is not loaded}%
    \renewcommand\transparent[1]{}%
  }%
  \providecommand\rotatebox[2]{#2}%
  \newcommand*\fsize{\dimexpr\f@size pt\relax}%
  \newcommand*\lineheight[1]{\fontsize{\fsize}{#1\fsize}\selectfont}%
  \ifx\svgwidth\undefined%
    \setlength{\unitlength}{574.77697754bp}%
    \ifx\svgscale\undefined%
      \relax%
    \else%
      \setlength{\unitlength}{\unitlength * \real{\svgscale}}%
    \fi%
  \else%
    \setlength{\unitlength}{\svgwidth}%
  \fi%
  \global\let\svgwidth\undefined%
  \global\let\svgscale\undefined%
  \makeatother%
  \begin{picture}(1,0.59406518)%
    \lineheight{1}%
    \setlength\tabcolsep{0pt}%
    \put(0,0){\includegraphics[width=\unitlength,page=1]{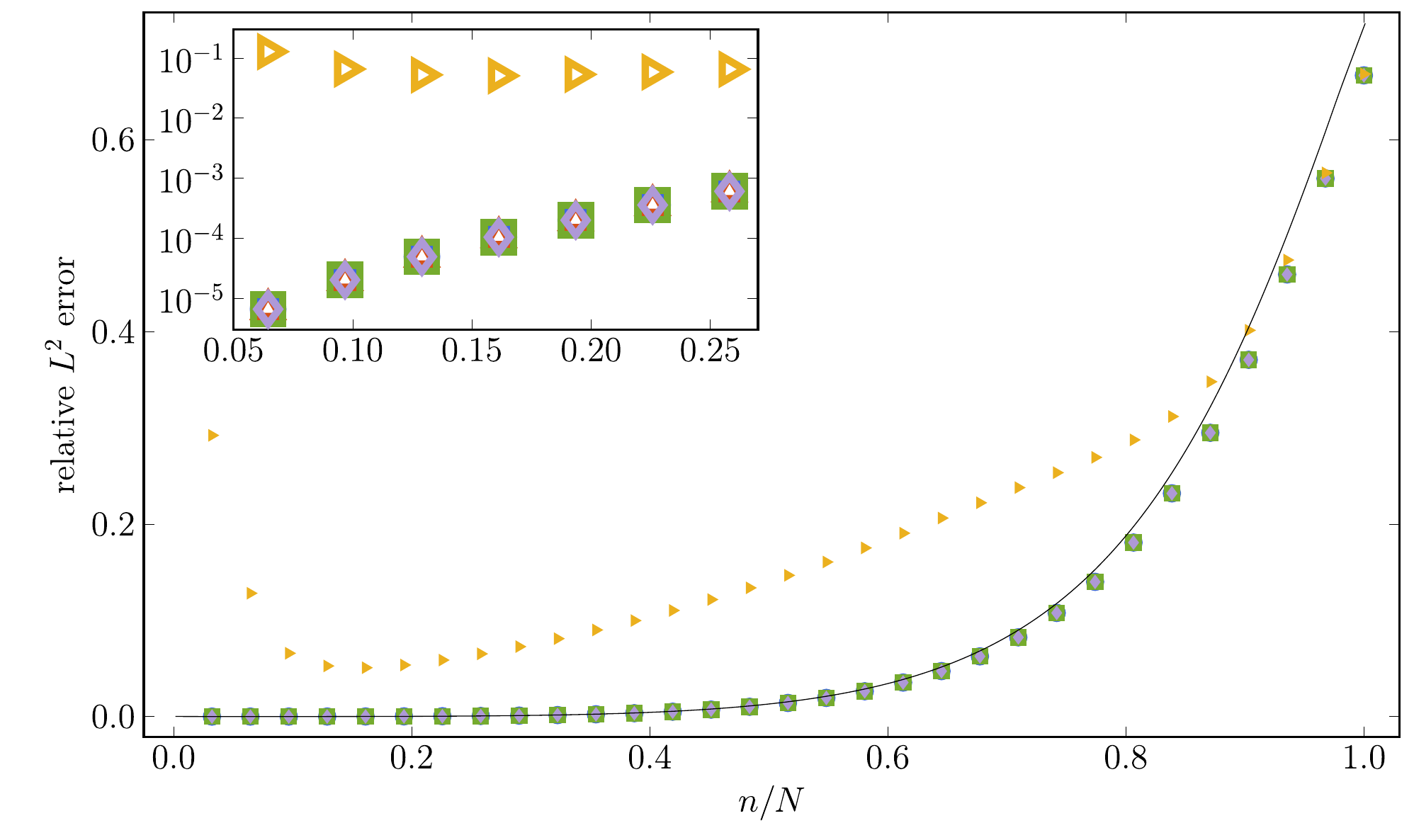}}%
    \put(0.78877766,0.42120455){\color[rgb]{0,0,0}\makebox(0,0)[lt]{\lineheight{1.25}\smash{\begin{tabular}[t]{l}$(\star, \dagger, \spadesuit)$\end{tabular}}}}%
    \put(0,0){\includegraphics[width=\unitlength,page=2]{eigen_ring_methods_L2mode2_t0.0015_p3_64ele.pdf}}%
    \put(0.54914306,0.5085116){\color[rgb]{0,0,0}\makebox(0,0)[lt]{\lineheight{1.25}\smash{\begin{tabular}[t]{l}\footnotesize{Overlapping}\end{tabular}}}}%
    \put(0,0){\includegraphics[width=\unitlength,page=3]{eigen_ring_methods_L2mode2_t0.0015_p3_64ele.pdf}}%
  \end{picture}%
\endgroup%
 \\
		\footnotesize{(b) Normalized $L^2$-norm error in circumferential mode shapes $\eigenvec_2^h$}
	\end{center}
	\begin{center}
		\def\svgwidth{0.7\textwidth}
		\input{sections/figures/svg/legend_methods.pdf_tex}
	\end{center}
	\caption{Normalized errors in eigenvalues $\lambda_2^h$ and \textbf{circumferential} mode shapes $\eigenvec_2^h$ computed with \textbf{cubic B-splines} ($p=3$) on a mesh of \textbf{64 elements} ($N=32$), for ``large'' slenderness ratio $R/t = 2000/3$.}
	\label{fig:spectra2_compare_formulations_p3}
\end{figure}

\subsubsection{DSG method and the circumferential modes}

Looking at the complete set of spectral plots, we observe that the issues we detected for the DSG method on a mesh with 64 elements do not vanish under mesh refinement. In Fig.~\ref{fig:spectra1_compare_formulations_h}b, we can see that the lowest transverse mode shapes are approximated at an accuracy level that is two orders of magnitude below the locking-free standard solution. In addition, several modes in the center of the spectrum seem completely inaccurate. Figures~\ref{fig:spectra2_compare_formulations_h}a and \ref{fig:spectra2_compare_formulations_h}b that illustrate the spectral behavior of the circumferential modes show that the error of the lowest eigenvalues and mode shapes are both four to five orders of magnitude larger than the error of the locking-free solution. These results support our notion that the DSG formulation itself is responsible for the increase in spectral error levels, and thus deteriorates the accuracy of the standard finite element formulation in part of the spectrum. 
As membrane unlocking is associated primarily with the proper behavior in the transverse eigenvalues, this shortcoming of the DSG method seems not to affect its analysis capabilities in this particular case, as demonstrated for our initial cantilever example in Fig.~\ref{fig:cantilever_convergence_methods}a.

\subsection{Sensitivity with respect to $p$-refinement}\label{sec:results_p_refine}

It has been often maintained that $p$-refinement constitutes an effective way to counteract locking phenomena, for instance in the context of the $p$-version of the finite element method \cite{leino1994membrane,suri1996analytical,rank2005high}. A presumed key argument in support of $p$-refinement is that it mitigates locking when applied within a standard displacement-based formulation and thus bypasses the derivation and implementation of special locking-free formulations. In the following, we will use spectral analysis to shed light on the efficiency of $p$-refinement with respect to mitigating membrane locking in the Euler-Bernoulli ring example. 

\subsubsection{Classical $p$-refinement: standard formulation}

We first move to cubic B-splines defined on 64 B\'ezier elements, re-computing the discrete eigenvalues and eigenmodes for each finite element formulation at hand. We note that due to periodicity, the cubic discretization of the circular ring exhibits the same number of spline basis functions and hence the same number of degrees of freedom as the quadratic discretization.
For polynomial degree $p=3$, Figures~\ref{fig:spectra1_compare_formulations_p3} and \ref{fig:spectra2_compare_formulations_p3} plot the relative eigenvalue errors \eqref{eve} and the relative $L^2$-norm mode errors \eqref{eme} across the normalized spectrum for the transverse and circumferential modes, respectively. They can be directly compared to Figs.~\ref{fig:spectra1_compare_formulations} and \ref{fig:spectra2_compare_formulations} that plot the equivalent results for quadratic B-splines.

We first consider the spectrum of eigenvalues $\lambda_1^h$ associated with the transverse modes, which is the relevant spectral quantity for membrane locking. Focusing on the lowest eigenvalues, we compare the corresponding error levels shown in the inset figures of Fig.~\ref{fig:spectra1_compare_formulations}a for quadratics and Fig.~\ref{fig:spectra1_compare_formulations_p3}a for cubics. We observe that the error level of the standard formulation improves by two orders of magnitude as a result of moving from quadratics to cubics. At the same time, however, we also see that the error levels of all locking-free formulations discretized with the same cubic B-splines improve by three to four orders of magnitude. The standard formulation thus lags far behind its true higher-order approximation power as a result of locking. Therefore, the reduction of locking with $p=3$ seems to be merely due to the increase of the approximation order, as also exemplified by higher convergence rates, but not to the mitigation of the locking phenomenon itself. We conclude that the standard formulation when discretized with cubic B-splines suffers from the effect of locking to (at least) the same extent as when it is discretized with quadratic B-splines. 
In addition, we observe that the standard formulation with selective reduced integration that employs $p$ quadrature points per B\'ezier element produces practically the same locking-prone results as the standard formulation with full integration. This observation indicates that reduced quadrature loses its locking-reducing effect when the polynomial degree is increased. 

\subsubsection{Increasing $p$ in locking-free formulations}
Comparing Fig.~\ref{fig:spectra1_compare_formulations}b and Fig.~\ref{fig:spectra1_compare_formulations_p3}b, we see that for the DSG method, the mode error in the transverse mode shapes significantly improves when we move from $p=2$ to $p=3$, and is now in the same range as the mode error of all other formulations. Figures~\ref{fig:spectra2_compare_formulations_p3}a and \ref{fig:spectra2_compare_formulations_p3}b show, however, that for the eigenvalues and mode shapes of the circumferential modes, the accuracy issues shown by the DSG methods remain and seem not to improve under $p$-refinement. 

As discussed above, the accuracy advantage of the Hellinger-Reissner formulation as a mixed method over purely displacement based formulations based on the B-bar and DSG methods reduces. We observe in Fig.~\ref{fig:spectra1_compare_formulations_p3}a that all locking-free formulations achieve very good accuracy, with advantages of the Hellinger-Reissner formulation in the lowest modes and advantages of the B-bar and DSG methods in the high modes. The effect of this observation is illustrated in Figs.~\ref{fig:cantilever_convergence_methods}b and \ref{fig:cantilever_convergence_methods}c for our initial cantilever example, where for cubic and quartic spline discretizations, all locking-free methods exhibit optimal convergence rates at practically the same accuracy level on both coarse and fine meshes.

\begin{figure}[h!]
	\centering
	\begin{center}
		\def\svgwidth{0.8\textwidth}
\begingroup%
  \makeatletter%
  \providecommand\color[2][]{%
    \errmessage{(Inkscape) Color is used for the text in Inkscape, but the package 'color.sty' is not loaded}%
    \renewcommand\color[2][]{}%
  }%
  \providecommand\transparent[1]{%
    \errmessage{(Inkscape) Transparency is used (non-zero) for the text in Inkscape, but the package 'transparent.sty' is not loaded}%
    \renewcommand\transparent[1]{}%
  }%
  \providecommand\rotatebox[2]{#2}%
  \newcommand*\fsize{\dimexpr\f@size pt\relax}%
  \newcommand*\lineheight[1]{\fontsize{\fsize}{#1\fsize}\selectfont}%
  \ifx\svgwidth\undefined%
    \setlength{\unitlength}{567.75299072bp}%
    \ifx\svgscale\undefined%
      \relax%
    \else%
      \setlength{\unitlength}{\unitlength * \real{\svgscale}}%
    \fi%
  \else%
    \setlength{\unitlength}{\svgwidth}%
  \fi%
  \global\let\svgwidth\undefined%
  \global\let\svgscale\undefined%
  \makeatother%
  \begin{picture}(1,0.60669695)%
    \lineheight{1}%
    \setlength\tabcolsep{0pt}%
    \put(0,0){\includegraphics[width=\unitlength,page=1]{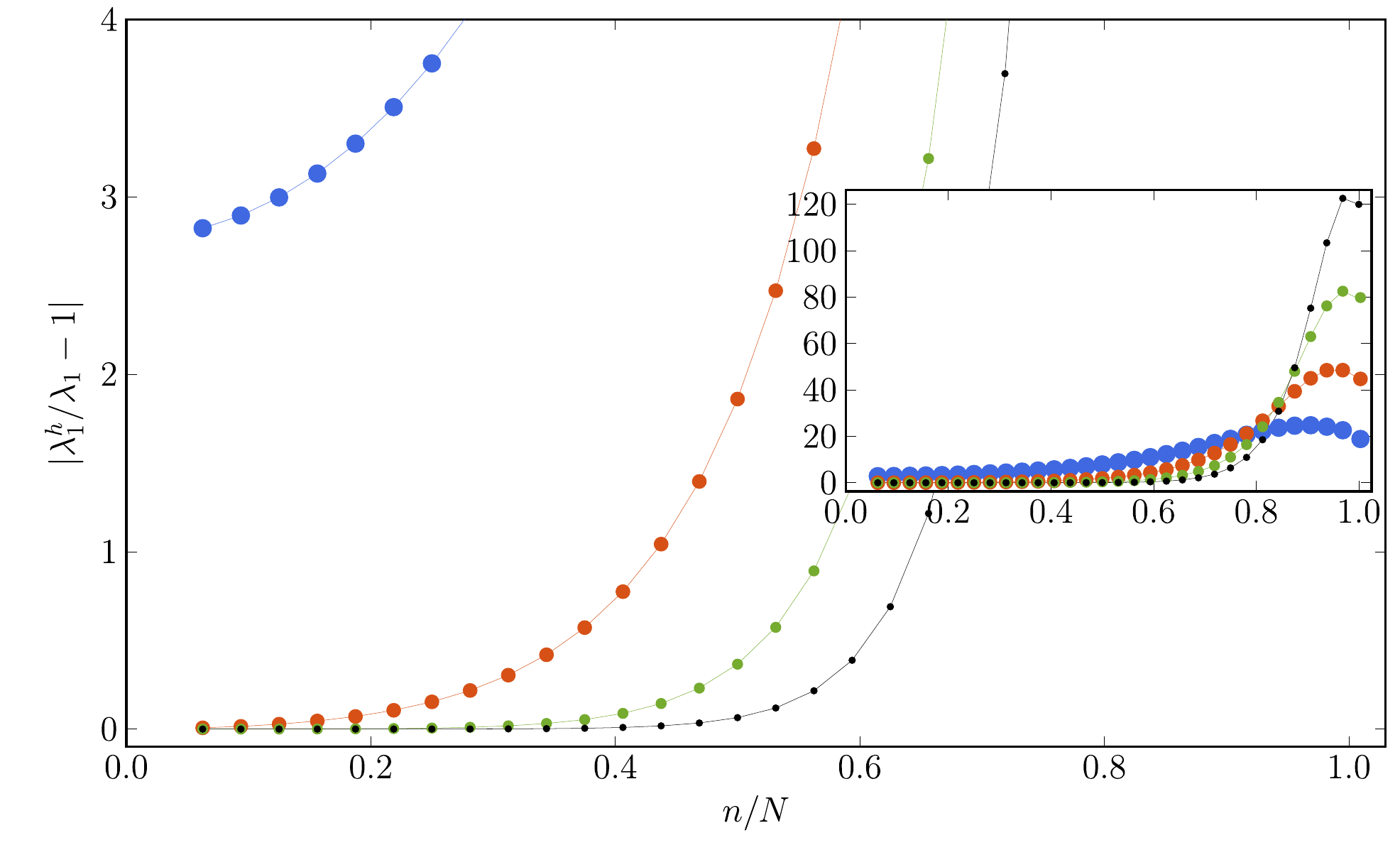}}%
    \put(0.2454338,0.46260224){\color[rgb]{0,0,0}\makebox(0,0)[lt]{\lineheight{1.25}\smash{\begin{tabular}[t]{l}\scriptsize{\textcolor{blue1}{$p = 2$}}\end{tabular}}}}%
    \put(0.37813323,0.19228147){\color[rgb]{0,0,0}\makebox(0,0)[lt]{\lineheight{1.25}\smash{\begin{tabular}[t]{l}\scriptsize{\textcolor{red1}{$p = 3$}}\end{tabular}}}}%
    \put(0.46619314,0.14592416){\color[rgb]{0,0,0}\makebox(0,0)[lt]{\lineheight{1.25}\smash{\begin{tabular}[t]{l}\scriptsize{\textcolor{green1}{$p = 4$}}\end{tabular}}}}%
    \put(0.63054757,0.12836844){\color[rgb]{0,0,0}\makebox(0,0)[lt]{\lineheight{1.25}\smash{\begin{tabular}[t]{l}\scriptsize{$p = 5$}\end{tabular}}}}%
  \end{picture}%
\endgroup%
 \\
		\footnotesize{(a) Overall behavior of the transverse eigenvalue error}
	\end{center}
	\begin{center}
		\def\svgwidth{0.8\textwidth}
\begingroup%
  \makeatletter%
  \providecommand\color[2][]{%
    \errmessage{(Inkscape) Color is used for the text in Inkscape, but the package 'color.sty' is not loaded}%
    \renewcommand\color[2][]{}%
  }%
  \providecommand\transparent[1]{%
    \errmessage{(Inkscape) Transparency is used (non-zero) for the text in Inkscape, but the package 'transparent.sty' is not loaded}%
    \renewcommand\transparent[1]{}%
  }%
  \providecommand\rotatebox[2]{#2}%
  \newcommand*\fsize{\dimexpr\f@size pt\relax}%
  \newcommand*\lineheight[1]{\fontsize{\fsize}{#1\fsize}\selectfont}%
  \ifx\svgwidth\undefined%
    \setlength{\unitlength}{567.75299072bp}%
    \ifx\svgscale\undefined%
      \relax%
    \else%
      \setlength{\unitlength}{\unitlength * \real{\svgscale}}%
    \fi%
  \else%
    \setlength{\unitlength}{\svgwidth}%
  \fi%
  \global\let\svgwidth\undefined%
  \global\let\svgscale\undefined%
  \makeatother%
  \begin{picture}(1,0.60669695)%
    \lineheight{1}%
    \setlength\tabcolsep{0pt}%
    \put(0,0){\includegraphics[width=\unitlength,page=1]{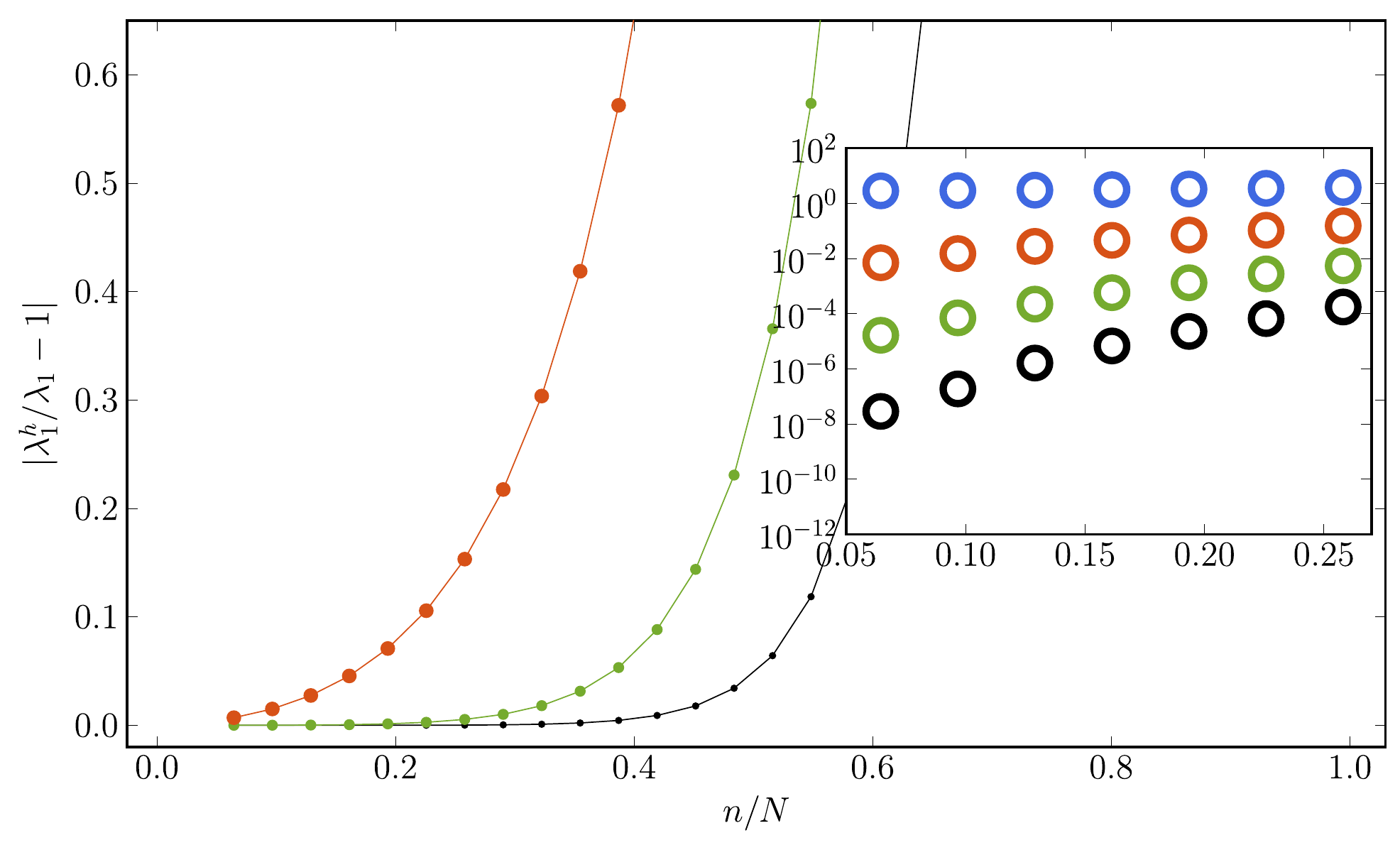}}%
    \put(0.26832386,0.21237618){\color[rgb]{0,0,0}\makebox(0,0)[lt]{\lineheight{1.25}\smash{\begin{tabular}[t]{l}\scriptsize{\textcolor{red1}{$p = 3$}}\end{tabular}}}}%
    \put(0.40819079,0.17083399){\color[rgb]{0,0,0}\makebox(0,0)[lt]{\lineheight{1.25}\smash{\begin{tabular}[t]{l}\scriptsize{\textcolor{green1}{$p = 4$}}\end{tabular}}}}%
    \put(0.57480189,0.13290623){\color[rgb]{0,0,0}\makebox(0,0)[lt]{\lineheight{1.25}\smash{\begin{tabular}[t]{l}\scriptsize{$p = 5$}\end{tabular}}}}%
  \end{picture}%
\endgroup%
 \\
		\footnotesize{(b) Behavior of the transverse eigenvalue error in the lower spectrum part}
	\end{center}
		\caption{\textbf{Standard formulation, $p$-refinement:} normalized errors in the transverse eigenvalues $\lambda_1^h$, computed on a fixed mesh of \textbf{64 elements} ($N=32$), for ``large'' slenderness ratio $R/t = 2000/3$.}	\label{fig:standard_p}
\end{figure}

\begin{figure}[h!]
	\centering
		\begin{center}
		\def\svgwidth{0.8\textwidth}
\begingroup%
  \makeatletter%
  \providecommand\color[2][]{%
    \errmessage{(Inkscape) Color is used for the text in Inkscape, but the package 'color.sty' is not loaded}%
    \renewcommand\color[2][]{}%
  }%
  \providecommand\transparent[1]{%
    \errmessage{(Inkscape) Transparency is used (non-zero) for the text in Inkscape, but the package 'transparent.sty' is not loaded}%
    \renewcommand\transparent[1]{}%
  }%
  \providecommand\rotatebox[2]{#2}%
  \newcommand*\fsize{\dimexpr\f@size pt\relax}%
  \newcommand*\lineheight[1]{\fontsize{\fsize}{#1\fsize}\selectfont}%
  \ifx\svgwidth\undefined%
    \setlength{\unitlength}{567.75299072bp}%
    \ifx\svgscale\undefined%
      \relax%
    \else%
      \setlength{\unitlength}{\unitlength * \real{\svgscale}}%
    \fi%
  \else%
    \setlength{\unitlength}{\svgwidth}%
  \fi%
  \global\let\svgwidth\undefined%
  \global\let\svgscale\undefined%
  \makeatother%
  \begin{picture}(1,0.60669695)%
    \lineheight{1}%
    \setlength\tabcolsep{0pt}%
    \put(0,0){\includegraphics[width=\unitlength,page=1]{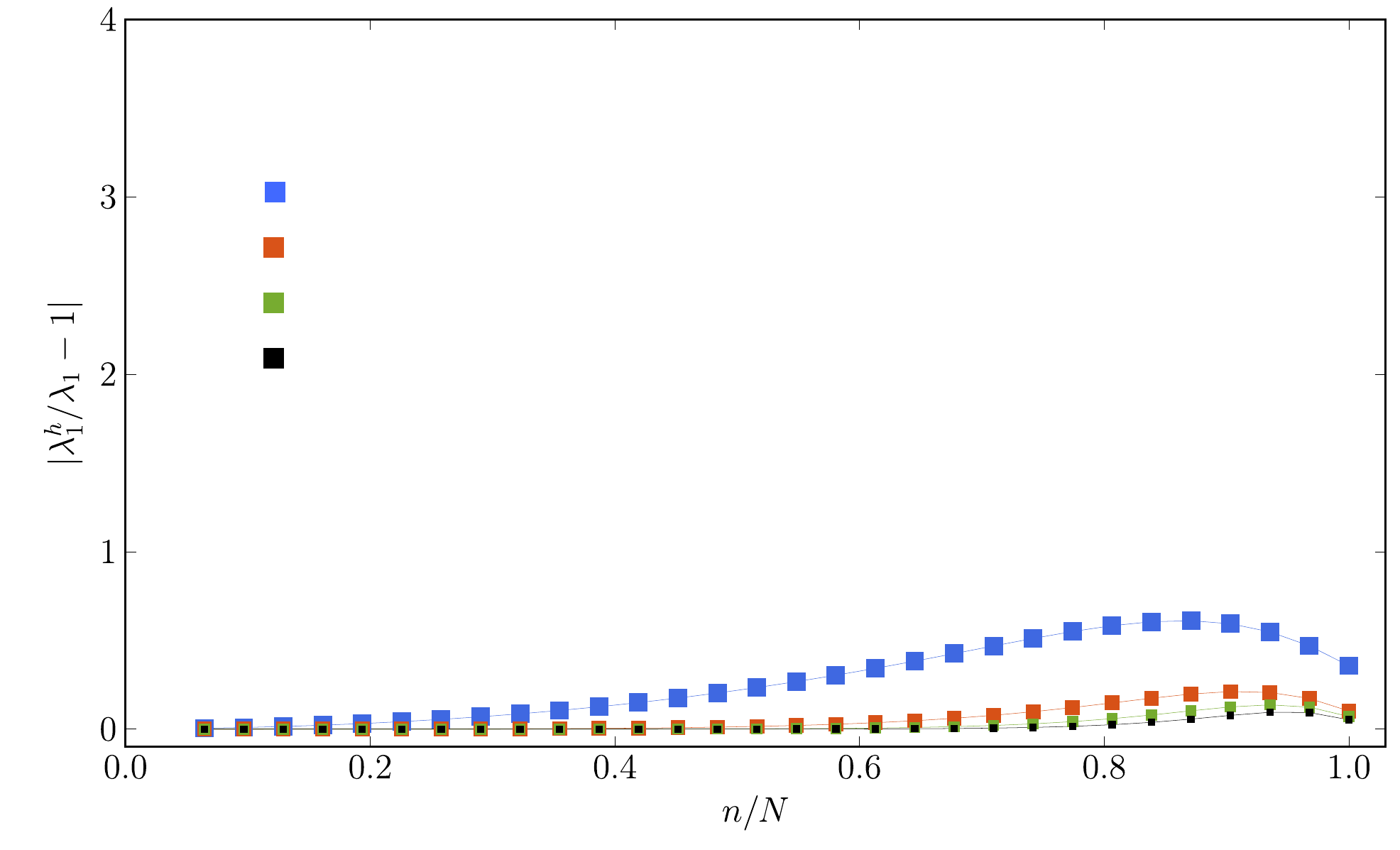}}%
    \put(0.21925034,0.46409012){\color[rgb]{0,0,0}\makebox(0,0)[lt]{\lineheight{1.25}\smash{\begin{tabular}[t]{l}\scriptsize{\textcolor{blue1}{$p = 2$}}\end{tabular}}}}%
    \put(0.21856528,0.38578317){\color[rgb]{0,0,0}\makebox(0,0)[lt]{\lineheight{1.25}\smash{\begin{tabular}[t]{l}\scriptsize{\textcolor{green1}{$p = 4$}}\end{tabular}}}}%
    \put(0.21869431,0.42441905){\color[rgb]{0,0,0}\makebox(0,0)[lt]{\lineheight{1.25}\smash{\begin{tabular}[t]{l}\scriptsize{\textcolor{red1}{$p = 3$}}\end{tabular}}}}%
    \put(0.21903383,0.34507467){\color[rgb]{0,0,0}\makebox(0,0)[lt]{\lineheight{1.25}\smash{\begin{tabular}[t]{l}\scriptsize{$p = 5$}\end{tabular}}}}%
  \end{picture}%
\endgroup%
 \\
		\footnotesize{(a) Overall behavior of the transverse eigenvalue error}
	\end{center}
	\begin{center}
		\def\svgwidth{0.8\textwidth}
\begingroup%
  \makeatletter%
  \providecommand\color[2][]{%
    \errmessage{(Inkscape) Color is used for the text in Inkscape, but the package 'color.sty' is not loaded}%
    \renewcommand\color[2][]{}%
  }%
  \providecommand\transparent[1]{%
    \errmessage{(Inkscape) Transparency is used (non-zero) for the text in Inkscape, but the package 'transparent.sty' is not loaded}%
    \renewcommand\transparent[1]{}%
  }%
  \providecommand\rotatebox[2]{#2}%
  \newcommand*\fsize{\dimexpr\f@size pt\relax}%
  \newcommand*\lineheight[1]{\fontsize{\fsize}{#1\fsize}\selectfont}%
  \ifx\svgwidth\undefined%
    \setlength{\unitlength}{567.75299072bp}%
    \ifx\svgscale\undefined%
      \relax%
    \else%
      \setlength{\unitlength}{\unitlength * \real{\svgscale}}%
    \fi%
  \else%
    \setlength{\unitlength}{\svgwidth}%
  \fi%
  \global\let\svgwidth\undefined%
  \global\let\svgscale\undefined%
  \makeatother%
  \begin{picture}(1,0.60669695)%
    \lineheight{1}%
    \setlength\tabcolsep{0pt}%
    \put(0,0){\includegraphics[width=\unitlength,page=1]{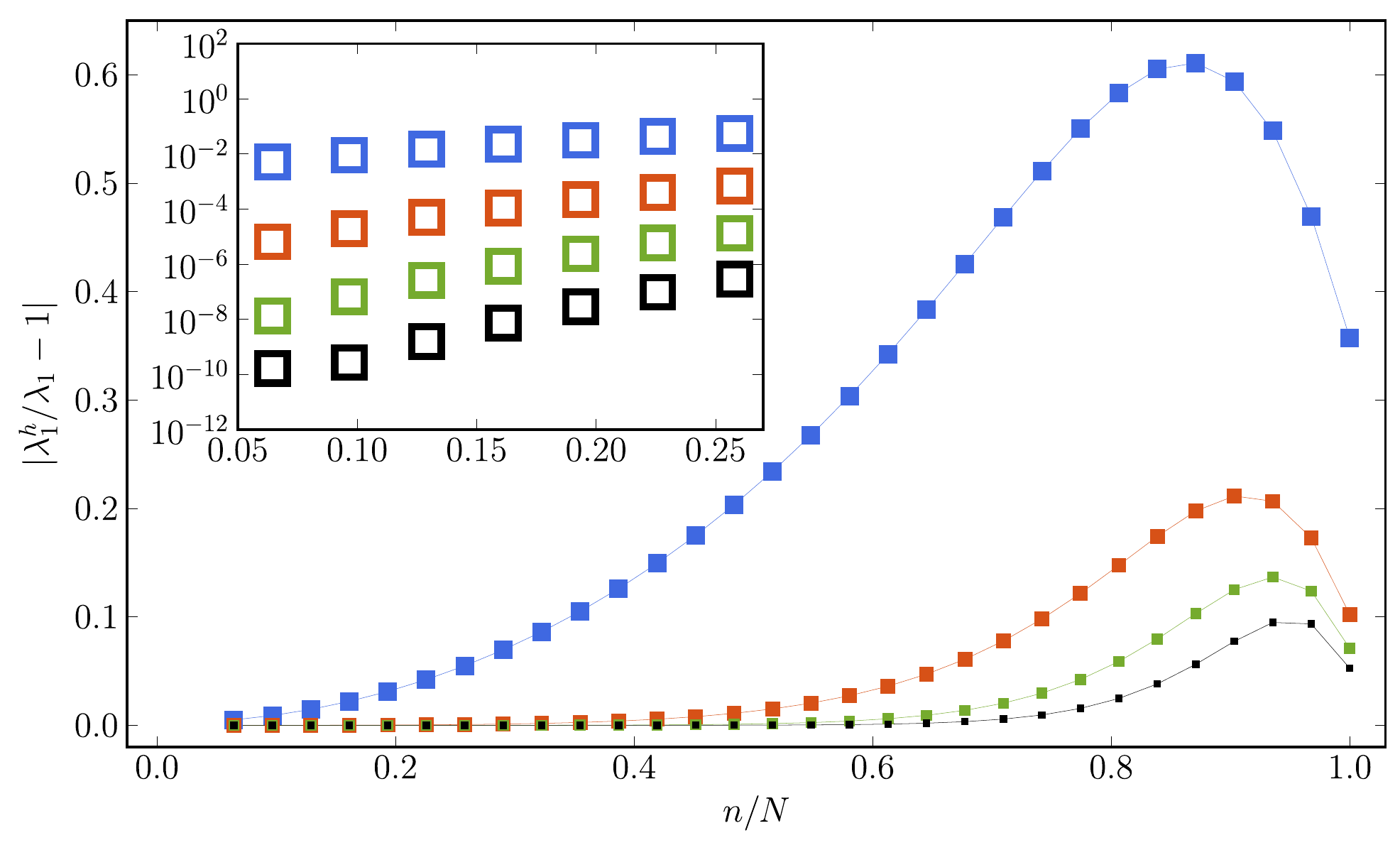}}%
    \put(0.68612858,0.38930061){\color[rgb]{0,0,0}\makebox(0,0)[lt]{\lineheight{1.25}\smash{\begin{tabular}[t]{l}\scriptsize{\textcolor{blue1}{$p = 2$}}\end{tabular}}}}%
    \put(0.71335173,0.1960289){\color[rgb]{0,0,0}\makebox(0,0)[lt]{\lineheight{1.25}\smash{\begin{tabular}[t]{l}\scriptsize{\textcolor{red1}{$p = 3$}}\end{tabular}}}}%
    \put(0.78869306,0.17013679){\color[rgb]{0,0,0}\makebox(0,0)[lt]{\lineheight{1.25}\smash{\begin{tabular}[t]{l}\scriptsize{\textcolor{green1}{$p = 4$}}\end{tabular}}}}%
    \put(0.84653701,0.11199184){\color[rgb]{0,0,0}\makebox(0,0)[lt]{\lineheight{1.25}\smash{\begin{tabular}[t]{l}\scriptsize{$p = 5$}\end{tabular}}}}%
  \end{picture}%
\endgroup%
 \\
		\footnotesize{(b) Behavior of the transverse eigenvalue error in the lower spectrum part}
	\end{center}
	\caption{\textbf{B-bar formulation, $p$-refinement:} normalized errors in the transverse eigenvalues $\lambda_1^h$, computed on a fixed mesh of \textbf{64 elements} ($N=32$), for ``large'' slenderness ratio $R/t = 2000/3$.}
	\label{fig:Bbar_p}
\end{figure}

\begin{figure}[h!]
	\centering
	\begin{center}
		\def\svgwidth{0.8\textwidth}
\begingroup%
  \makeatletter%
  \providecommand\color[2][]{%
    \errmessage{(Inkscape) Color is used for the text in Inkscape, but the package 'color.sty' is not loaded}%
    \renewcommand\color[2][]{}%
  }%
  \providecommand\transparent[1]{%
    \errmessage{(Inkscape) Transparency is used (non-zero) for the text in Inkscape, but the package 'transparent.sty' is not loaded}%
    \renewcommand\transparent[1]{}%
  }%
  \providecommand\rotatebox[2]{#2}%
  \newcommand*\fsize{\dimexpr\f@size pt\relax}%
  \newcommand*\lineheight[1]{\fontsize{\fsize}{#1\fsize}\selectfont}%
  \ifx\svgwidth\undefined%
    \setlength{\unitlength}{567.75299072bp}%
    \ifx\svgscale\undefined%
      \relax%
    \else%
      \setlength{\unitlength}{\unitlength * \real{\svgscale}}%
    \fi%
  \else%
    \setlength{\unitlength}{\svgwidth}%
  \fi%
  \global\let\svgwidth\undefined%
  \global\let\svgscale\undefined%
  \makeatother%
  \begin{picture}(1,0.60669695)%
    \lineheight{1}%
    \setlength\tabcolsep{0pt}%
    \put(0,0){\includegraphics[width=\unitlength,page=1]{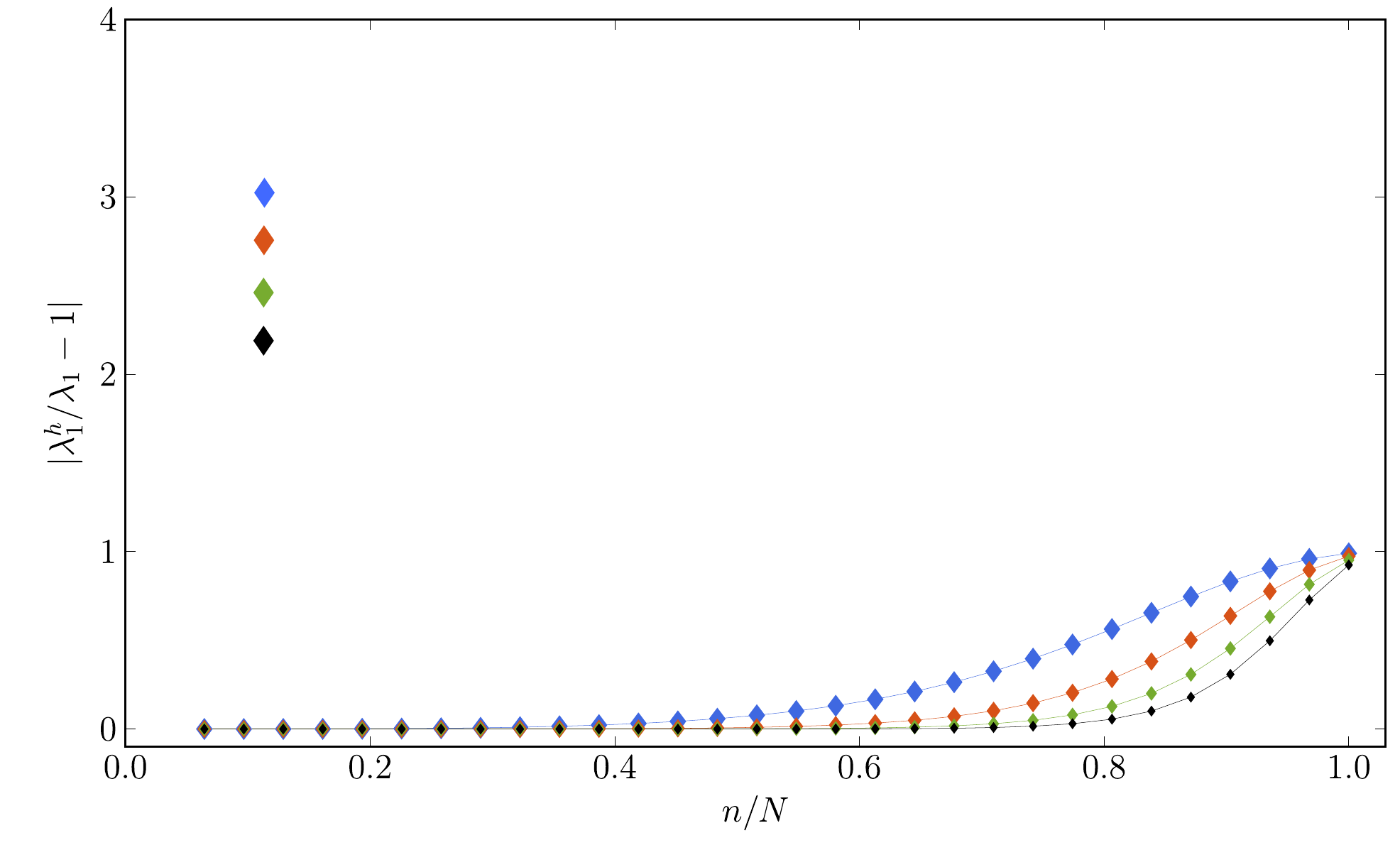}}%
    \put(0.21154438,0.46376469){\color[rgb]{0,0,0}\makebox(0,0)[lt]{\lineheight{1.25}\smash{\begin{tabular}[t]{l}\scriptsize{\textcolor{blue1}{$p = 2$}}\end{tabular}}}}%
    \put(0.21113393,0.39266394){\color[rgb]{0,0,0}\makebox(0,0)[lt]{\lineheight{1.25}\smash{\begin{tabular}[t]{l}\scriptsize{\textcolor{green1}{$p = 4$}}\end{tabular}}}}%
    \put(0.21135981,0.42986999){\color[rgb]{0,0,0}\makebox(0,0)[lt]{\lineheight{1.25}\smash{\begin{tabular}[t]{l}\scriptsize{\textcolor{red1}{$p = 3$}}\end{tabular}}}}%
    \put(0.21002217,0.35766635){\color[rgb]{0,0,0}\makebox(0,0)[lt]{\lineheight{1.25}\smash{\begin{tabular}[t]{l}\scriptsize{$p = 5$}\end{tabular}}}}%
  \end{picture}%
\endgroup%
 \\
		\footnotesize{(a) Overall behavior of the transverse eigenvalue error}
	\end{center}
	\begin{center}
		\def\svgwidth{0.8\textwidth}
\begingroup%
  \makeatletter%
  \providecommand\color[2][]{%
    \errmessage{(Inkscape) Color is used for the text in Inkscape, but the package 'color.sty' is not loaded}%
    \renewcommand\color[2][]{}%
  }%
  \providecommand\transparent[1]{%
    \errmessage{(Inkscape) Transparency is used (non-zero) for the text in Inkscape, but the package 'transparent.sty' is not loaded}%
    \renewcommand\transparent[1]{}%
  }%
  \providecommand\rotatebox[2]{#2}%
  \newcommand*\fsize{\dimexpr\f@size pt\relax}%
  \newcommand*\lineheight[1]{\fontsize{\fsize}{#1\fsize}\selectfont}%
  \ifx\svgwidth\undefined%
    \setlength{\unitlength}{567.75299072bp}%
    \ifx\svgscale\undefined%
      \relax%
    \else%
      \setlength{\unitlength}{\unitlength * \real{\svgscale}}%
    \fi%
  \else%
    \setlength{\unitlength}{\svgwidth}%
  \fi%
  \global\let\svgwidth\undefined%
  \global\let\svgscale\undefined%
  \makeatother%
  \begin{picture}(1,0.60669695)%
    \lineheight{1}%
    \setlength\tabcolsep{0pt}%
    \put(0,0){\includegraphics[width=\unitlength,page=1]{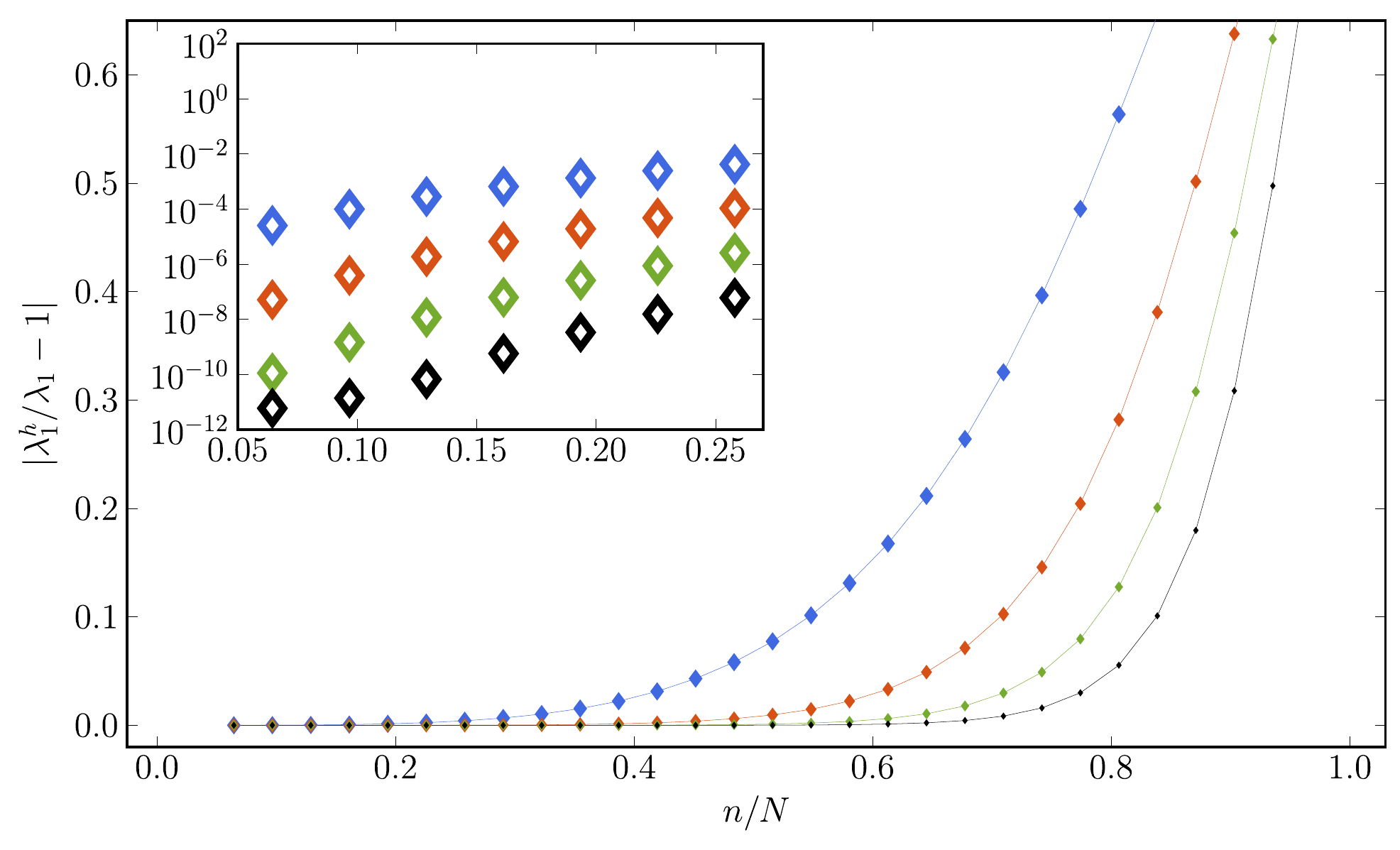}}%
    \put(0.62838734,0.31739719){\color[rgb]{0,0,0}\makebox(0,0)[lt]{\lineheight{1.25}\smash{\begin{tabular}[t]{l}\scriptsize{\textcolor{blue1}{$p = 2$}}\end{tabular}}}}%
    \put(0.6911296,0.22521837){\color[rgb]{0,0,0}\makebox(0,0)[lt]{\lineheight{1.25}\smash{\begin{tabular}[t]{l}\scriptsize{\textcolor{red1}{$p = 3$}}\end{tabular}}}}%
    \put(0.73168991,0.17051701){\color[rgb]{0,0,0}\makebox(0,0)[lt]{\lineheight{1.25}\smash{\begin{tabular}[t]{l}\scriptsize{\textcolor{green1}{$p = 4$}}\end{tabular}}}}%
    \put(0.82021647,0.13298457){\color[rgb]{0,0,0}\makebox(0,0)[lt]{\lineheight{1.25}\smash{\begin{tabular}[t]{l}\scriptsize{$p = 5$}\end{tabular}}}}%
  \end{picture}%
\endgroup%
 \\
		\footnotesize{(b) Behavior of the transverse eigenvalue error in the lower spectrum part}
	\end{center}
	\caption{\textbf{Hellinger-Reissner formulation, $p$-refinement:} normalized errors in the transverse eigenvalues $\lambda_1^h$, computed on a fixed mesh of \textbf{64 elements} ($N=32$), for ``large'' slenderness ratio $R/t = 2000/3$.}
	\label{fig:HR_p}
\end{figure}

\subsubsection{Classical versus locking-free $p$-refinement}

To corroborate our observations, we carry out a $p$-refinement study (i.e. increasing polynomial degree and smoothness) that drives the polynomial degree beyond cubics on a fixed mesh of 64 B\'ezier elements. We compare the spectrum of eigenvalues $\lambda_1^h$ associated with the transverse modes, computed via the standard formulation (``classical'' $p$-refinement) and via a-priori locking-free $p$-refinement based on the B-bar and the Hellinger-Reissner formulations. 

In Figs.~\ref{fig:standard_p}, \ref{fig:Bbar_p} and \ref{fig:HR_p} , we plot the transverse eigenvalue errors for each formulation at two different scales. We first focus on the overall behavior of the eigenvalue error across the complete spectrum. We observe in Fig.~\ref{fig:standard_p}a that the spectral accuracy of the standard formulation significantly improves with $p$-refinement in the lower part of the spectrum. At the same time, however, the high modes seem to diverge with increasing polynomial degree $p$ of the basis functions. 
Figure~\ref{fig:Bbar_p}a plots the corresponding eigenvalue error obtained with the B-bar formulation. We observe that the overall error levels are significantly smaller across the complete spectrum, and in particular for the high modes. We conclude that in contrast to the standard formulation, the B-bar method converges with increasing $p$ in the high modes. Figure~\ref{fig:HR_p}a plots the corresponding eigenvalue error obtained with the Hellinger-Reissner formulation. The overall error levels are significantly smaller as well. The highest modes, however, seem not to converge, but remain at the same error level.

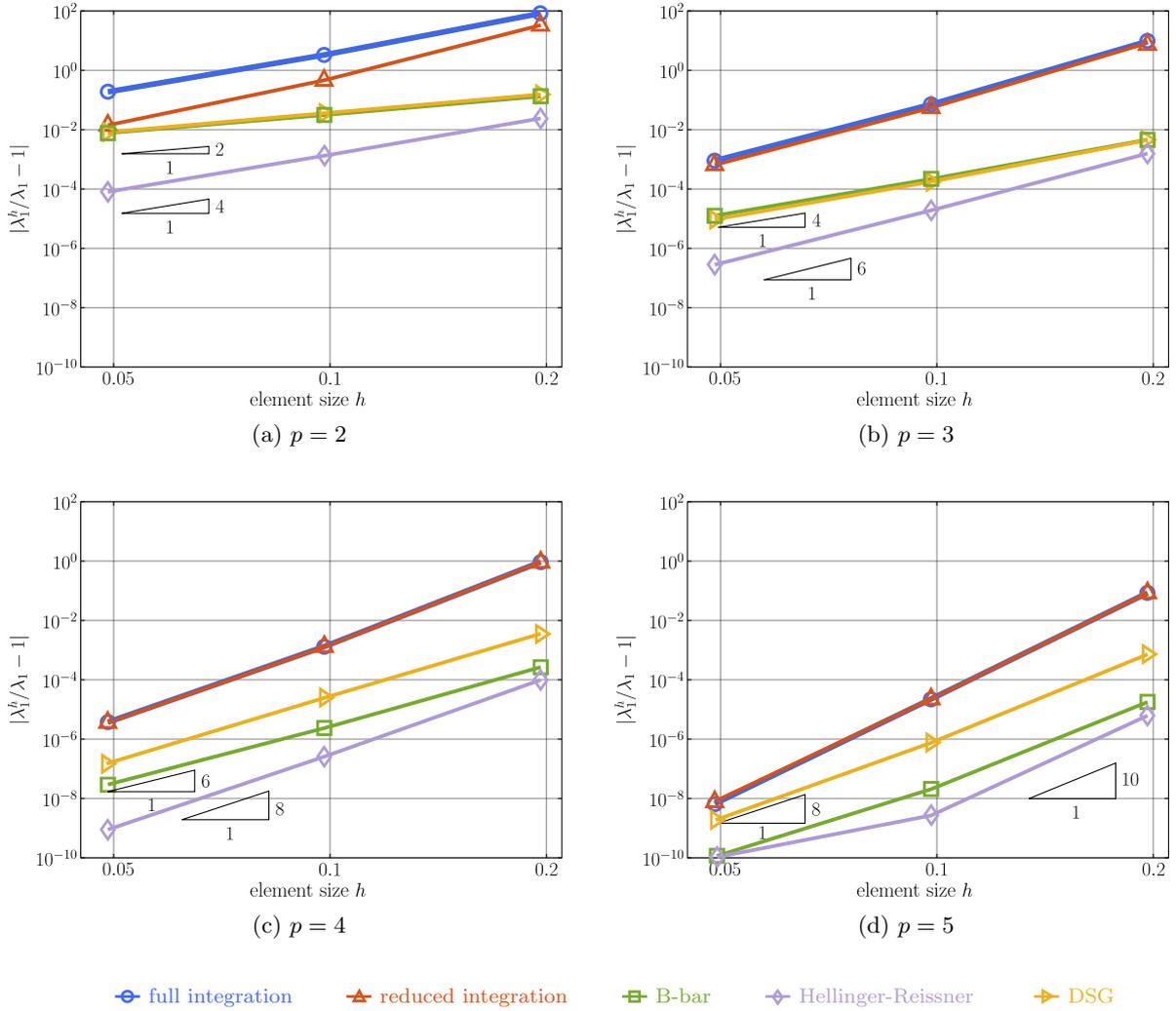
\begin{figure}[t!]
	\centering
	\begin{minipage}[b]{0.49\textwidth}
	\centering
	\includegraphics[width=1.0\textwidth]{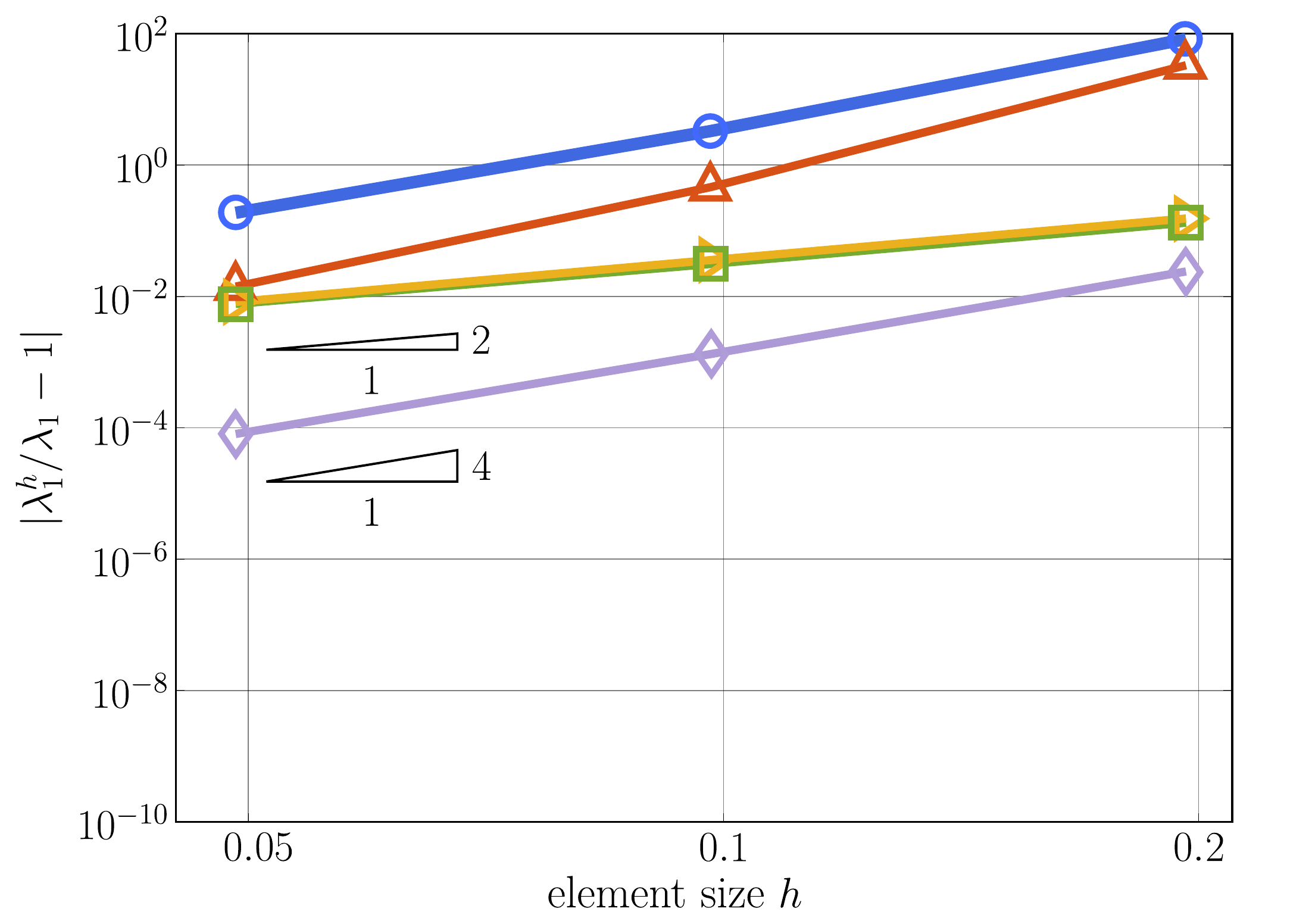}
	\footnotesize{(a) $p = 2$}
	\end{minipage}
	\begin{minipage}[b]{0.49\textwidth}
	\centering
	\includegraphics[width=1.0\textwidth]{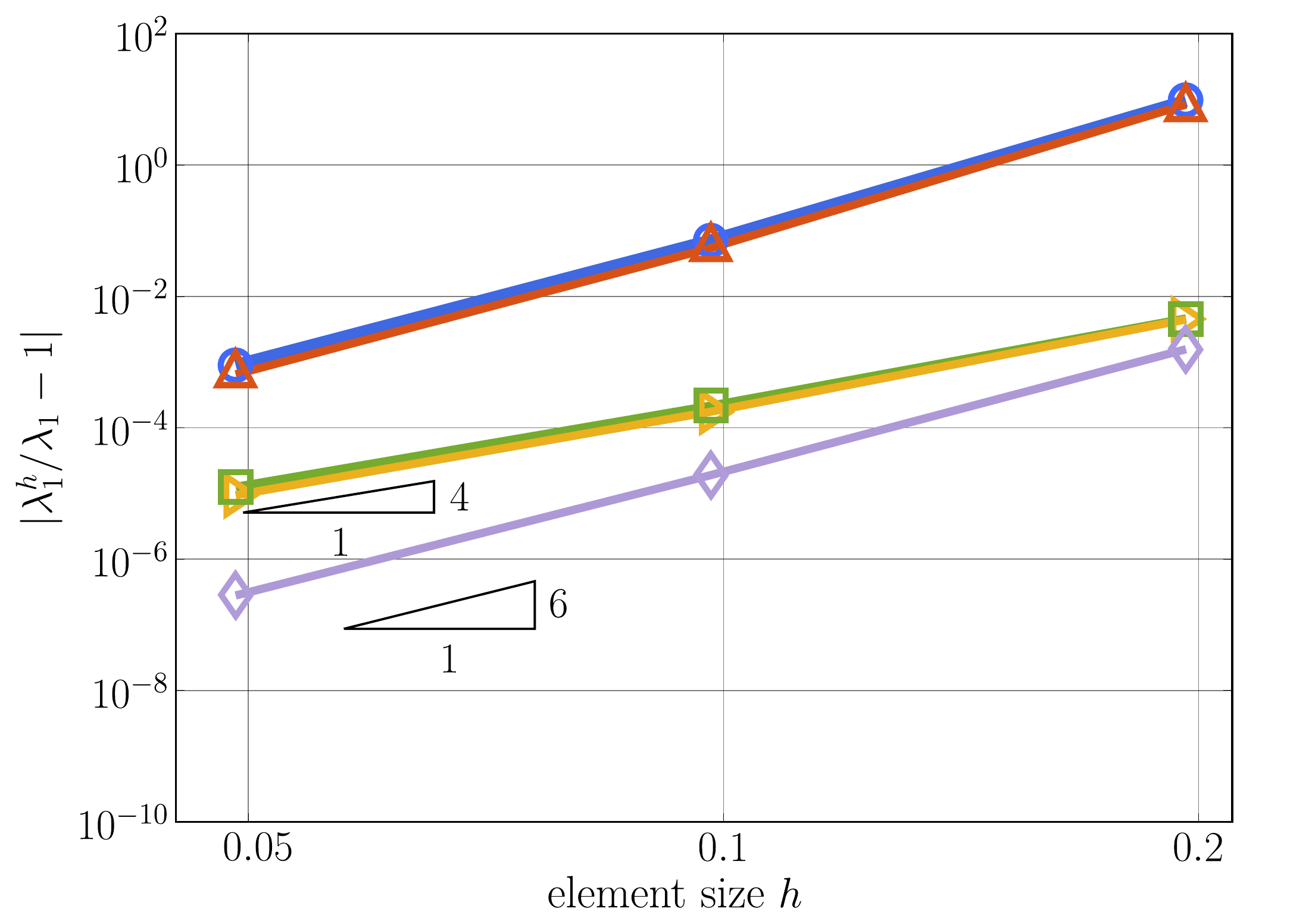}
	\footnotesize{(b) $p = 3$} \\
	\end{minipage}
	\begin{minipage}[b]{0.49\textwidth}
	\centering
	\vspace{0.5cm}
	\includegraphics[width=1.0\textwidth]{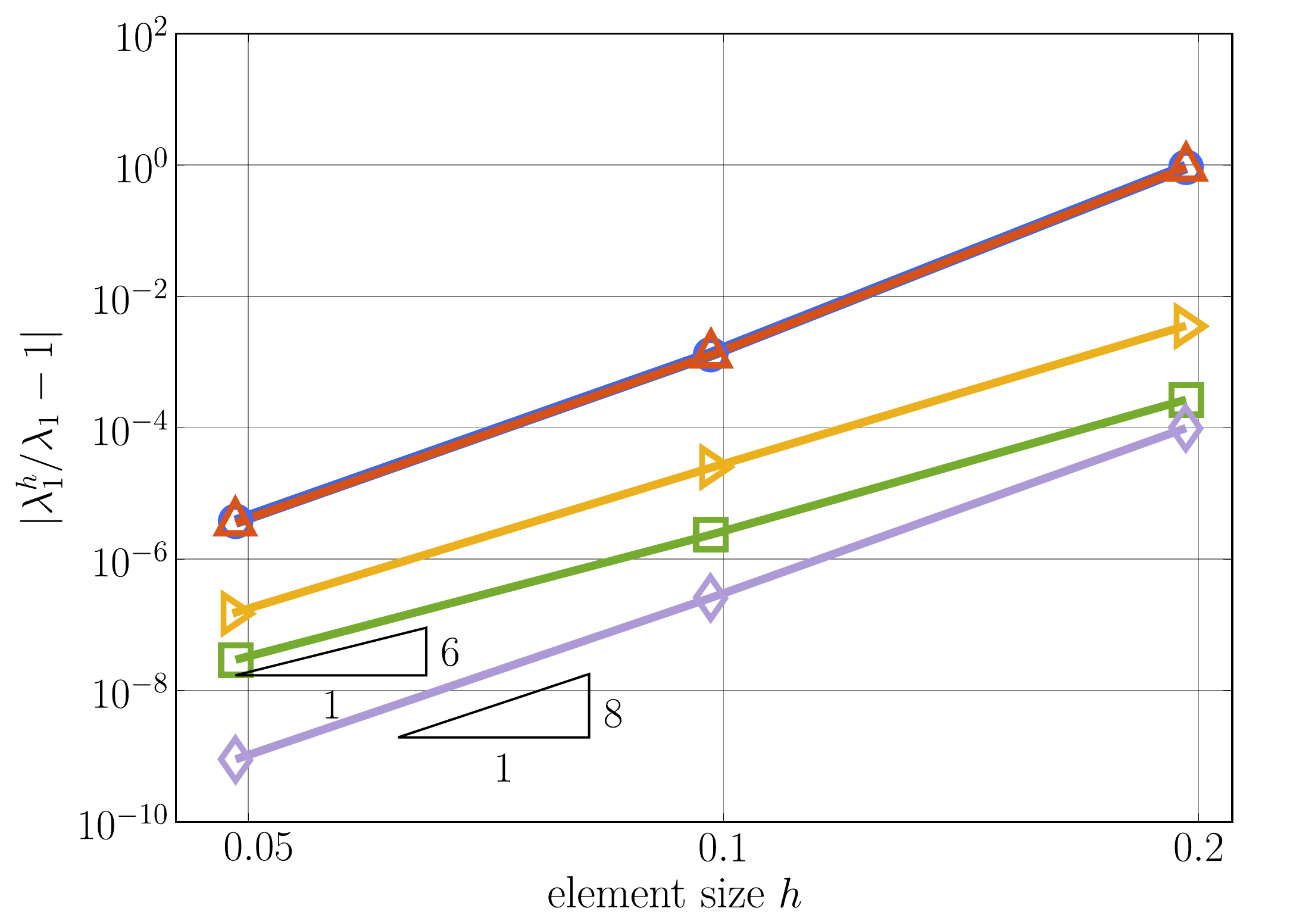}
	\footnotesize{(c) $p = 4$}
	\end{minipage}
	\begin{minipage}[b]{0.49\textwidth}
	\centering
	\vspace{0.5cm}
	\includegraphics[width=1.0\textwidth]{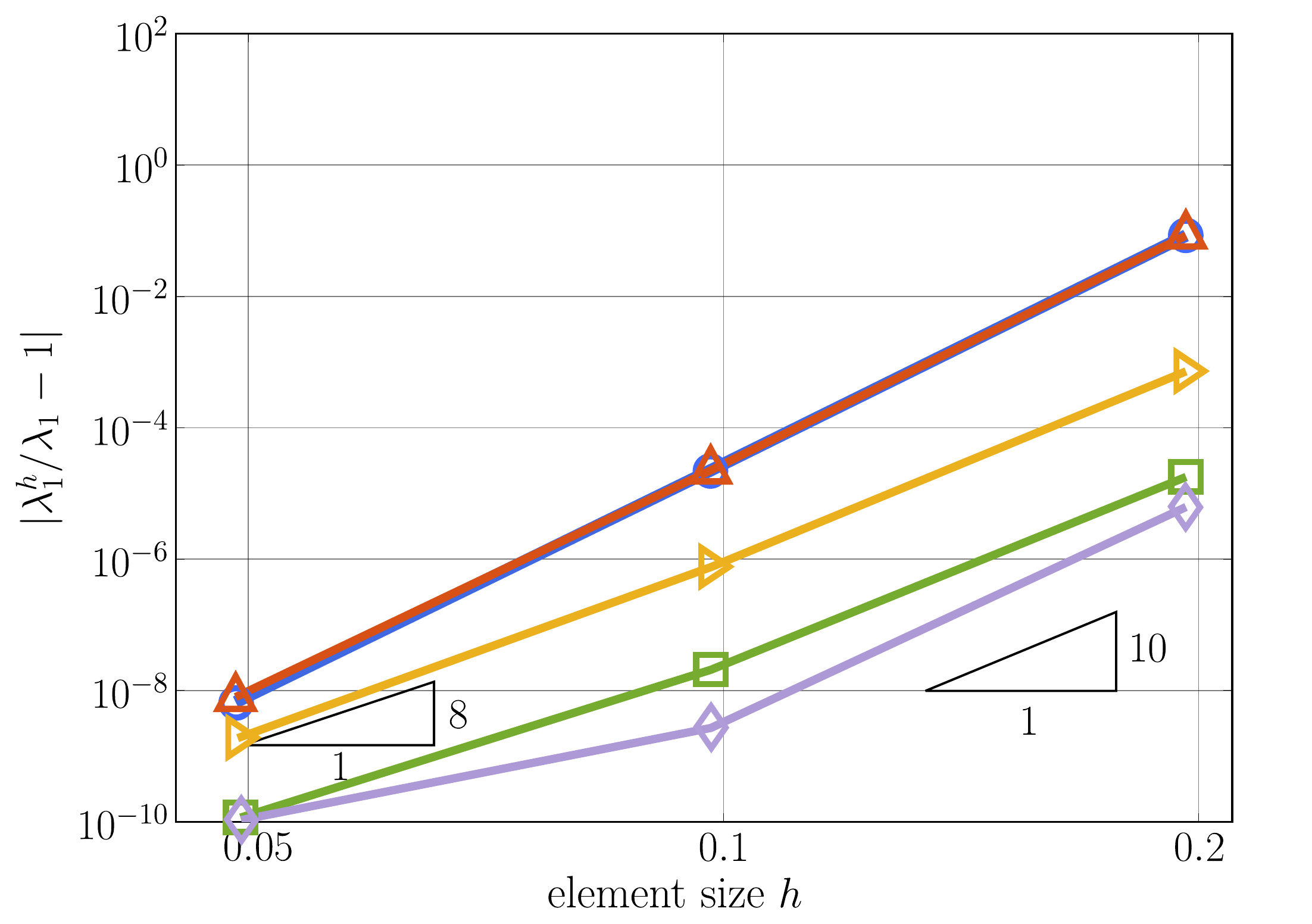}
	\footnotesize{(d) $p = 5$} \\
	\end{minipage}
	\begin{center}
		\vspace{0.2cm}
		\def\svgwidth{0.8\textwidth}
\begingroup%
  \makeatletter%
  \providecommand\color[2][]{%
    \errmessage{(Inkscape) Color is used for the text in Inkscape, but the package 'color.sty' is not loaded}%
    \renewcommand\color[2][]{}%
  }%
  \providecommand\transparent[1]{%
    \errmessage{(Inkscape) Transparency is used (non-zero) for the text in Inkscape, but the package 'transparent.sty' is not loaded}%
    \renewcommand\transparent[1]{}%
  }%
  \providecommand\rotatebox[2]{#2}%
  \newcommand*\fsize{\dimexpr\f@size pt\relax}%
  \newcommand*\lineheight[1]{\fontsize{\fsize}{#1\fsize}\selectfont}%
  \ifx\svgwidth\undefined%
    \setlength{\unitlength}{822.04724409bp}%
    \ifx\svgscale\undefined%
      \relax%
    \else%
      \setlength{\unitlength}{\unitlength * \real{\svgscale}}%
    \fi%
  \else%
    \setlength{\unitlength}{\svgwidth}%
  \fi%
  \global\let\svgwidth\undefined%
  \global\let\svgscale\undefined%
  \makeatother%
  \begin{picture}(1,0.01724138)%
    \lineheight{1}%
    \setlength\tabcolsep{0pt}%
    \put(0.03466904,0.00245947){\color[rgb]{0,0,0}\makebox(0,0)[lt]{\lineheight{1.25}\smash{\begin{tabular}[t]{l}\scriptsize{\textcolor{blue1}{full integration}}\end{tabular}}}}%
    \put(0.55459263,0.00228257){\color[rgb]{0,0,0}\makebox(0,0)[lt]{\lineheight{1.25}\smash{\begin{tabular}[t]{l}\scriptsize{\textcolor{green1}{B-bar}}\end{tabular}}}}%
    \put(0.2718277,0.00242293){\color[rgb]{0,0,0}\makebox(0,0)[lt]{\lineheight{1.25}\smash{\begin{tabular}[t]{l}\scriptsize{\textcolor{red1}{reduced integration}}\end{tabular}}}}%
    \put(0.70154869,0.00244378){\color[rgb]{0,0,0}\makebox(0,0)[lt]{\lineheight{1.25}\smash{\begin{tabular}[t]{l}\scriptsize{\textcolor{purple1}{Hellinger-Reissner}}\end{tabular}}}}%
    \put(0.97715175,0.00253347){\color[rgb]{0,0,0}\makebox(0,0)[lt]{\lineheight{1.25}\smash{\begin{tabular}[t]{l}\scriptsize{\textcolor{orange1}{DSG}}\end{tabular}}}}%
    \put(0,0){\includegraphics[width=\unitlength,page=1]{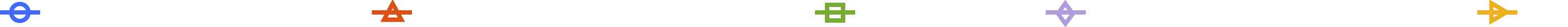}}%
  \end{picture}%
\endgroup%

	\end{center}
	\caption{Convergence of the relative error in the fifth transverse eigenvalue, obtained with different formulations and quadratic, cubic, quartic and quintic B-spline basis functions, for ``large'' slenderness ratio $R/t = 2000/3$.}
	\label{eig_conv}
\end{figure}

Due to its importance for the approximation power of the basis (see Section~\ref{sec:lowmode}), we then focus on the accuracy of the lower part of the spectrum. We observe in Fig.~\ref{fig:standard_p}b that $p$-refinement in the standard formulation continuously improves the accuracy of the lowest modes. A comparison with the results of the locking-free B-bar and Hellinger-Reissner formulations plotted in Figs.~\ref{fig:Bbar_p}b and \ref{fig:HR_p}b, however, clearly demonstrates that the negative impact of membrane locking persists with increasing $p$ in the standard formulation. For instance, for $p=3$, the eigenvalue error level of the lowest modes obtained with the B-bar method is three orders of magnitude smaller, and for the Hellinger-Reissner method even five orders of magnitude smaller, than the one obtained with the standard formulation. We note that for the B-bar and Hellinger-Reissner methods at $p=5$, the eigenvalue solver hits the level of machine accuracy, preventing the further decrease of the eigenvalue error of the lowest mode.

Our observations confirm that $p$-refinement in a standard displacement-based finite element formulation reduces the effect of membrane locking with respect to a low-order locking-prone discretization. A comparison with the true approximation power of the higher-order basis obtained in a locking-free formulation, however, clearly shows that membrane locking continues to heavily affect the accuracy of the standard formulation at high polynomial degrees. 
Therefore, we conclude that $p$-refinement is not an effective way to mitigate the effect of locking, at least not for the curved Euler-Bernoulli beam model.
In addition, our results indicate that the divergence of the higher transverse modes with increasing $p$ is another negative effect of membrane locking that, to our knowledge, has not been reported before. We note that inaccurate and divergent high modes can have significant negative effects in explicit dynamics and nonlinear analysis. 

\begin{figure}[t!]
	\centering
	\begin{minipage}[b]{0.49\textwidth}
	\centering
	\includegraphics[width=1.0\textwidth]{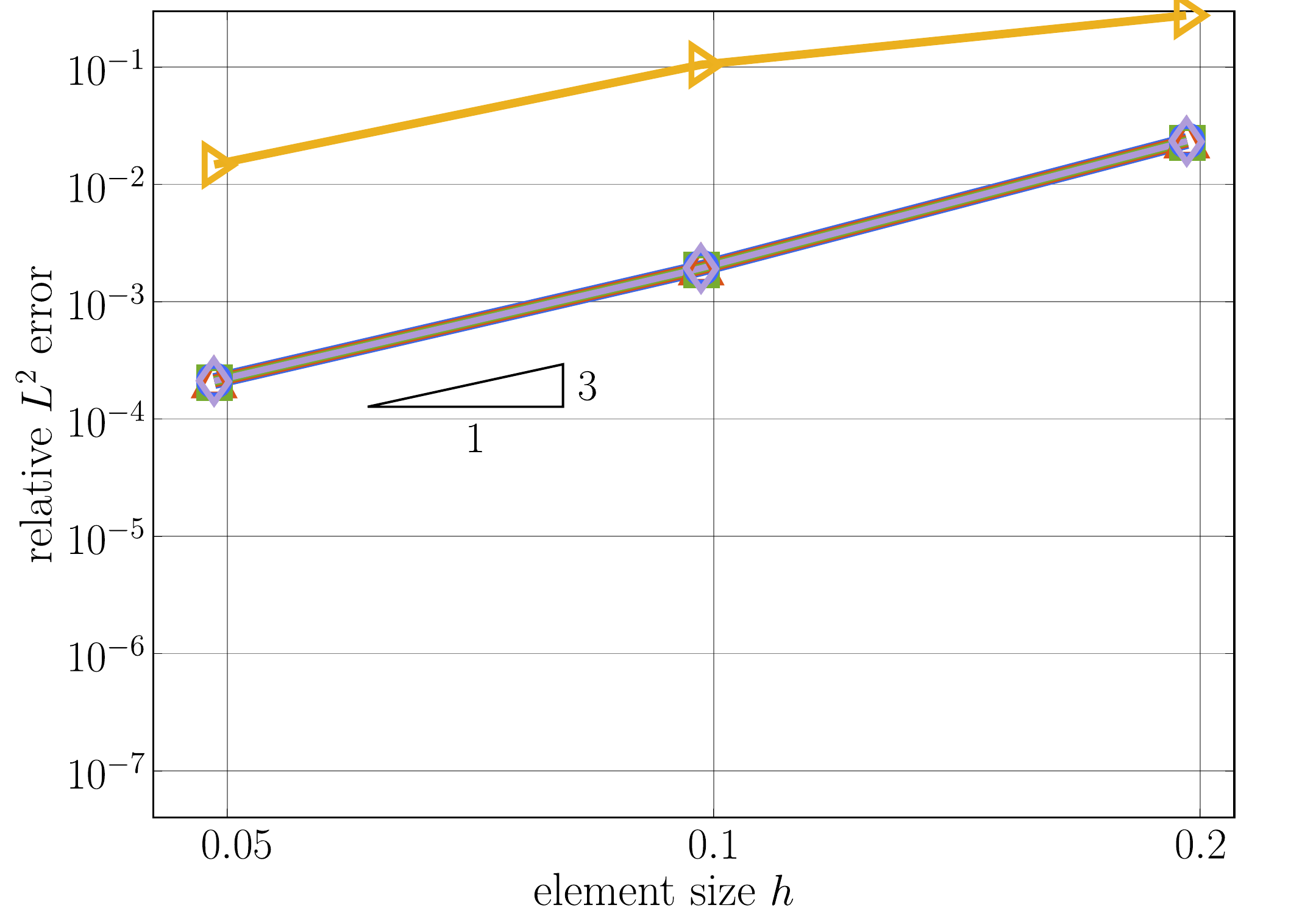}
	\footnotesize{(a) $p = 2$}
	\end{minipage}
	\begin{minipage}[b]{0.49\textwidth}
	\centering
	\includegraphics[width=1.0\textwidth]{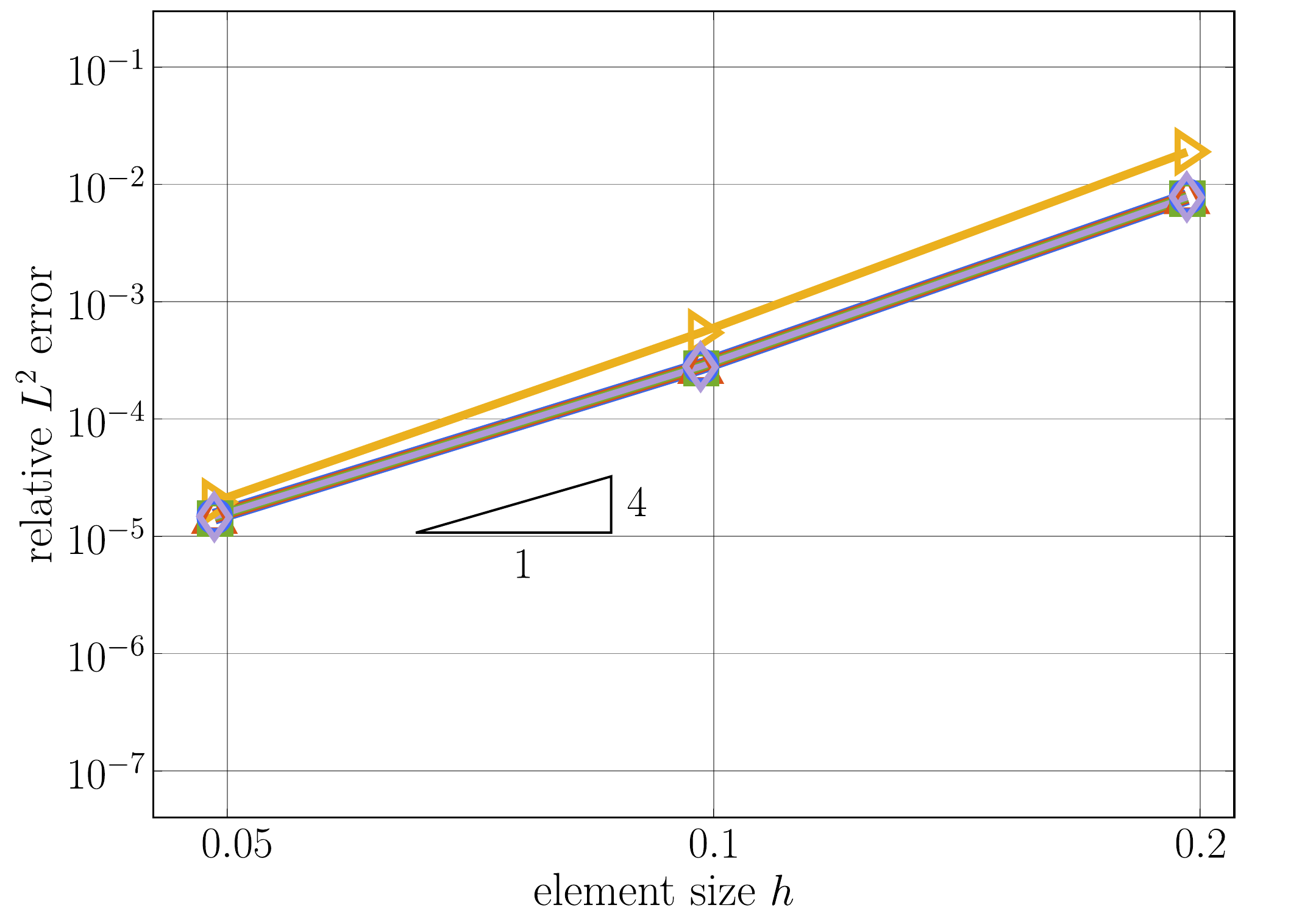}
	\footnotesize{(b) $p = 3$} \\
	\end{minipage}
	\begin{minipage}[b]{0.49\textwidth}
	\centering
	\vspace{0.5cm}
	\includegraphics[width=1.0\textwidth]{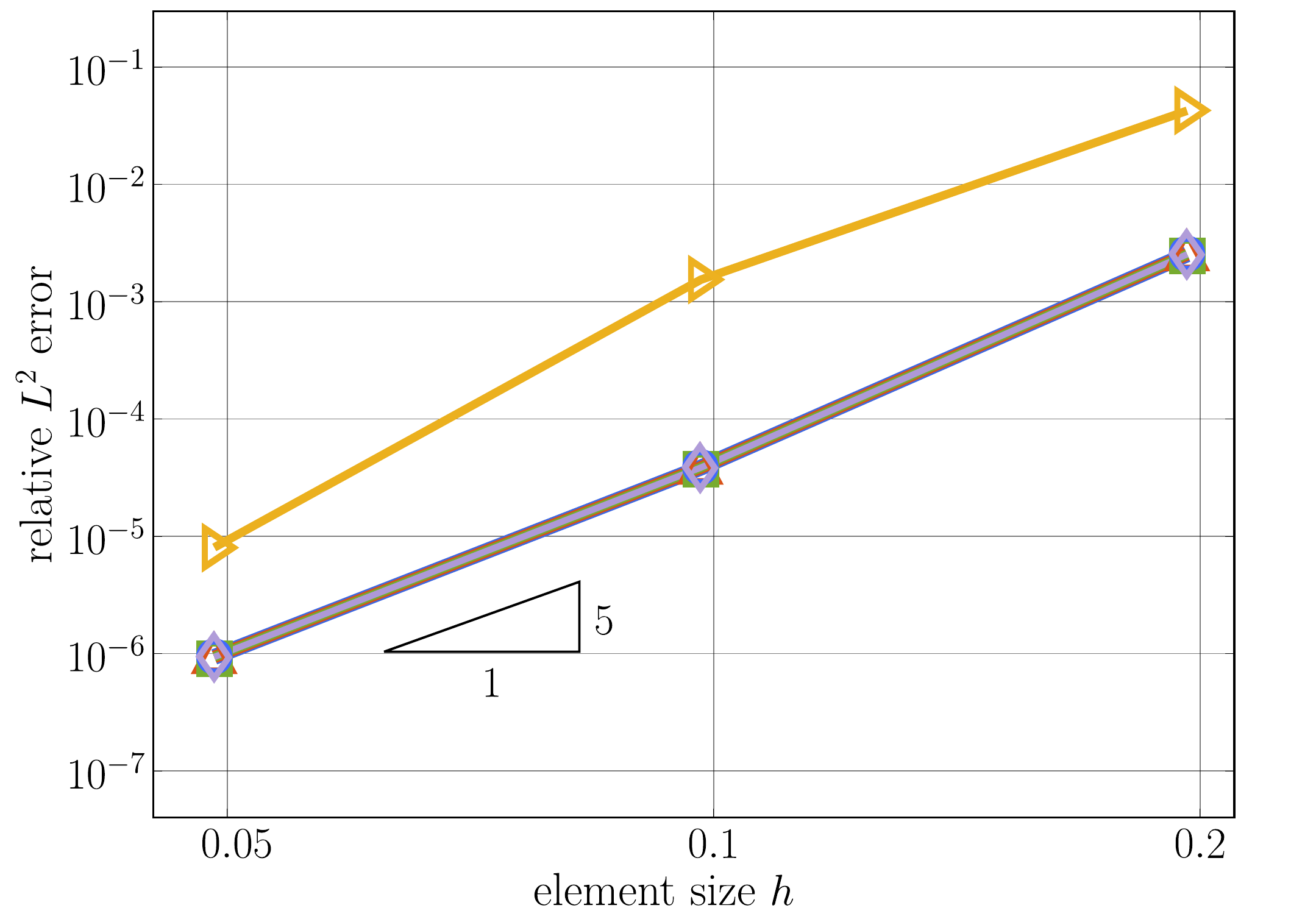}
	\footnotesize{(c) $p = 4$}
	\end{minipage}
	\begin{minipage}[b]{0.49\textwidth}
	\centering
	\vspace{0.5cm}
	\includegraphics[width=1.0\textwidth]{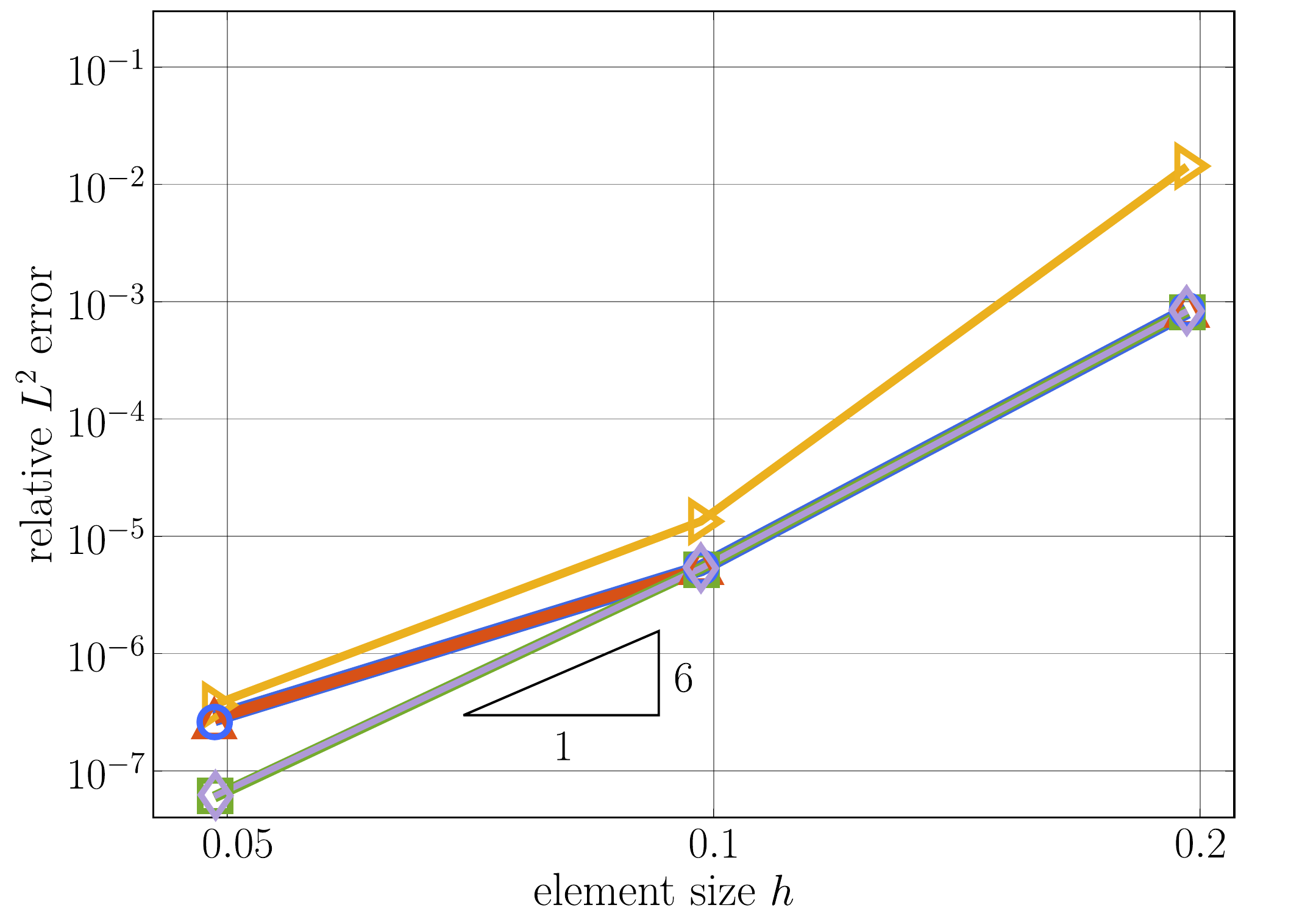}
	\footnotesize{(d) $p = 5$} \\
	\end{minipage}
	\begin{center}
		\vspace{0.2cm}
		\def\svgwidth{0.8\textwidth}
		
	\end{center}
	\caption{Convergence of the relative error in the fifth transverse eigenmode, obtained with different formulations and quadratic, cubic, quartic and quintic B-spline basis functions, for ``large'' slenderness ratio $R/t = 2000/3$.}
	\label{mod_conv}
\end{figure}

\subsection{Convergence of the lowest eigenvalues and mode shapes}

As outlined in Section~\ref{sec:lowmode}, the accuracy of the lowest eigenvalues and mode shapes directly relate to the accuracy of the approximation that can be achieved with a specific finite element discretization. 
It is therefore worthwhile to take a closer look at the accuracy and rate of convergence of the lower eigenvalues and mode shapes that are obtained with the different formulations. 
Figures~\ref{eig_conv} and \ref{mod_conv} plot the convergence of the relative error of the fifth transverse eigenvalue and the convergence of the relative $L^2$-norm error of the fifth transverse eigenmode obtained with 32, 64 and 128 B\'ezier elements and polynomial degrees $p=2$ through 5.

For the eigenvalue error, an eigenvalue problem with fourth-order differential operators achieves optimal rates of convergence of $\mathcal{O}(2(p-1))$, while an eigenvalue problem with second-order differential operators achieves optimal rates of $\mathcal{O}(2p)$ \cite{Cottrell2006,weeger2013isogeometric}. We observe in Fig.~\ref{eig_conv} that the standard formulation with full and reduced integration exhibits a significantly increased level of eigenvalue error. In particular, we can see that the error gap to the locking-free B-bar formulation decreases with each mesh refinement step, but does not decrease when the polynomial degree is increased on a fixed mesh. The locking-free B-bar and DSG methods based on a displacement-based formulation achieve optimal rates of convergence for all polynomial degrees. We observe that the eigenvalue error of the DSG method for $p=4$ and $p=5$ is slightly larger than the one for the B-bar method. The Hellinger-Reissner formulation also achieves optimal rates, which are consistently higher than the ones for the B-bar and DSG methods due to its mixed-method formulation. This confirms the increased accuracy of the eigenvalues obtained with the Hellinger-Reissner formulation that we observed in many of the inset figures of the previous plots. 

For the $L^2$-norm mode error, the optimal convergence is always $\mathcal{O}(p+1)$. We observe in Fig.~\ref{mod_conv} that all methods with the exception of the DSG method achieves practically the accuracy in the mode shapes, indicated by indistinguishable mode errors that converge optimally. The mode error of the DSG method, however, is significantly larger.

\section{Summary and conclusions}\label{sec:conclusion}

In this paper, we have taken first steps towards establishing spectral analysis as a tool for understanding and assessing locking phenomena in finite element formulations and comparing their effectivity with respect to unlocking. We proposed to ``measure'' locking (or unlocking) from a spectral analysis viewpoint as follows. For the finite element formulation in question, eigenvalue and mode errors are computed on a coarse discretization and corresponding asymptotic eigenvalue and mode errors are computed with an ``overkill'' discretization. The eigenvalue and mode errors from the coarse mesh are related to the corresponding asymptotic eigenvalue and mode errors by plotting both sets with respect to the normalized mode number $n/N$, where $N$ denotes the total number of modes in each discretization. The finite element formulation is locking-free, if the corresponding spectral error curves are matching up irrespective of the mesh size, and the finite element formulation is locking-prone, if the corresponding spectral error curves are different, implying that the spectral error curve changes with mesh refinement.

To illustrate the validity and significance of spectral analysis in the context of assessing locking, we employed the example of a circular ring discretized with curved Euler-Bernoulli beam elements, which are susceptible to membrane locking. We showed that for the Euler-Bernoulli circular ring, membrane locking heavily affects the accuracy of the eigenvalues of the transverse modes, while the transverse mode shapes and both the eigenvalues and mode shapes of the circumferential modes do not lock. We assessed and compared the effectivity of the standard displacement-based formulation with full and selective reduced integration as well as three representative locking-free formulations (B-bar method, DSG method, Hellinger-Reissner formulation) in terms of their accuracy in the eigenvalues and eigenmodes. Our study showed that spectral analysis can rigorously characterize membrane locking. 
With respect to mitigating membrane locking in curved Euler-Bernoulli beams, we summarize the essential results of our study in Table \ref{tab:compare_formulations}.

\begin{table}[ht]
\begin{tabularx}{\linewidth}{| >{\raggedright\arraybackslash\hsize=.4\hsize} X | >{\centering\arraybackslash\hsize=.14\hsize}X | >{\centering\arraybackslash\hsize=.14\hsize}X | >{\centering\arraybackslash\hsize=.1\hsize}X | >{\centering\arraybackslash\hsize=.1\hsize}X | >{\centering\arraybackslash\hsize=.12\hsize}X | }
    \hline
       &  \multicolumn{2}{ c |}{\textbf{Standard formulation}} & \multicolumn{3}{ c |}{\textbf{Locking-free formulation}} \\\cline{2-6}
    & Full integration & Reduced integration & B-bar & DSG & Hellinger-Reissner \\
    \hline \hline
    Locking-free on coarse meshes (accuracy low transverse modes) & {\color{red}\ding{55}} &  {\color{red}\ding{55}} & {\color{Green}\ding{51}} &  {\color{Green}\ding{51}} &  {\color{Green}\ding{51}} \\
    \hline 
    Locking-free with increasing $p$ (accuracy low transverse modes) & {\color{red}\ding{55}} &  {\color{red}\ding{55}} &  {\color{Green}\ding{51}} &  {\color{Green}\ding{51}}& {\color{Green}\ding{51}} \\
    \hline 
    Upper transverse modes converge with increasing $p$ &{\color{red}\ding{55}} & {\color{red}\ding{55}} & {\color{Green}\ding{51}} & {\color{Green}\ding{51}} &{\color{red}\ding{55}} \\
    \hline 
    No negative effect on accuracy of circumferential modes & {\color{Green}\ding{51}} & {\color{Green}\ding{51}} & {\color{Green}\ding{51}} & {\color{red}\ding{55}} &{\color{Green}\ding{51}} \\
    \hline 
   Convergence rate $\mathcal{O}(p+1)$ for quadratic FE approximations & {\color{red}\ding{55}} & {\color{red}\ding{55}} & {\color{red}\ding{55}} & {\color{red}\ding{55}} & {\color{Green}\ding{51}} \\
    \hline 
    No additional cost, e.g., due to static condensation or projection & {\color{Green}\ding{51}} &{\color{Green}\ding{51}} & {\color{red}\ding{55}} & {\color{red}\ding{55}} & {\color{red}\ding{55}} \\
    \hline 
\end{tabularx}
\caption{Summary of the comparative spectral analysis study for the Euler-Bernoulli circular ring problem.}
    \label{tab:compare_formulations}
\end{table}

Our spectral analysis results illustrate that the standard formulation with full integration is severely affected by membrane locking and does not enable efficient and accurate finite element solutions, even when the mesh is heavily refined. The standard formulation with selective reduced integration removes membrane locking for finer meshes for quadratic discretizations, but do not remove membrane locking on coarse meshes and for polynomial degrees larger than quadratics. 

The B-bar, DSG and Hellinger-Reissner methods all enable effective locking-free finite element formulations for curved Euler-Bernoulli beams, leading to accurate results on coarse meshes. Due to its mixed-method character, the Hellinger-Reissner formulation 
hits a sweet spot for quadratic basis functions, since it converges with $\mathcal{O}(3)$ in the $L^2$ displacement norm unlike the purely displacement-based B-bar and DSG formulations that achieve only $\mathcal{O}(2)$ for quadratic basis functions. 
For polynomial degrees larger than two, all methods achieve the same optimal convergence rates, so that practically, all locking-free formulations achieve the same accuracy. 

For the DSG method, we observed an increased level of error across large spectrum parts for the transverse mode shapes, the circumferential mode shapes and the eigenvalue error of the circumferential modes. When we refined the mesh or increased the polynomial degree $p$, this issue only improved for the transverse mode error, but persisted for the circumferential eigenvalue and mode errors. We hypothesize that the DSG formulation itself is responsible for this issue, since the accuracy with respect to the standard formulation decreases in parts of the spectrum. As membrane unlocking is associated primarily with the proper behavior in the eigenvalues of the transverse modes, this issue seems not to affect the unlocking capability of the DSG approach.

Classical $p$-refinement (in our study we use splines of maximum smoothness), where the polynomial degree is driven beyond cubics on a fixed coarse mesh, reduces the effect of membrane locking with respect to a low-order locking-prone discretization. Membrane locking, however, continues to heavily affect the accuracy of the standard formulation with respect to a locking-free formulation at high polynomial degrees. Therefore, $p$-refinement by itself is not an effective way to mitigate the effect of locking. In addition, we observed that the higher transverse modes diverge with increasing $p$ as a result of membrane locking. Divergent high modes deteriorate the conditioning of the system matrix, and can seriously affect the approximation accuracy and robustness in structural dynamics \cite{Hughes2000}. In contrast, we showed that locking-free formulations unlock the full potential of higher-order accurate discretizations. We observed that the low modes obtained with the locking-free formulations consistently were several orders of magnitude more accurate than the ones obtained with standard formulations, also at high polynomial degrees beyond cubics, and the high modes converged for the B-bar formulation and did not diverge for the Hellinger-Reissner formulation.

In summary, the results presented in this paper demonstrate the potential of spectral analysis as a tool to help assess locking phenomena in finite element formulations. In the future, we plan to extend the approach and corroborate its potential for finite element formulations of more complex structural models, in particular Kirchhoff-Love and Reissner-Mindlin shells.

\section*{Acknowledgments}
The authors gratefully acknowledge financial support from the German Research Foundation (Deutsche Forschungsgemeinschaft) through the DFG Emmy Noether Grant SCH 1249/2-1.

\appendix

\section{Analytical solution of the freely vibrating circular ring}\label{sec:analytic_eigen_ring}

Soedel solved the eigenvalue problem of a circular Euler-Bernoulli ring analytically \cite[p.82-85]{Soedel2004} using the equations of motion in curvilinear coordinates. He assumed the mode shapes of a free floating closed ring, that are
	\begin{align}\label{eq:ring_mode_shape}
		& v_n(\theta) = A_{1n} \sin(n\theta), \\
		& w_n(\theta) = A_{2n} \cos(n\theta),
	\end{align}
where $n \in \mathbb{N}$ and $\theta$ are the mode number and the angular coordinate, respectively. The $n^{\text{th}}$ pair of the analytical eigenvalues are then
\begin{align}\label{eq:exact_eigen_ring}
\lambda_{1n} = \omega_{1n}^2 = \frac{k_{1n}}{\rho A} \text{, } \quad \lambda_{2n} = \omega_{2n}^2 = \frac{k_{2n}}{\rho A}
\end{align}
where $\rho$ and $A$ are the density and the cross section area of the ring, and the parameters $k_{in}$ are defined as
	\begin{align}
		& k_{1n} = \frac{C - B}{2 R^4} \text{, } \quad k_{2n} = \frac{C + B}{2 R^4}, \;\;\; \text{with}\\
		& C = (\text{EA} \, R^2 + \text{EI} \, n^2) (n^2 + 1), \nonumber \\
		& B = \sqrt{ (\text{EA}^2 \, R^4 + \text{EI}^2 \, n^4) (n^2 + 1)^2 + 2 \, \text{EA} \, R^2 \, \text{EI} \, n^2 (6n^2 - n^4 - 1) } \nonumber 
	\end{align}
$\text{EA}$ and $\text{EI}$ denote the membrane and bending stiffness, and $R$ the radius of the ring. The two eigenvalues $\lambda_{1n}$ and $\lambda_{2n}$ correspond to different values of the relative amplitude between the corresponding circumferential and radial modes
\begin{align}\label{eq:relative_amplitude_analytic}
	r_{in} = \frac{A_{1n}}{A_{2n}} = \frac{\frac{\text{EA}}{R^2} n + \frac{\text{EI}}{R^4} n^3}{ \rho A \omega^2_{in} -\frac{\text{EA}}{R^2} n^2 - \frac{\text{EI}}{R^4} n^2 }, \quad i=1,2.
\end{align}
The analytical eigenvalues and relative amplitudes of the first twenty modes of the Euler-Bernoulli circular ring used in this study are listed in Table \ref{tab:analytic_eigen_ring}. For each mode number $n$, one eigenvalue corresponds to the transverse-deflection-dominating modes, i.e. $A_{1n} \leq A_{2n} \, (|r_{in}| \leq 1.0)$, and one corresponds to the circumferential-deflection-dominating modes $(|r_{in}| \geq 1.0)$ (see Table \ref{tab:analytic_eigen_ring} and also \cite[p.82-85]{Soedel2004}).

\begin{table}[ht]
\centering
{\footnotesize
\begin{tabular}{ c c c c c }
	 \hline
	 \multirow{2}{*}{$n$} & \multicolumn{2}{c}{eigenvalue pair} & \multicolumn{2}{c}{amplitude ratio} \\\cline{2-5}
	  	  & $\lambda_{1n}$ 		  & $\lambda_{2n}$ 	  	  & $r_{1n}$ 			    & $r_{2n}$ \\
	 \hline
	 $0$  & $0$ 				  & $1.200000000000000\cdot10^6$ & $0$ 					& $0$ \\
	 $1$  & $0$ 				  & $2.400000450000000\cdot10^6$ & $-1$ 					& $1$ \\
	 $2$  & $1.619999222176224\cdot10^0$ & $6.000002880000779\cdot10^6$ & $-5.000004499999865\cdot10^{-1}$ & $1.999998200001673\cdot10^0$ \\
	 $3$  & $1.295999212658217\cdot10^1$ & $1.200000729000787\cdot10^7$ & $-3.333342333340016\cdot10^{-1}$ & $2.999991900015852\cdot10^0$ \\
	 $4$  & $4.764702716878825\cdot10^1$ & $2.040001355297283\cdot10^7$ & $-2.500013235320360\cdot10^{-1}$ & $3.999978823599535\cdot10^0$ \\
	 $5$  & $1.246152982048443\cdot10^2$ & $3.120002163470180\cdot10^7$ & $-2.000017307755196\cdot10^{-1}$ & $4.999956730986444\cdot10^0$ \\
	 $6$  & $2.681754852697320\cdot10^2$ & $4.440003152451474\cdot10^7$ & $-1.666687950571240\cdot10^{-1}$ & $5.999923378921967\cdot10^0$ \\
	 $7$  & $5.080316340600650\cdot10^2$ & $6.000004321836594\cdot10^7$ & $-1.428596628775543\cdot10^{-1}$ & $6.999876521178022\cdot10^0$ \\	
	 $8$  & $8.792855145050755\cdot10^2$ & $7.800005671448547\cdot10^7$ & $-1.250029077240125\cdot10^{-1}$ & $7.999813909991993\cdot10^0$ \\
	 $9$  & $1.422437983402840\cdot10^3$ & $9.840007201201658\cdot10^7$ & $-1.111144038404168\cdot10^{-1}$ & $8.999733296829945\cdot10^0$ \\
	 $10$ & $2.183389483803921\cdot10^3$ & $1.212000891105162\cdot10^8$ & $-1.000036758074283\cdot10^{-1}$ & $9.999632432768221\cdot10^0$ \\
	 $11$ & $3.213440252163006\cdot10^3$ & $1.464001080097479\cdot10^8$ & $-9.091314837369126\cdot10^{-2}$ & $1.099950906869468\cdot10^1$ \\
	 $12$ & $4.569290413144246\cdot10^3$ & $1.740001287095869\cdot10^8$ & $-8.333777137925940\cdot10^{-2}$ & $1.199936095541999\cdot10^1$ \\
	 $13$ & $6.313040026479182\cdot10^3$ & $2.040001512099735\cdot10^8$ & $-7.692789471743249\cdot10^{-2}$ & $1.299918584374570\cdot10^1$ \\
	 $14$ & $8.512189111741010\cdot10^3$ & $2.364001755108883\cdot10^8$ & $-7.143376831426732\cdot10^{-2}$ & $1.399898148450718\cdot10^1$ \\
	 $15$ & $1.123963766205331\cdot10^4$ & $2.712002016123379\cdot10^8$ & $-6.667224211689353\cdot10^{-2}$ & $1.499874562860407\cdot10^1$ \\
	 $16$ & $1.457368565216181\cdot10^4$ & $3.084002295143479\cdot10^8$ & $-6.250595358651052\cdot10^{-2}$ & $1.599847602702302\cdot10^1$ \\
	 $17$ & $1.859803304334188\cdot10^4$ & $3.480002592169566\cdot10^8$ & $-5.882986078224645\cdot10^{-2}$ & $1.699817043085363\cdot10^1$ \\
	 $18$ & $2.340177978620280\cdot10^4$ & $3.900002907202137\cdot10^8$ & $-5.556226441714824\cdot10^{-2}$ & $1.799782659130378\cdot10^1$ \\
	 $19$ & $2.907942582313202\cdot10^4$ & $4.344003240241770\cdot10^8$ & $-5.263866505443284\cdot10^{-2}$ & $1.899744225971415\cdot10^1$ \\
	 $20$ & $3.573087108895835\cdot10^4$ & $4.812003591289111\cdot10^8$ & $-5.000746314489048\cdot10^{-2}$ & $1.999701518756431\cdot10^1$ \\
\end{tabular}}
\caption{The first twenty exact eigenvalue pairs of the Euler-Bernoulli circular free floating ring with a slenderness ratio $R/t = 2000/3$}
\label{tab:analytic_eigen_ring}
\end{table}

Another set of mode shapes exists,
\begin{align}
	& v_n(\theta) = A_{3n} \cos(n\theta), \label{analytical_mode2a} \\
	& w_n(\theta) = A_{4n} \sin(n\theta) \label{analytical_mode2b} \, ,
\end{align}
which results in exactly the same analytical eigenvalues. This explains repeated eigenvalues in numerical computations. In our study, we only considered the free-floating modes in \eqref{eq:ring_mode_shape}, which satisfy
\begin{align}\label{eq:free-floating_BCs_curv}
	v_n(\theta = 0) = 0, \quad w_{n,\theta}(\theta = 0) = 0.
\end{align}
These constraints are also built into the spline trialspaces in order to remove the arbitrary phase shift in the numerical mode shapes; allowing direct comparisons to be made between the discrete and analytical eigenmodes. Because the trialspaces are defined in the Cartesian frame a rotation is necessary, see \ref{sec:variational_form_cartesian}. The free-floating modes and their constraints in Cartesian coordinates are
\begin{align}
	& \eigenvec_x(\theta) = A_{2n} \cos(n\theta) \, \cos(\theta) - A_{1n} \sin(n\theta) \, \sin(\theta) \, , \\
	& \eigenvec_y(\theta) = A_{2n} \cos(n\theta) \, \sin(\theta) + A_{1n} \sin(n\theta) \, \cos(\theta)\, ,
\end{align}
and		 
\begin{align}\label{eq:free-floating_BCs_cartesian}
	\eigenvec_{x,\theta}(\theta=0) = 0 \text{, } \quad \eigenvec_y(\theta=0) = 0. 
\end{align}

\begin{algorithm}
\textbf{Input}:
	$r_{in}$ ($i=1,2$, $n = 1, 2, \ldots, N$) (see equation \eqref{eq:relative_amplitude_analytic}) \\
\textbf{Output}: $\text{transverse\textunderscore mode\textunderscore numbers\textunderscore lambda1, circumferential\textunderscore mode\textunderscore numbers\textunderscore lambda1,}$ \\
$\text{transverse\textunderscore mode\textunderscore numbers\textunderscore lambda2, circumferential\textunderscore mode\textunderscore numbers\textunderscore lambda2}$
\begin{algorithmic}[1]
	\State $\text{transverse\textunderscore mode\textunderscore numbers\textunderscore lambda1 = findall}(|r_{1n}| \leq 1)$
	\State $\text{circumferential\textunderscore mode\textunderscore numbers\textunderscore lambda1 = findall}(|r_{1n}| > 1)$
	\State
	\State $\text{transverse\textunderscore mode\textunderscore numbers\textunderscore lambda2 = findall}(|r_{2n}| < 1)$
	\State $\text{circumferential\textunderscore mode\textunderscore numbers\textunderscore lambda2 = findall}(|r_{2n}| \geq 1)$	
\caption{Identify mode type corresponding to $\lambda_1$ and $\lambda_2$}\label{alg:find_switching_mode_number}
\end{algorithmic}
\end{algorithm}

\begin{algorithm}
\textbf{Input}:
	$\lambda_i^h$,
	$\mat{\eigenvec}_i^h$ ($i=1,2, \ldots, 2N$)\\
\textbf{Output}: $\text{transverse\textunderscore mode\textunderscore numbers, circumferential\textunderscore mode\textunderscore numbers}$
\begin{algorithmic}[1]
	\For{$i$ in $1:2N$}
	\State $v^h_i, w^h_i = \text{rotate\textunderscore axes}(\mat{\eigenvec}_i^h) $
	\State $|r_i^h| = \sqrt{ \left( \int_0^{2\pi} (v^h_i)^2 \, R \, \mathrm d \theta \right) / \left( \int_0^{2\pi} (w^h_i)^2 \, R \, \mathrm d \theta \right) }$ 
	\If{$|r_i^h| = 1 \And \lambda_i^h = 0$}
	\State $\text{transverse\textunderscore mode\textunderscore numbers}[1] = i$ 
	\ElsIf{$|r_i^h| = 1 \And \lambda_i^h \neq 0$} 
	\State $\text{circumferential\textunderscore mode\textunderscore numbers}[1] = i$ 
	\ElsIf{$|r_i^h| < 1$}
	\State $\text{append(transverse\textunderscore mode\textunderscore numbers}, i)$
	\ElsIf{$|r_i^h| > 1$} 	
	\State $\text{append(circumferential\textunderscore mode\textunderscore numbers}, i)$
	\EndIf
	\EndFor
\caption{Sort discrete transverse and circumferential modes}\label{alg:sort_transverse_vs_circumferential_mode}
\end{algorithmic}
\end{algorithm}

\begin{algorithm}
\textbf{Input}:
	$\mat{\eigenvec_i^h}$ ($i=1,2, \ldots, 2N$),
	$\text{transverse\textunderscore mode\textunderscore numbers}$,
	$\text{circumferential\textunderscore mode\textunderscore numbers}$ \\
\textbf{Output}: $\text{transverse\textunderscore mode\textunderscore numbers, circumferential\textunderscore mode\textunderscore numbers}$
\begin{algorithmic}[1]
	\For{$\mat{n}$ in $(\text{transverse\textunderscore mode\textunderscore numbers, circumferential\textunderscore mode\textunderscore numbers})$}
	\For{$i$ in $\mat{n}$}
	\State $v^h_{i,0}, w^h_{i,0} = \text{rotate\textunderscore axes}(\mat{\eigenvec}_i^h(\theta=0)) $
	\If{$v^h_{i,0} \neq 1 \, || \, w^h_{i,0} \approx 0$} 
	\State $\mat{n} \setminus \{i\}$ 		\Comment{Remove index $i$ from current vector of mode numbers}
	\EndIf
	\EndFor
	\EndFor
\caption{Verify free-floating constraints}\label{alg:check_free-floating_BCs}
\end{algorithmic}
\end{algorithm}

\begin{algorithm}
\textbf{Input}:
$\mat{\eigenvec}^h_i$,
$\mat{\eigenvec}_i$ ($i=1,2, \ldots, 2N$),
$\text{transverse\textunderscore mode\textunderscore numbers}$,
$\text{circumferential\textunderscore mode\textunderscore numbers}$, \\
$\text{transverse\textunderscore mode\textunderscore numbers\textunderscore lambda1, circumferential\textunderscore mode\textunderscore numbers\textunderscore lambda1,}$ \\
$\text{transverse\textunderscore mode\textunderscore numbers\textunderscore lambda2, circumferential\textunderscore mode\textunderscore numbers\textunderscore lambda2}$ \\
\textbf{Output}: $\lambda_1$\textunderscore mode\textunderscore numbers, $\lambda_2$\textunderscore mode\textunderscore numbers
\begin{algorithmic}[1]
	\For{$i$ in transverse\textunderscore mode\textunderscore numbers}
		\For{$n$ in transverse\textunderscore mode\textunderscore numbers\textunderscore lambda1}
			\State $e_1[n] = \text{compute\textunderscore} L2 \text{\textunderscore error}(\mat{\eigenvec}^h_i, \mat{\eigenvec}_{n})$
		\EndFor
		\For{$n$ in transverse\textunderscore mode\textunderscore numbers\textunderscore lambda2}
			\State $e_2[n] = \text{compute\textunderscore} L2 \text{\textunderscore error}(\mat{\eigenvec}^h_i, \mat{\eigenvec}_{n})$
		\EndFor
		\If{$\text{minimal}(e_1) < \text{minimal}(e_2)$}
			\State $\text{append}(\lambda_1\text{\textunderscore mode\textunderscore numbers}, i \Rightarrow \text{argmin}(e_1))$
		\Else
			\State $\text{append}(\lambda_2\text{\textunderscore mode\textunderscore numbers}, i \Rightarrow \text{argmin}(e_2))$
		\EndIf
	\EndFor		
	\State $\text{repeat with circumferential\textunderscore mode\textunderscore numbers}$
\caption{Assign discrete to the correct analytical modes}\label{alg:mode_assigment_via_mode-L2-error}
\end{algorithmic}
\end{algorithm}

\section{Postprocessing of numerical eigenvalues and modes}\label{sec:postprocessing}

The numerical eigenvalues and modes obtained from a finite element discretization are not automatically ordered in the same way as the analytical solution. To correctly identify and assign the numerical solutions to the analytical reference, we compute the $L^2$-norm error in the mode shape of each discrete mode with respect to all analytical modes and assign pairs based on the smallest error. We first identify the transverse and circumferential modes, using the criterion of the relative amplitude (p. 82-85, \cite{Soedel2004}). We describe the identification scheme in Algorithms~\ref{alg:find_switching_mode_number} and \ref{alg:sort_transverse_vs_circumferential_mode}. Since we only consider free-floating modes in this paper, we verify whether each numerical mode satisfies \eqref{eq:free-floating_BCs_cartesian}, see Algorithm~\ref{alg:check_free-floating_BCs}. We then assign each numerical transverse and circumferential mode to the correct analytical counterpart, see Algorithm~\ref{alg:mode_assigment_via_mode-L2-error}. The last step is to arrange the numerical eigenvalues and modes in ascending order. Except for the mixed formulation, this ordering scheme results in ascending discrete eigenvalues.

\bibliographystyle{elsarticle-num}
\bibliography{elsarticle-template}

\end{document}